\documentclass[10pt]{report}

\usepackage{times}
\usepackage{epsf}
\usepackage[numbers,sort&compress]{natbib}
\usepackage{amsmath}
\usepackage{amsfonts}
\usepackage{amssymb}
\usepackage{bm}
\usepackage{bbm}
\usepackage{graphicx}
\usepackage{epsfig}
\usepackage{latexsym}
\usepackage{nicefrac}
\usepackage{xr}

\usepackage{fancyhdr}
\renewcommand{\chaptermark}[1]%
{\markboth{\MakeUppercase{\chaptername~\thechapter: #1\\[2ex]}}{}}

\def\corresponds{{\lower.2ex\hbox{=}}{\rm\kern-.75em^\triangle}}
\def\succsim{\succ\kern-.9em_\sim\kern.3em}
\def\precsim{\prec\kern-1em_\sim\kern.3em}
\def\slantfrac#1#2{\kern1em^{#1}\kern-.3em/\kern-.1em_{#2}}
\def\lfrac#1#2{{}^{#1\!}\kern-.0em/_{#2}}

\def\buildrel#1\under#2{\mathrel{\mathop{\kern0pt #2}\limits_{#1}}}

\def\tr{{\rm Tr}\,}
\def\d{{\rm d}}
\def\e{{\rm e}}
\def\ud{{\textstyle{1\over 2}}}

\begin{document}

\pagestyle{empty}

{\bf 

\Large

\vspace*{1.5in}

\flushright{\Large \bf \sf Multi--Instantons and Exact Results II:}
\flushright{\Large \bf \sf
Specific Cases, Higher--Order Effects, and Numerical Calculations}

\vspace{.5in}

\large
\flushright{Jean Zinn--Justin\\
{\em DAPNIA/CEA Saclay}\\
{\em F-91191 Gif-sur-Yvette, France and}\\
{\em Institut de Math\'{e}matiques}\\
{\em de Jussieu--Chevaleret,}\\
{\em Universit\'{e} de Paris VII, France}\\[2ex]
Ulrich D.~Jentschura \\
{\em University of Freiburg, Germany and} \\
{\em NIST Division 842: Atomic Physics,} \\
{\em Physics Laboratory, USA} }

}

\vfill

\newpage

\pagestyle{empty}

\vspace*{0.0cm}
\begin{center}
\begin{tabular}{c}
\hline
\rule[-3mm]{0mm}{12mm}
{\large \sf Multi--Instantons and Exact Results II:}\\
\rule[-5mm]{0mm}{12mm}
{\large \sf Specific Cases, Higher--Order Effects, and Numerical Calculations}\\
\hline
\end{tabular}
\end{center}
\vspace{0.0cm}
\begin{center}
Jean Zinn--Justin\\
\vspace{0.2cm}
\scriptsize
{\em DAPNIA/DSM$^{*,**}$\\
Commissariat \`{a} l'\'{E}nergie Atomique, Centre de Saclay,
F-91191 Gif-sur-Yvette, France and\\
Institut de Math\'{e}matiques
de Jussieu--Chevaleret, Universit\'{e} de Paris VII, France}
\end{center}

\normalsize
\begin{center}
Ulrich D. Jentschura\\
\vspace{0.2cm}
\scriptsize
{\em 
Physikalisches Institut der Universit\"{a}t Freiburg,\\
Hermann--Herder--Stra\ss{}e 3, 79104 Freiburg im Breisgau, Germany and\\
National Institute of Standards and Technology,
Gaithersburg, MD20899-8401, Maryland}
\end{center}
\vspace{0.3cm}
\begin{center}
\begin{minipage}{14.0cm}
{\underline{Abstract}}
In this second part of the treatment of
instantons  in quantum mechanics, the focus is on specific
calculations related to a number of quantum mechanical
potentials with degenerate minima. We calculate the leading multi-instanton
constributions to the partition function,
using the formalism introduced in the
first part of the treatise [J. Zinn-Justin and U. D. Jentschura,
e-print quant-ph/0501136]. The following potentials
are considered: (i) asymmetric potentials with degenerate
minima, (ii) the periodic cosine potential,
(iii) anharmonic oscillators with radial symmetry, and
(iv) a specific potential which bears an analogy with the
Fokker--Planck equation.
The latter potential has the peculiar property
that the perturbation series for the
ground-state energy vanishes to all orders and is thus
formally convergent (the ground-state energy, however, is
nonzero and positive).
For the potentials (ii), (iii), and~(iv), we calculate the perturbative
$B$-function as well as the instanton $A$-function
to fourth order in $g$.
We also consider the double-well potential in detail,
and present some higher-order analytic as well as
numerical calculations to verify explicitly the related conjectures
up to the order of three instantons.
Strategies analogous to those outlined here
could result in new conjectures for problems where our present
understanding is more limited.
\end{minipage}
\end{center}

\vspace{0.6cm}

Asymptotic problems and properties\\
\vfill
\begin{center}
\begin{minipage}{15cm}
\begin{center}
\hrule
{\bf \scriptsize
\noindent electronic mail: zinn@spht.saclay.cea.fr.\\[2ex]
\noindent ${}^{*}$D\'{e}partment d'astrophysique, de physique des particules,
de physique nucl\'{e}aire et de l'instrumentation associ\'{e}e\\
\noindent ${}^{**}$ Laboratoire de la Direction des
Sciences de la Mati\`ere du Commissariat \`{a} l'Energie Atomique}
\end{center}
\end{minipage}
\end{center}

\newpage

\tableofcontents

\newpage

\clearpage\fancyhead[R]{\normalsize \rightmark}
\pagestyle{fancy}

\setcounter{chapter}{4}

We continue here the investigations~\cite{ZJJe2004I}
on multi-instantons and exact results.
Sections, equations, tables and figures are numbered
consecutively after those of~\cite{ZJJe2004I},
while the numbering of bibliographic items is independent
of~\cite{ZJJe2004I}. 
For relevant definitions and conventions, the reader is referred
to chapters~\ref{intro} and~\ref{BSqf}, which are
contained in the the first part of the 
treatise~\cite{ZJJe2004I}. We will often refer to the 
conjectures summarized in
chapter~\ref{sSummary}, and start the discussion 
here with chapter~\ref{ssninstg}. Many of the instanton
calculations rely on generalizations of the ideas and
methods introduced in chapter~\ref{ssninstdw} 
(which is again part of the first paper~\cite{ZJJe2004I}
of this series).

%
% General potentials with degenerate minima 
%
\chapter{Instantons in General Potentials with Degenerate Minima}
\label{ssninstg}

%
% Orientation
%
\section{Orientation}
\label{sOrientationGeneral}

We now consider a general analytic potential possessing two degenerate
minima located at the origin and another point $ q_{0}>0$:
\begin{subequations}
\label{edegenminima}
\begin{eqnarray}
V(q) & = & \ud q^2 + {\mathcal O}\left(q^3 \right)\,, 
\\
V(q) & = & \ud \, \omega^2 \, \left(q-q_{0} \right)^2
+ {\mathcal O}\left( \left(q-q_{0} \right)^3 \right)\,.
\end{eqnarray}
\end{subequations}
For definiteness we assume $\omega > 1$.
 
In such a situation the classical equations of motion have instanton solutions
connecting the two minima of the potential. 
However, there is no ground state degeneracy beyond the classical limit. 
Therefore, the one-instanton solution
does not contribute anymore to the path integral. Only periodic
classical paths are relevant, the leading contribution coming now from the
two-instanton configuration~\cite{BrPaZJ1977} in the sense
of the discussion in chapter~\ref{ssBSminst}.
Indeed, in the asymmetric case $\omega > 1$, it is useful 
to redefine the instanton order $n$ in the sense that an
instanton configuration of order $n$ describes a trajectory
in which the quantal particle returns to the original 
minimum $n$ times. 

To calculate the potential between instantons and the normalization of 
the path integral, it is convenient to first calculate the contribution at 
$\beta$ finite of a trajectory described $n$ times and 
take the large $\beta $ limit of this expression. One finds  
\begin{eqnarray} 
\label{eEnbeta} 
\left\{ \tr \e^{-\beta H} \right\}_{(n)} &=&
\left(-1 \right)^{n}{\beta \over n \sqrt{ \pi g}} \,
\nonumber\\
& & \times \sqrt{{ \omega \, C_\omega \over n \left(1+\omega \right)}} \,
\exp\left(-\frac{\omega \, \beta}{2 \, n \, (1+\omega)}\right)\,
\exp\left(- \frac{nA(\beta )}{g} \right) \,,
\end{eqnarray}
with the definitions [see also 
equations~(\ref{edefCgen}) and~(\ref{etildeCii})]
\begin{equation} 
\label{edefCgenii}
C_\omega = q^2_{0} \, \omega^{2/(1+\omega)} \,
\exp \left\{{ 2\omega \over 1+\omega} \,
\left[ \int^{q_{0}}_{0} \d q \,
\left({ 1 \over \sqrt{ 2V (q )}}-{1 \over q}-
{1 \over\omega (q_{0}-q )} \right)\right] \right\}\,.
\end{equation}  
Also,
\begin{equation} 
A(\beta) = 2 \, \int^{q_{0}}_{0} \sqrt{ 2V(q )} \,\, \d q 
- 2 \, C_\omega \, { \left(1+\omega \right) \over \omega} \,
\e^{-(\beta /n) \, \omega/(1+\omega )} + \cdots \, . 
\label{eAbeta} 
\end{equation}
Note that $n$ has not the same 
meaning here as in chapter~\ref{ssninstdw}.
Since $\omega$ is different from 1, the partition function may 
be described, in the path-integral formalism, by 
trajectories that return to the starting point. 
Thus, the 
``one-instanton''---in the sense of chapter~\ref{ssBSminst}---configuration 
does not contribute to the path integral, 
and $n$ instead counts the number of instanton 
anti-instanton pairs in the language of chapter~\ref{ssninstdw}. 
That is to say, the natural ``one-instanton'' configuration 
in the asymmetric potential corresponds to the ``two-instanton''
trajectory in the conventions of chapter~\ref{ssninstdw}.
Therefore, $n$ here actually corresponds to $2n$ 
in the limit $\omega \to 1$. 

%
% The $n$-instanton action 
%
\section{The $n$--Instanton Action}

We now call $ \theta_{i} $ the successive amounts of time
the classical trajectory spends near $ q_{0} $, and $ \varphi_{i} $ near
the origin. The {$ n $-instanton} action then takes the form
\begin{equation} 
A (\theta_{i},\varphi_{j} ) = n \, a - 2\,
\sum^{n}_{i=1} \left(C_1 \, \e^{-\omega \theta_{i}} +
C_2 \, \e^{-\varphi_{i}} \right)\,, 
\label{eAgenpot}
\end{equation}
with $ \sum^{n}_{i=1} \left(\theta_{i}+\varphi_{i} \right) = \beta$ and
[in agreement with~(\ref{edefA})]
\begin{equation} 
a = 2 \int^{q_{0}}_{0} \sqrt{ 2V (q)} \, \d  q\,.  
\label{einstAdegen}
\end{equation}
By comparing the value of the action at the saddle point,
\begin{equation} 
\theta_{i} = {\beta\over n(1+\omega)}\,,\qquad 
\varphi_{i}={\omega \beta\over n(1+\omega)} \,,
\end{equation} 
with the expression (\ref{eAbeta}), 
one verifies that one can choose
[see also (\ref{egeneral})]
\begin{equation} 
C_1=C_\omega\,, \qquad C_2 = C_\omega/\omega\,, 
\label{eCunCde}
\end{equation}
by adjusting the definitions of $ \theta $ and $ \varphi $. 

%
% The $n$-instanton contribution
%
\section{The $n$--Instanton Contribution to the Partition Function}
\label{ssContribPartition}

The $n$-instanton contribution to the partition function,
in the case of asymmetric wells, has the form 
\begin{eqnarray} 
\left\{ \tr\e^{-\beta H} \right\}_{(n)} & = &
\beta \, \e^{-\beta /2}{ \e^{-na/g} \over \left(\pi g \right)^{n}}\,
N_{n} \, 
\int_{\theta_{i}\,,\varphi_{i}\geq 0}\delta \,
\left( \sum_{i}\theta_{i}+\varphi_{i}-\beta \right) 
\nonumber\\
& & \quad \times \exp  \left[ \sum^{n}_{i=1}{1 \over 2}
\left(1-\omega \right)\theta_{i}-{1 \over g}A (\theta ,\varphi )
\right] \,.
\label{eninstc} 
\end{eqnarray}
The additional term 
$ \sum_{i}{1 \over 2} \left(1-\omega \right)\theta_{i} $ 
in the integrand comes from the determinant generated by 
the Gaussian integration around the classical path. The normalization can
be
obtained by performing a steepest descent 
integration over the variables $ \theta_{i} $ and $ \varphi_{i} $ 
and comparing the result with expression
(\ref{eEnbeta}). The result is 
\begin{equation} 
N_{n}={ \left(C_\omega \sqrt{ \omega} \right)^{n} \over n}. 
\end{equation}
The factor $ 1 /n  $ comes from the symmetry of the
action under cyclic permutations of the 
$ \theta_{i} $ and $ \varphi_{i}$. 

We now set
\begin{equation} 
\lambda = { \e^{-a/g} \over \pi g}C_\omega \sqrt{ \omega}\, , \quad
\mu_1 = -{2C \over g} \quad \mu_2= -{2C \over \omega g}\,,
\end{equation}
in such a way that
\begin{equation} 
{ \e^{-a/g} \over \pi g}\,
C_\omega \sqrt{ \omega} = \lambda \, \omega \, \sqrt{\mu_1\mu_2}\,.
\end{equation}
As in chapter~\ref{ssninstdw}, we introduce the Laplace 
transform $G^{(n)}(E)$ of 
$\left\{ \tr\e^{-\beta H} \right\}_{(n)}$. 
It involves now two integrals, $ {\mathcal I}(E - \ud, \mu_1)$ and 
\begin{equation}
\omega\sqrt{\mu_2} \,
\int^{+\infty}_{0} 
\exp \left\{ \left[\ud ( 1-\omega ) +E- \ud \right]\theta 
- \mu_2 \, \e^{-\omega \theta}  \right\} 
\d \theta \sim {\mathcal I}(E/\omega -1/2,\mu_2 )  ,   
\label{eninstJE}  
\end{equation}
where ${\mathcal I}(s,\mu)$ has been defined in 
(\ref{einstIsmu}) and evaluated for $\mu\to\infty $ 
[equation (\ref{einstIsmub})]:  
\begin{equation}
{\mathcal I}(s,\mu)\sim \mu^{s+1/2} \Gamma (-s).
\end{equation}
Then, $G^{(n)}(E)$ can be written as 
\begin{equation}
G^{(n)}(E)= {1 \over n} \lambda^n \,
{\partial \over \partial  E} 
\left[{\mathcal I} (E-\ud,\mu_1 ) \,
{\mathcal I}(E/\omega -\ud,\mu_2)  \right]^{n}\,. 
\end{equation}
From the sum over $ n$, one then infers $\Delta(E)$, 
the sum of the leading multi-instanton contributions to the 
Fredholm determinant ${\mathcal D} (E)$. One finds
\begin{equation}
\Delta(E) = {1\over \Gamma (\ud- E/\omega ) \, \Gamma (\ud-E)}
+ \left(-{2C_\omega \over g}\right)^{E}\,
\left(-{2C_\omega \over \omega \, g} \right)^{E/\omega} \, 
{\e^{-a/g}\over 2\pi}  \,,
\label{egenpole}  
\end{equation}
where now we have added the harmonic oscillator contributions 
corresponding to the two wells. A generalization 
of this equation, including effects of higher order in $g$,
leads to (\ref{egeneral}).

This equation $\Delta(E)=0$ has two series of energy eigenvalues, 
close for $g\to0$ to the poles 
\begin{subequations}
\begin{eqnarray} 
E_{N} & = & N+ \ud + {\mathcal O}(\lambda)\,, 
\\
E_{N} & = & \left(N+\ud \right)\omega 
+ {\mathcal O}(\lambda) \,,
\end{eqnarray}
\end{subequations}
of the two $ \Gamma $-functions. 
The same expression contains the instanton contributions to the 
two different sets of eigenvalues. 
One  verifies that multi-instanton contributions are
singular for $ \omega =1$. However, if one directly sets $ \omega =1 $ in
the expression (\ref{egenpole}), one obtains 
\begin{equation}
\Delta(E) = {1\over \Gamma^2 (\ud - E)}
+ \left(-{2C \over g}\right)^{2E} \, 
{\e^{-a/g}\over 2\pi}\,,
\label{einstsyge} 
\end{equation}
an expression consistent with the product $\Delta_+(E) \, \Delta _-(E)$
of the results (\ref{epoles}) found for the double-well potential,
and likewise consistent with (\ref{equantization}).

\chapter{Instantons in the Periodic Cosine Potential}
\label{ssninstcos}

%
% Orientation
%
\section{Orientation}

Analytic periodic potentials lead to additional complications, 
which the example of the cosine potential illustrates.
We consider the Hamiltonian
\begin{equation} 
H= -{g \over 2} \, \left({ \d  \over \d  q} \right)^2+
{1 \over 16g}\,\left(1- \cos 4q   \right)\,, 
\end{equation}
where the peculiar normalization has been chosen to avoid the 
proliferation of big integer factors. 

The unitary operator $\mathcal T$ that translates wave functions by
one period $T=\pi /2 $ of the potential,
\begin{equation} 
{\mathcal T}\psi(q)\equiv \psi (q+T ) ,
\end{equation}
commutes with the Hamiltonian.
Because the adjoint of ${\mathcal T}$
fulfills ${\mathcal T}^+\psi(q)\equiv \psi (q-T)$,
the eigenvalues of ${\mathcal T}$ have unit modulus.
Indeed, ${\mathcal T}$ can be diagonalized simultaneously
with the Hamiltonian,
\begin{equation}
{\mathcal T}\psi(q)=\e^{{\rm i}\,\varphi}\psi(q).
\end{equation}
This corresponds to decomposing the initial Hilbert space $\mathcal H$ into
a sum of spaces ${\mathcal H}_\varphi$. The coordinate $q$ 
can then be restricted to 
one period or more conveniently considered as an angular variable 
parameterizing a circle. 

In ${\mathcal H}_\varphi$ the Hamiltonian has a discrete spectrum with 
eigenvalues $E_{N}(g,\varphi)$ and eigenfunctions
$\psi_{N,\varphi}(q)$:
\begin{equation} 
{\mathcal T}\psi_{N,\varphi}= \e^{{\rm i}\,\varphi} \,
\psi_{N,\varphi}\ ,\quad 
H \, \psi_{N,\varphi} = 
E_{N}(g,\varphi) \, \psi_{N,\varphi}\ .
\end{equation} 
The eigenvalues $E_{N}(g,\varphi)$ are periodic functions of the angle
$\varphi$ and $E_{N}(g,\varphi)=N + \ud + {\mathcal O}(g)$.

To each state of the harmonic oscillator is thus associated in 
$\mathcal H$,  for $ g $ small, a band when $\varphi$ varies.  

%
% The partition function in the $\varphi$-sector 
%
\section{The Partition Function in the $\varphi$--Sector}

We now introduce the partition function in  ${\mathcal H}_\varphi$: 
\begin{equation}
{\mathcal Z}(\beta,g,\varphi)=\sum_N \e^{- \beta \, E_N(g,\varphi)}.
\label{eZphi}
\end{equation} 
In the initial Hilbert space $\mathcal H$ we also define the quantity  
\begin{eqnarray}
{\mathcal Z}_l(\beta,g)&=&
\tr\left({\mathcal T}^{l} \, \e^{-\beta H}\right)
\nonumber\\
& \equiv &
\int_0^T \d q\,\left<q\right|{\mathcal T}^{l} \e^{-\beta
H}\left|q\right> =\int_0^T\d q\,\left<q+l\,T\right| \,
\e^{-\beta H}\left|q\right> 
\nonumber\\
& = & {1\over2\pi}\int_0^{2\pi}\d\varphi \,
\sum_N \e^{- \beta E_N(g,\varphi)}\int\d q\,
\psi^*_{N,\varphi}(q+lT)\psi_{N,\varphi}(q ) 
\nonumber\\
& = & 
{1\over2\pi} \, 
\int_0^{2\pi}\d\varphi \sum_N \e^{- \beta E_N(g,\varphi)} \,
\e^{-{\rm i}\,l\,\varphi} \nonumber\\
& = & 
{1\over2\pi} \,
\int_0^{2\pi}\d\varphi\, 
{\mathcal Z}(\beta,g,\varphi) \, \e^{-{\rm i}\,l\,\varphi}\,.
\end{eqnarray}
Inverting this last relation, one finds
\begin{equation}
{\mathcal Z}(\beta,g,\varphi) = 
\sum^{+\infty}_{l=-\infty} \e^{{\rm i}\,l\,\varphi}
{\mathcal Z}_l(\beta,g). 
\label{esumtrTl} 
\end{equation}
This is a generalized partition function with twisted
boundary conditions, which depends on the rotation angle $\phi$.

The path integral representation of ${\mathcal Z}_l(\beta,g)$ can be written as
\begin{equation} 
{\mathcal Z}_l(\beta,g) = 
\int_{q(\beta /2) = q(-\beta /2)+l\pi /2}
\left[ \d  q (t) \right] \exp\left[-{\mathcal S}(q)/g\right]
\end{equation}
with
\begin{equation} 
{\mathcal S} ( q) = 
\int^{\beta /2}_{-\beta /2} \d t \,
\left[{ 1 \over 2} \, \dot q (t)^2+{1 \over 16} \,
\left(1 - \cos 4q  \right) \right] . 
\label{einstZlcos}
\end{equation}
The integration variable $q$ parameterizes a circle and the 
boundary condition is periodic with the constraint that $q(t)$ 
belongs to the topological sector
of trajectories turning $l$ times around the circle.

Note that since 
\begin{equation} 
\label{ephi}
{\rm i}\, l \, \varphi= 
{2 {\rm i} \over\pi} \, \varphi \, \bigl(q(\beta /2)-q(-\beta /2)\bigr) = 
{2\,{\rm i}\over\pi} \, \varphi \,
\int^{+\beta/2}_{-\beta/2} \d  t\, \dot q (t)\,, 
\end{equation} 
the factor $\exp[ {\rm i}\,l\,\varphi ]$ can be incorporated into the path
integral by adding a topological term, the integral of a local density, 
to the action:
\begin{equation}
\frac{{\mathcal S}(q)}{g} \mapsto 
\frac{{\mathcal S}(q)}{g} - 
{2\,{\rm i} \over \pi} \, \varphi \,
\int^{+\beta/2}_{-\beta/2} \, \d  t\, \dot q (t)\,. 
\end{equation}
The sum (\ref{esumtrTl}) can thus be written as
\begin{equation} 
{\mathcal Z}(\beta,g,\varphi) = 
\int \left[\d q(t) \right] \, 
\exp\left[- \frac{{\mathcal S} (q)}{g}
+{ 2 \, {\rm i} \over \pi} \, \varphi \,
\int^{+\beta/2}_{-\beta/2} \d  t\, \dot q (t)\right]  \,,
\label{ecosvaci} 
\end{equation}
with now unrestricted periodic boundary conditions 
on the circle: $ q(\beta /2)=q(-\beta /2)\pmod {\pi/2}$.

%
% Perturbation Theory and Instantons
%
\section{Perturbation Theory and Instantons}

In an expansion for $\beta \to\infty $, due to the boundary conditions 
in the expression (\ref{einstZlcos}), ${\mathcal Z}_0(\beta ,g)$ is dominated 
by the perturbative expansion and  ${\mathcal Z}_l(\beta ,g)$ for $l\ne0$ 
by instantons: in a band all eigenvalues are degenerate to all orders in  
perturbation theory. Only for $l=\pm 1$ do instantons correspond to 
solutions of the classical equation
of motion:
\begin{equation}
q_c(t)=\arctan \e^{\pm t}\ \Rightarrow \ {\mathcal S}(q_c)=\ud\,. 
\end{equation}
For $|l|>1$, ${\mathcal Z}_l(\beta ,g)$ is dominated by multi-instantons.

The energy eigenvalue  $ E_N(g,\varphi) $ can be expanded in a Fourier
series: 
\begin{equation}  
E_N(g,\varphi) = 
\sum^{+\infty}_{l=-\infty} E_{N,l}(g) \, 
\e^{{\rm i}\,l\,\varphi}, \qquad  
E_{N,l}= E_{N,-l}\,.
\end{equation} 
For $ g $ small, $ E_{N,l} (g) $ is
dominated by $ l $-instanton contributions. In particular, for the ground
state energy $ E_{0}(g,\varphi) $ in the $\varphi$ sector, the $l=1$ term
behaves like 
\begin{equation} 
E_{0,l=1}(g)\sim{ 1 \over \sqrt{ \pi g}} \e^{-1/2g}.
\end{equation}

%
% Multi--Instantons
%
\section{Multi--Instantons}

There is one important difference between the double well 
and the cosine potentials. In the case of the double well potential, 
each configuration is a succession of alternatively instantons and 
anti-instantons. Here, by contrast, the paths consist in an arbitrary 
succession of turns in the positive and the negative direction, that 
is an arbitrary succession of instantons and anti-instantons. 
Therefore, we  assign a sign $ \varepsilon =+1 $ to an instanton and 
a sign $ \varepsilon =-1 $ to an anti-instanton. A straightforward 
calculation, similar to the calculation presented above 
(for details see Appendix \ref{ssninstint}) 
yields the following interaction term between two consecutive instantons of
types $ \varepsilon_{1} $ and 
$ \varepsilon_2 $ separated by a distance $ \theta_{12} $:
\begin{equation} 
{ 2 \,\varepsilon_{1}\varepsilon_2 \over g} \,
\e^{-\theta_{12}}. 
\end{equation}
The interaction between instantons is repulsive, while it
is attractive between instantons and anti-instantons. 

We redefine the parameters $\lambda$ and $\mu$ previously 
introduced in equation (\ref{einstmulampar}) in the context of 
the double-well problem,
\begin{equation} 
\label{einstcoslampar}
\lambda = {1 \over \sqrt{2 \pi  }} \, \e^{-1/2g} \,,
\qquad \mu={2 \over g}\, .
\end{equation}
The 
one-instanton contribution at leading order can then be written as
\begin{equation}
{1 \over \sqrt{ \pi g}} \, \e^{-1/2g} =\lambda \, \sqrt{\mu}\,. 
\end{equation}
With this notation, the {$ n $-instanton} contribution reads
\begin{equation}  
{\mathcal Z}^{(n)}(\beta,g,\varphi) =
\beta \, \e^{-\beta /2} \, {\lambda^{n} \over n} \,
\int_{\theta_{1} \geq 0} \delta \,
\left( \sum^{n}_{i=1}\theta_{i}-\beta \right) \, J_{n}(\theta) 
\end{equation} 
with
\begin{equation} 
J_{n} (\theta )= \sum_{\varepsilon_{i}=\pm 1}\,
\exp \left( \sum^{n}_{i=1}
- {2 \over g} \, \varepsilon_{i} \, \varepsilon_{i+1} \,
\e^{-\theta_{i}} + {\rm i}\,\varepsilon_{i} \, \varphi \right)\,. 
\end{equation}
The additional term $ {\rm i}\,\varepsilon_{i}\varphi $ comes from the 
topological term in the expression (\ref{ecosvaci}). 
We have identified $ \varepsilon_{n+1} $ and $\varepsilon_{1}$. 

Since the interaction
between instantons contains both attractive and repulsive contributions, 
we sum series with $ g $ imaginary and then perform the analytic continuation 
of both the Borel sums and the instanton contributions. 

Following the same steps as in the case of the double-well potential,
we introduce the contribution $ G ^{(n)}(E,g,\varphi)$
to the resolvent, Laplace transform of $ {\mathcal Z}^{(n)}(\beta,g,\varphi)$.
The integral over the $\theta _i$ again involves only the function 
(\ref{einstIsmu}), which we replace by its asymptotic form 
(\ref{einstIsmub}).
We obtain 
\begin{eqnarray} 
G ^{(n)}( E,g,\varphi) &=&
{\lambda^{n} \over n} \,
{\partial \over \partial E} \left\{ [\Gamma( \ud-E)]^n \, \mu^{nE} \,
\right.
\nonumber\\
& & \left. \times \sum_{ \{ \varepsilon_{i}=\pm 1 \}} \,
\exp\left[\sum^{n}_{i=1} {\rm i}\,\varepsilon_{i} \,\varphi 
+ (E - \ud)\, 
\ln \left(\varepsilon_{i} \, \varepsilon_{i+1} \right) \right]\right\}\,, 
\label{eEinstsum} 
\end{eqnarray}  
where the choice in the determination of 
$\ln(\varepsilon_i \varepsilon _{i+1})$ depends  on the initial phase 
of $g$. We choose
\begin{equation} 
\ln \left( \varepsilon_{i} \, \varepsilon_{i+1} \right) =  
- \ud \, {\rm i}\, \pi \,
\left(1 - \varepsilon_{i} \, \varepsilon_{i+1} \right)\,. 
\end{equation}
The expression (\ref{eEinstsum}) can then be rewritten as
\begin{eqnarray} 
G^{(n)}(E,g,\varphi) &\sim& {\lambda^{n} \over n} 
{\partial \over \partial  E} 
\left[ \Gamma(\ud-E ) \mu^  E  \right]^{n} \,
\nonumber\\
& & \times 
\sum_{\left\{ \varepsilon_{i}=\pm 1 \right\}} 
\exp\left[\sum^{n}_{i=1} {\rm i}\,\varepsilon_{i}\, \varphi 
- \ud \, {\rm i}\, \pi \,
(E-\ud) \left(1-\varepsilon_{i} \, \varepsilon_{i+1} \right)
\right]\,. 
\label{eEinsumb} 
\end{eqnarray}  
The summation over the set $ \left\{ \varepsilon_{i}
\right\} $ corresponds to calculating the partition function of a
one-dimensional Ising model with the transfer matrix   
\begin{equation}{\bf M} =
\begin{pmatrix}  
\e^{{\rm i}\,\varphi} &
\e^{- {\rm i}\,\pi\,(E-1/2)} \cr  
\e^{-{\rm i}\,\pi\,(E-1/2)} &
\e^{- {\rm i}\,\varphi} 
\end{pmatrix}\,. 
\end{equation}
The sum then is simply $\tr{\bf M}^{n}$. 
The expression (\ref{eEinsumb}) then becomes 
\begin{equation}
G^{(n)}(E,g,\varphi)\sim 
{ \lambda^{n} \over n}  \,
{\partial \over \partial  E} 
\left\{\left[ \Gamma (\ud-E ) \mu^E  \right]^{n}
\tr{\bf M}^n\right\}. 
\end{equation}  
The sum ${\mathcal G} (E ,g,\varphi ) $ of all leading order
multi-instanton contributions can now be calculated. 
One finds (using $\ln\det=\tr\ln$)
\begin{equation} 
{\mathcal G} (E ,g,\varphi ) = 
-{\partial \over \partial  E} 
\ln \det \left[ 1 - \lambda \, \Gamma(\ud-E) \, \mu^E \, {\bf M} \right]\,. 
\label{einsumres}  
\end{equation}    
The contribution to the Fredholm determinant 
${\mathcal D}(E,\varphi )$ in the $\varphi$-sector thus is
\begin{eqnarray} 
& & \Delta (E,\varphi) =
\det  \left[ 1-\lambda \, \Gamma (\ud - E) \, \mu \, E - 
{\bf M} \right]   
\nonumber\\ 
& & = 1 - 2 \, \lambda \, \cos\varphi\,
\Gamma(\ud - E) \, \mu^E  +
\lambda^2 \, 
\left[\Gamma(\ud - E)\right]^2 \, \mu^{2E} 
\, \left[ 1+\e^{-2\pi {\rm i}\,E }\right]\,. 
\end{eqnarray}
Again, after addition of the partition function of the 
harmonic oscillator, the expression becomes
\begin{eqnarray}
\Delta  (E,\varphi) &=&
{1\over \Gamma (\ud-E  )}-
{2\cos \varphi\over\sqrt{2\pi}} \,
\left(2\over g\right)^E \e^{-1/2g} 
\nonumber\\
& &  + \left(2\over g\right)^E  \e^{-1/2g} \left(-{2\over g}\right)^E  
 {\e^{-1/2g}\over\Gamma(\ud+E) } \,.
\end{eqnarray}
Remarkably enough, the equation $\Delta  (E,\varphi)=0$ can also be written
in a form that is symmetric in the exchange $g,E\mapsto -g,-E$:
\begin{equation}
\label{ecospole}
\left({2 \over g}\right)^{-E}{\e^{1/2g}\over \Gamma(\ud - E)} +
\left(-{2 \over g}\right)^E{\e^{-1/2g}\over \Gamma(\ud + E)} ={2 \cos\varphi
\over \sqrt{2\pi}}\,, 
\end{equation} 
a property that depends explicitly on the normalization of the 
one-instanton contribution. This symmetry, 
however, is slightly fictitious because the equation is actually quadratic in
$\Gamma(\ud - E)$ and only one root  is relevant for
$g>0$. The spectral condition (\ref{ecospole})
is fully compatible with (\ref{eQuantCos})
at leading order in $g$. Indeed, the generalization to higher
orders in $g$ is
\begin{equation}
\left({2 \over g}\right)^{-B_{\rm pc}(E,g)}{\e^{A_{\rm pc}(E,g)/2}\over
\Gamma[\ud - B_{\rm pc}(E,g)]} +
\left({-2 \over g}\right)^{B_{\rm pc}(E,g)}\,
{\e^{-A_{\rm pc}(E,g)/2}\over \Gamma[\ud + B_{\rm pc}(E,g)]} =
{2 \cos\varphi \over \sqrt{2\pi}}.
\end{equation}
We supply here a few more terms in comparison to (\ref{ecos}),
\begin{eqnarray}
\label{eBEgpcg4}
\lefteqn{B_{\rm pc}(E, g)  \; = \; 
E + \left( E^2 + \frac14 \right) \, g }
\nonumber\\
& & + \left( 3\, E^3 + \frac{5}{4}\, E \right) \, g^2
+ \left(\frac{25}{2}\, E^4 + \frac{35}{4}\,E^2 + 
\frac{17}{32} \right)\,g^3
\nonumber\\
& & + \left( \frac{245}{4} \, E^5 
+ \frac{525}{8} \, E^3
+ \frac{721}{64} \, E \right) \, g^4 
+  {\mathcal O}(g^5).
\end{eqnarray}
The first few terms of the expansion of the 
instanton $A_{\rm pc}$-function are:
\begin{eqnarray}  
\label{eAEgpcg4}
\lefteqn{A_{\nu}(E,g,j) = 
\frac{1}{g} + 
\left(3\,E^2 + \frac{3}{4}\right) \, g}
\nonumber\\
& & + \left( 11 \, E^3 + \frac{23}{4}\,E  \right) \, g^2 
+ \left(-\frac{199}{4}\,E^4 + \frac{341}{8}\,E^2
+ \frac{215}{64}\right)\, g^3
\nonumber\\
& & + \left(\frac{1021}{4}\,E^5 
+ 326\,E^3 
+ \frac{4487}{64}\,E \right)\,g^4
+  {\mathcal O}(g^5)\,.
\end{eqnarray}
%

%
% Instantons in Radially Symmetric Oscillators 
%
\chapter{Instantons in Radially Symmetric Oscillators}
\label{ssOnuanhar}

%
% Orientation
%
\section{Orientation}
\label{ssOnuOrientation}

It is interesting to consider a last example, the analytic continuation of
the energy eigenvalues of the $ {\mathcal O}(\nu) $ symmetric anharmonic 
oscillator corresponding to the Hamiltonian
\begin{equation} 
H = - \ud \, \bm{\nabla}^2 + \ud \, {\bf q}^2 + 
g \, \left( {\bf q}^2 \right)^2 
\end{equation} 
from $g>0$ to $g<0$. The radial one-dimensional Hamiltonian is 
given in (\ref{eradialschr}),
\begin{equation}
\label{eradialschrii}
H_l(g) = - {1\over 2}\left(\d\over\d r\right)^2
- {1\over 2}\,{\nu-1\over r}{\d\over\d r} 
+ {1\over 2}\,{l\,(l+\nu-2)\over r^2}
+ {1\over 2}\,r^2 + g\,r^4.
\end{equation}
For $g > 0$, the potential is bound from below, and there are 
no degenerate minima. Therefore, the quantization condition for 
$g > 0$, which is given in (\ref{eAOperturbative}),
can be considered as rather trivial: only the perturbative
$B$-function enters, and since there are no degenerate minima,
there are no nontrivial saddle points of the Euclidean
action, and thus no instantons to consider.

For $g < 0$, there are two possibilities: one may either 
endow the Hamiltonian with a self-adjoint extension, as discussed 
in chapter~\ref{ssAOFP} and
below in chapter~\ref{sselfadjoint}, 
or one may consider resonances, as in chapter~\ref{ssAOresonances}
and below in~\ref{sAOresonances}.

%
% Self--Adjoint Extension 
%
\section{Self--Adjoint Extension}
\label{sselfadjoint}

%
% Double-well and ${\mathcal O}(\nu)$ symmetric quartic potentials 
%
\subsection{Double--Well, Symmetry Breaking and ${\mathcal O}(\nu)$--Potentials}
\label{ssOnudw}

One particular aspect facilitates considerably the 
analysis of the self-adjoint extension of (\ref{eradialschrii}):
There is a general connection~\cite{SeZJ1979,An1982,DaPrMa1984,BuGr1993} 
between the double-well potential with an additional 
symmetry breaking term (``broken-double-well'') on the 
one side and anharmonic oscillators with 
${\mathcal O}(\nu)$-symmetry on the other side (acting in 
${\mathbbm R}^\nu$).
Typically, the sign convention in this investigation is chosen 
such that the double-well potential is bounded from below,
and the anharmonic ${\mathcal O}(\nu)$-symmetric oscillator is 
formulated for $g < 0$.
We here briefly explain the relevant method
and recall the radial Hamiltonian
(\ref{eradialschr}) with negative coupling 
\begin{equation}
\label{eradialschrrecalled}
H_l(-g) = - {1\over 2}\left(\d\over\d r\right)^2
- {1\over 2}\,{\nu-1\over r}{\d\over\d r}  
+ {1\over 2}\,{l\,(l+\nu-2)\over r^2}
+ {1\over 2}\,r^2 -g\,r^4\,.
\end{equation}
For illustrational purposes, we have explicitly 
replaced $g \to -g$ in comparison to (\ref{eradialschr}).
The potential, in this convention, is not bounded from 
below for a (redefined) positive $g$.

We now try to find a self-adjoint extension of this Hamiltonian;
although the coupling is negative, a self-adjoint extension implies
that the spectrum will be real.
The Hamiltonian (\ref{eradialschrrecalled}) 
leads to a radial Schr\"odinger equation
\begin{equation}
\label{eOnuquart}
-\psi''(r) - \frac{\nu -1}{r} \,\psi'(r)
+ \frac{l \, (l+\nu-2)}{r^2} \, \psi(r) 
+ (r^2 - 2 \, g\,r^4) \, \psi(r)
= 2 \, E \, \psi(r)\,. 
\end{equation}
In order to show the connection between the 
``broken-double-well'' potential and the anharmonic oscillator, 
we use here a derivation based on differential 
equations, but the same results can be obtained by path 
integral methods (for an overview of related issues see~\cite{An1982}).
We consider first the more general Hamiltonian (\ref{eOnuhamg})
\begin{equation}
H = {g\over 2}\,\left[-\left(\d\over\d r\right)^2- 
{\nu-1\over r}{\d\over\d r} + 
\frac{l \, (l+\nu-2)}{r^2}\right]+
\frac{1}{g} \, V(r)\,. 
\end{equation}
We have used here the scaling $x \to x/\sqrt{g}$.
In the case of (\ref{eradialschr}), we would 
have $V(r) = r^2/2 + r^4$.
Because $V(r)$ is an even function, we now set
\begin{equation}
V(r) = W(r^2) = \frac{r^2}{2} + {\mathcal O}(r^4)\,.
\end{equation}
We first eliminate the $1/r^2$ term by the change
of variables
\begin{equation}
\psi (r)=r^l \,\chi(r), 
\end{equation} 
and obtain
\begin{equation}
-{g\over2} \, \left[\chi''(r) - \frac{2j + 1}{r}\,\chi'(r) \right]
+{1\over g} \, W(r^2) \, \chi(r) =  E \, \chi(r)\,,
\end{equation}
where we have set [see also (\ref{edefj})]
\begin{equation}
\label{edefjii}
j=l+\nu/2-1\,.
\end{equation}
We then take $r^2=x$ as a new variable,
\begin{equation}
\chi(r)=\xi (r^2), 
\end{equation}
and find
\begin{equation}
\label{eOnuhamgii} 
-g \left[2 \, x \, \xi''(x) + 2 \, (j+1) \, \xi'(x) \right]+
{1\over g} \, W(x) \, \xi(x) = E \, \xi(x) \,.
\end{equation}
We now write $\xi $ as a Laplace transform 
\begin{equation}
\xi (x)=\int\d p\, \e^{px/g} \, \eta (p).
\end{equation}
With proper boundary conditions, the equation becomes
\begin{equation}
\frac{1}{g} \, \left[ W\left( -g \frac{\d}{\d p} \right) +
\frac{g}{2}\, \frac{\d}{\d p} \right] \, \eta(p)  +
\frac12 \, (4p^2-1) \, \eta'(p) - 
2 \, (j-1) \, p \, \eta(p) =
E \, \eta(p) \,.
\end{equation}
In the special example $W(x) = x/2 - x^2$ given in (\ref{eOnuquart}), 
to which we restrict the discussion in the sequel,
the equation is again a second-order differential equation.
After a last transformation  to eliminate the $\eta '$ term,
\begin{equation}
\eta(p)=a(p) \, \varphi(p),\quad {\rm with}\quad 
{a ' \over a}={1-4p^2\over 4g} \,,
\end{equation}
one finally obtains
\begin{equation}
\label{ecentral}
-{ g\over2} \, \varphi''(p) +
{1\over 2g} \, \left(p^2 - \frac{1}{4}\right)^2 \, \varphi(p) - 
j \, p \, \varphi(p) =
{E\over 2} \, \varphi(p) \,. 
\end{equation}
This is equivalent to the Schr\"{o}dinger equation for a
particle moving in a double-well potential with minima and 
$p \pm 1/2$, and an additional symmetry-breaking term 
$j \, p$. The equation has been derived by appropriate 
substitutions in (\ref{eOnuquart}) and naturally 
provides a self-adjoint extension of the 
Hamiltonian~(\ref{eradialschrrecalled}). 
This concludes the derivation of the 
correspondence between the anharmonic 
oscillator at negative coupling and 
the double-well potential with a linear breaking term,
which is expressed by the 
equations (\ref{ediffstart}) and (\ref{ediffeq}). 

For $j=0$, the symmetry-breaking term vanishes.
In this case, an comparison of (\ref{ecentral}) 
with (\ref{ehamdw}) leads one automatically to the 
$A$ and $B$ functions of the transformed problem 
(\ref{ecentral}) . These have to be the same as those 
for the ${\mathcal O}(\nu)$ anharmonic oscillator 
obtained in (\ref{eBEgj}) and (\ref{eAEgj}) as we set
$j = 0$. We thus recover the correspondence 
between the double-well potential on the one side
and the anharmonic oscillator with radial symmetry,
considered in the case of negative 
coupling and endowed with a self-adjoint extension,
on the other side,
\begin{equation}
\label{enudw}
B_\nu(E,g,j=0) = 2 \, B_{\rm dw}(E/2,-g), \quad  
A_\nu(E,g,j=0) = A_{\rm dw}(E/2,-g).
\end{equation}
This has already been mentioned 
in chapter~\ref{ssAOFP} [see equation~(\ref{ecorrespondence})].
This correspondence is valid on the perturbative 
level ($B$ function) as well as on the level of
instanton effects ($A$ function).
For $j\ne 0$, the reflection symmetry
$p \to -p$ is broken at the 
same order as the first quantum correction,
so that the degeneracy is lifted. 

%
% Degenerate minima and symmetry breaking
%
\subsection{Degenerate Minima and Symmetry Breaking}
\label{ssinstdegb}

We now consider a slight generalization of the 
situation encountered in the Fokker--Planck equation~(\ref{ssAOFP}),
where the potential has a symmetric structure broken 
at relative order $g$, and may allow for a  vanishing 
perturbative expansion. Indeed, in several situations, 
for instance in the case $j = \pm 1$ in (\ref{ecentral}),
the potential may have the typical form [we assume
the general structure~(\ref{econvention}) for the 
Hamiltonian]
\begin{equation}
V_{\rm tot}(q)=V (q) + g \, V_{\rm pert}(q)\,,
\end{equation}
where the form $V$ is a symmetric potential [$V(q) = V(q_0-q)$]
with degenerate 
minima and $V_{\rm pert}(q)$ breaks the symmetry at order $g$. 
When the breaking term is treated at leading order, it is only 
the difference between the values at the minima of the symmetric 
potential which is important. Therefore, as a simplifying feature but 
without loss of generality at leading order, we consider the example 
of a linear symmetry breaking potential
\begin{equation}
V_{\rm tot}(q) = V(q) + g \, \eta \, \frac{q}{q_0}\,,
\end{equation}
where we assume $V(0)=0$, $\eta  >0$. 

The analysis is, in many respects, analogous to the 
calculations of chapter~\ref{ssContribPartition}
(see also reference~\cite{BrPaZJ1977}). 
Eigenvalues are only degenerate at leading order, and the 
degeneracy is lifted at order $g$, the ground state corresponding 
to the well at $q=0$.
At leading order, the instanton contributions can be derived from chapter
\ref{ssninstg} with $\omega =1$. Again, we call 
$ \varphi_i$ the times spent near
$q=0$ and $\theta _i$ near $q=q_0$. Then, the $n$-instanton action reads
\begin{equation} 
A(\theta_{i},\varphi_{j} )= n \, a - 
2 \,C \, \sum^{n}_{i=1} 
\left(\e^{-\theta_{i}} + \e^{-\varphi_{i}} \right)\,,
\label{eAgenbpot}
\end{equation}
with $ \sum^{n}_{i=1} \left(\theta_{i}+\varphi_{i} \right) = \beta$ and
[see also (\ref{edefC}) and (\ref{edefA})]
\begin{subequations}
\begin{eqnarray} 
a & = & 2 \int^{q_{0}}_{0} \, \sqrt{ 2 \, V (q)} \,\, \d  q\,, 
\\
C & = & q^2_{0} \, \exp\left[ 
\int^{q_{0}}_{0} \d q \, \left({ 1 \over \sqrt{ 2 \, V (q )}}
- {1 \over q} - {1 \over q_{0} - q } \right)\right]\,.
\label{econstbC} 
\end{eqnarray}  
\end{subequations}
The $ n $-instanton contribution then has the form 
\begin{eqnarray} 
\left\{ \tr\e^{-\beta H} \right\}_{(n)} & = &
{\beta \over n} \, \e^{-\beta /2} \, 
\left({C \over \pi g }\right)^n \,
\e^{-na/g} \,
\int_{\theta_{i},\,\varphi_{i}\geq 0}
\delta \left( \sum_{i}\theta_{i}
+\varphi_{i}-\beta \right) 
\nonumber\\
& & \quad \times \exp  \left[-  \eta \sum^{n}_{i=1}
 \theta_{i}-{1 \over g}A (\theta ,\varphi )
\right]\,. 
\label{eninstbc} 
\end{eqnarray}
The additional term $-\eta\sum_{i} \theta_{i} $ in the 
integrand comes from  the value of the potential at $q=q_0$. 

The only required integral [equation (\ref{einstIsmu})] 
has already been evaluated. 
The sum of leading order multi-instanton contributions reads
\begin{equation}
\label{egenbpole}
\Delta(E) = 
{1\over \Gamma (\ud - E)\, \Gamma (\ud +\eta -E)}
+ \left(-{2C\over g}\right)^{2E - \eta} \,
{\e^{-a/g}\over 2\pi}\,.
\end{equation}
Note that when $\eta$ is a positive integer,  
the one-instanton contribution
to the states $E_N=N+\ud$ is real, indicating that the 
behaviour of the perturbative expansion at large order $k$ is 
at least smaller by a factor $1/k$ than naively expected.
For $N<\eta$, the instanton contribution is of 
order $\e^{-a/g}$, while for $N\ge\eta$
it is of order $\e^{-a/2g}$.

%
% The Fokker--Planck Hamiltonian
%
\subsection{The Fokker--Planck Hamiltonian} 
\label{sfp}

A simple example illustrating the remarks of
chapter~\ref{ssinstdegb} is provided by the Fokker--Planck 
Hamiltonian (see chapter~\ref{ssAOFP}): 
the stationary solution, as shown below, is not normalizable,
and instanton effects determine the energy of the ground 
state~\cite{HeSi1978plbremark}.

We recall the Riccati equation~(\ref{eRiccat}), 
\begin{equation}
g\,S'(q) - S^2(q) + 2 \, V(q) - 2\,g\,E=0\,.
\end{equation}
This equation formally allows for a solution with $E=0$
if the potential $V(q)$ has the following structure,
\begin{equation}
V(q) = \ud\, \left[ U^2(q) - g \, U'(q) \right]\,.
\end{equation}
Indeed, this structure automatically
leads to a class of potentials for which the perturbative
expansion of at least one eigenvalue vanishes identically
to all orders in the coupling. 
We assume $U(q)$ to be a polynomial that has two zeros 
at $q=0$ and $q=q_0>0$, such that
\begin{equation}
U(q)= q + {\mathcal O}(q^2), \quad U(q_0-q)=U(q),
\end{equation}
and assume that it is an exact solution of the Riccati 
equation (\ref{eRiccat}) with $E=0$. 

We now specialize the treatment to the Fokker--Planck potential 
\begin{equation}
\label{eVp}
V_{\rm FP}(p) = \frac12 \, \left( p^2 - \frac14 \right)^2 + g \, p\,,
\end{equation}
which is obtained in a natural way by setting $j = - 1$ in (\ref{ecentral}).
In a perturbative expansion around the well $p=-\ud$, 
one finds a ground state with $E=0$ to all orders. 
The equation may alternatively be written as ($p = q - 1/2$),
\begin{equation}
\label{eVq}
V_{\rm FP}(q = p + \ud) = \frac12 \, q^2 \left( 1 - q \right)^2 + g \, 
\left( q - \frac{1}{2} \right)\,.
\end{equation}
At leading order in $g$,
$V(q)$ is symmetric with degenerate minima. The breaking term
lifts the degeneracy at order $g$ and implies that the perturbative ground 
state, corresponding to the well at $q=0$, has $E=0$ while the lowest 
state in the other well has $E=1$. 

In fact the issue is more complicated because the wave function
\begin{equation}
\label{epseudo}
\psi (q) = \exp\left[ -{1\over g}\int^q \d q'U(q') \right] =
\exp\left[ \frac{1}{g} \, \left( \frac{q^3}{3} - \frac{q^2}{2} \right)
\right]
\propto \exp\left[{1\over g} \,
\left(\frac{1}{3} \, p^3-\frac{1}{4} \, p\right)\right]
\end{equation}
is not normalizable, and thus is not an eigenfunction. 
An analogy with the Fokker--Planck equation 
suggests that the case $E=0$ be identified with an equilibrium 
probability distribution. 
Therefore, the non-normalizable wave function (\ref{epseudo}) 
may naturally be identified with 
a ``pseudo-equilibrium'' distribution.

The true ground state has $E>0$
and at leading order for $g\to0$ is dominated by 
the one-instanton contribution.
Shifting $E$ in the expression 
(\ref{egenbpole}) as $E \to E + 1/2$
and setting $\eta =1$, one finds
\begin{equation}
\label{eninstFP}
\Delta(E) = {1\over \Gamma (- E)\,\Gamma (1 -E)}
+ \left(-{2\over g}\right)^{2E} \, {\e^{-1/3 g}\over 2\pi}  \,,
\end{equation}
the generalization being (\ref{equantfp})
\begin{equation}
{1\over \Gamma \left( - B_{\rm FP}(E, g)\right)\, 
\Gamma \left(1 - B_{\rm FP}(E, g) \right)}
+ \left(-{2\over g}\right)^{2 B_{\rm FP}(E, g)} \,
\frac{\exp\left(- A_{\rm FP}(E, g)\right)}{2 \pi} = 0 \,.
\end{equation}
At leading order, for the ground state,
\begin{equation}
\label{efp}
E_0(g) \sim {\e^{-1/3 g}\over 2\pi}\,.
\end{equation}
An imaginary part appears only at two-instanton order and governs 
the large-order behaviour of the non-Borel summable 
one-instanton expansion, where for the classification 
of the instanton order we follow the convention as 
outlined in chapter~\ref{sOrientationGeneral}.
The arguments presented here can be easily generalized to
the non-symmetric situation.

The perturbation series, in the 
Fokker--Planck potential, vanishes to all orders
for the ground state only. For excited states, the
leading-order as well as corrections of relative 
order $g, g^2, g^3, \dots$ are nonvanishing.
The following statements illustrate this phenomenon.
First, we supplement the perturbative $B_{\rm FP}$-function
given in (\ref{equantpertFP}) by an expression
valid up to the order $g^4$,
\begin{eqnarray} 
\label{eBEgFPO4}
\lefteqn{B_{\rm FP}\left(E, g\right) = E + 
3 \, E^2 \, g + \left( 35\, E^3 + \frac52\,E \right)\, g^2 }
\nonumber\\
& & 
+ \left( \frac{1155}{2}\, E^4 + 105\,E^2  \right)\, g^3 
+ \left( \frac{45045}{4}\, E^5 + \frac{15015}{4}\,E^3 +
\frac{1155}{8}\,E \right)\, g^4 .
\end{eqnarray}
Inverting the perturbative quantization condition~(\ref{equantpertFP})
\begin{equation}
B_{\rm FP}(E, g) = N
\end{equation}
for general $N$, we obtain the following
perturbative expansion up to ${\mathcal O}(g^4)$,
\begin{eqnarray} 
\lefteqn{E_N(g) \sim N - 3 \, N^2 \, g 
- \left( 17\, N^3 + \frac52\,N \right)\, g^2 }
\nonumber\\
& & 
- \left( \frac{375}{2}\, N^4 + \frac{165}{2}\,N^2  \right)\, g^3 
- \left( \frac{10689}{4}\, N^5 + \frac{9475}{4}\,N^3 +
\frac{1105}{8}\,N \right)\, g^4 \,.
\end{eqnarray}
For the ground state ($N=0$), all the terms vanish,
whereas for excited states with $N=1,2,\dots$, the perturbation series
is manifestly nonvanishing.
For completeness, we also supplement the 
instanton $A_{\rm FP}$-function (\ref{eAEgFP}) for the 
Fokker--Planck potential by a few more terms,
\begin{eqnarray} 
\label{eAEgFPO4}
\lefteqn{A_{\rm FP}\left(E, g\right) = \frac{1}{3 g} +
\left( 17 \, E^2 + \frac56 \right)\, g
+ \left( 227 \, E^3 + \frac{55}{2} \, E \right)\, g^2 }
\nonumber\\
& & 
+ \left( \frac{47431}{12} \, E^4 
+ \frac{11485}{12}\,E^2 
+ \frac{1105}{72} \right)\, g^3 
\nonumber\\
& & 
+ \left( \frac{317629}{4} \, E^5 
+ \frac{64535}{2}\,E^3 
+ \frac{4109}{2}\,E \right)\, g^4 
+ {\mathcal O}(g^5)\,.
\end{eqnarray}
The functions $B_{\rm FP}$ and $A_{\rm FP}$ determine the perturbative
expansion, and the perturbative expansion about the instantons, in higher
order (see also chapter~\ref{Explicit} for an application of related
ideas to the double-well problem).

({\em Remark.}) When investigating a slight 
generalization of the potential (\ref{eVp}),
namely [see also (\ref{ediffeq})]
\begin{equation}
\label{eVpgen}
V_{\rm FP}(p) = 
\frac12 \, \left( p^2 - \frac14 \right)^2 + g \, j\, p\,,
\end{equation}
it may be shown that for negative integer $j$, the 
perturbation series of at least one energy level 
terminates.
Assuming that the solution of equation
(\ref{eOnuhamgii}) is regular at $x=0$, we expand it, 
as well as the potential, in a Taylor series, after changing variables
$x\mapsto gx$,  
\begin{equation}
W(x) = \sum_{n=1} W_n \, g^n \, x^n\,,\qquad 
\xi (x) = \sum_{n=0} \xi_n \, g^n \, x^n\,.
\end{equation}
We then obtain the recursion relations
\begin{equation}
\label{easymj}
-2\,g\,(n+1) \, (j+1+n) \, \xi_{n+1}
+\sum_{p=0}^n g^{n-p-1} \, W_{n-p} \, \xi_p = 
E \, \xi_n\,. 
\end{equation}
If $j$ is a negative integer: $j=-N-1$ (with $N\ge 0$),
the set of linear equations with 
$0\le n\le N$ is closed. {\em Idem est},
the coefficient $j+1+n$ vanishes in this case for $n = N$,
and $\xi_n = 0$ for all $n > N$. Alternatively, one observes that 
the determinant of the $[N+1]\times [N+1]$
system of linear equations defined by equation (\ref{easymj}) 
has to vanish, a condition that determines $E$ 
as a solution of an
algebraic equation of degree $N-1$. This property  reflects the equivalence
of ${\mathcal O}(\nu)$ models with $\nu$ even and negative
with fermion Hamiltonians. For
$N=0$ and thus $j=-1$, one recovers the known result
\begin{equation} 
E=0\,,
\end{equation}
which holds in a strict sense only on the level of perturbation theory
[see also equation (\ref{efp})].
For $j=-2$, one finds $E=\pm 1$. The instanton corrections follow directly
from equations~(\ref{egenbpole}) and (\ref{eninstFP}).
As an illustration, at next order, and thus, for $j=-3$, one finds
\begin{equation}
E^3 - 4 \, E -8 \, g \, W''(0)=0\,,
\end{equation}
an equation that has three solutions for small $g$.
Also, for $g=0$, the solutions are $E=1-|j|+2k$ with $k$ integer,
$0\le k\le |j|-1$. We conclude this chapter by noting that
the Hamiltonian (\ref{ecentral}) implies
a spectral symmetry with respect to the sign change $j \to -j$.
However, here this symmetry is broken in the discussion
following (\ref{easymj}), because the potential (\ref{eVp}) has a
unique global minimum at $p = -1/2$, not $p=+1/2$
[analogous statements hold for the generalization (\ref{eVpgen})].
For this reason, the system of equations is closed for 
negative integer $j$ (but the closure does not hold for 
positive $j$). 

For a more detailed treatment of the intriguing 
aspects related to the Fokker--Planck potential,
the reader is referred to~\cite{JeZJ2004}.

%
% ${\mathcal O}(\nu)$ symmetric quartic potentials
%
\subsection{${\mathcal O}(\nu)$--Symmetric Quartic Potentials}

We have related in chapter 
\ref{ssOnudw} perturbative expansions of 
${\mathcal O}(\nu)$ symmetric quartic potentials and double-well potentials 
with linear symmetry breaking. 
% Before calculating directly the 
% instanton contributions in chapter \ref{ssOnuanhar}, we apply
% the results of the chapter to the double-well potential. 
The additional contribution to the potential $-jp$ has the 
effect of adding a contribution $\pm j/2$ to the action depending 
whether the instanton
is close to $1/2$  or $-1/2$, respectively~\cite{BrZJunpublished}.
% It bears strong similarities to the analysis of chapter
% \ref{ssninstg}, where also quantum corrections lift the 
% classical degeneracy. 
We  now call $ \theta_{i} $ the successive amounts of time
the classical trajectory spends near $-1/2$, and $ \varphi_{i} $ near
$1/2$. The {$ n $-instanton} action then takes the form
\begin{equation} 
A(\theta_{i},\varphi_{j} ) = 
{n \over 3} - 2 \, \sum^{n}_{i=1} 
\left(\e^{-  \theta_{i}} + \e^{-\varphi_{i}} \right)
\end{equation}
with 
$ \sum^{n}_{i=1} \left(\theta_{i}+\varphi_{i} \right) = \beta$. We set
\begin{equation}
\lambda = {\e^{-1/3g}\over 2\pi}\,,\quad 
\mu=-{2\over g}\,.
\end{equation}
The $ n $-instanton contribution then takes the form 
\begin{eqnarray} 
\left\{ \tr\e^{-\beta H} \right\}_{(n)} & = &
{ \beta \over n} \, \e^{-\beta/2} \, 
(-\lambda \, \mu )^n \,
\int_{\theta_{i}\,,\varphi_{i}\geq 0} \delta \,
\left( \sum_{i} \theta_{i} + \varphi_{i} - \beta \right)   
\nonumber\\
& & 
\quad \times 
\exp\left[ \sum^{n}_{i=1} {1 \over 2} \, j \, (\theta _i - \varphi_i)
- {1 \over g} \, A(\theta, \varphi)
\right]\,. 
\end{eqnarray}
The sum then involves the integrals (taking into account $E\mapsto E/2$)
\begin{equation} 
\sqrt{\mu} \int^{+\infty}_{0} \,
\exp\left\{ \left[ \pm \ud j +\ud E-\ud) \right] \,
\theta - \mu \, \e^{-\theta} \right\} \d \theta
= {\mathcal I}\bigl(\ud(\pm j+E-1),\mu\bigr)\,. 
\end{equation}
Using the asymptotic form (\ref{einstIsmub}),
\begin{equation}
I(s,\mu) \sim \mu^{s+1/2} \, \Gamma (-s),
\end{equation} 
we obtain the spectral equation
\begin{equation}
\label{eOnuAnharmonicSelfAdjoint}
\Delta_j(E) = 
{1\over \Gamma \bigl (\ud( 1+j-E)  \bigr) \Gamma \bigl (\ud(1-j-E) \bigr) }
+ \left(-{2\over g}\right)^{ E} {\e^{-1/3g}\over 2\pi} =0 \,.
\end{equation}
The generalization to higher orders in $g$ is given by
(\ref{equantAnharmOsc}) and reads
\begin{eqnarray}
& & {1\over \Gamma \left[ \ud( 1+j-B_{\nu}(E, -g, j)) \right] \,
\Gamma \left[ \ud(1-j-B_{\nu}(E, -g, j)) \right] } \nonumber\\
& & \qquad \qquad \qquad \qquad \qquad
+ \left(-{2\over g}\right)^{B_{\nu}(E, -g, j)} 
\frac{\exp\left(- A_{\nu}(E, -g)\right)}{2 \pi} = 0 \,.
\end{eqnarray}
We supplement here a few terms in comparison
to (\ref{eABnu}). The perturbative $B_\nu$-function
reads [see also equation (\ref{einstBgengij})]
\begin{eqnarray}
\label{eBEgjg4}
\lefteqn{B_\nu(E, g, j)  \; = \; 
E+ \left( -\frac32 \, E^2 + \frac{j^2}{2} - \frac12 \right) \, g }
\nonumber\\
& & + \left( 
\frac{35}{4}\, E^3 + \frac{25}{4}\, E - \frac{15}{4}\, j^2\, E
\right) \, g^2
\nonumber\\
& & + \left(-\frac{1155}{16}\, E^4 - \frac{735}{8}\,E^2 + 
\frac{315}{8} \, j^2\,E^2 - \frac{35}{16}\, j^4 +
\frac{105}{8}\,j^2 - \frac{175}{16} \right) \, g^3 
\nonumber\\
& & + \left(\frac{45054}{64} \, E^5 + \frac{45045}{32} \, E^3 
+ \frac{31185}{64} \, E 
- \frac{15015}{32} \, j^2 \, E^3 \right.
\nonumber\\
& & \qquad \left. + \frac{3465}{64} \, j^4 \, E 
- \frac{12705}{32} \, j^2 \, E \right) \, g^4 
+  {\mathcal O}(g^5).
\end{eqnarray}
The first few terms of the expansion of the 
instanton $A_\nu$-function are:
\begin{eqnarray}  
\label{eAEgjg4}
\lefteqn{A_{\nu}(E,g,j) = 
-{1 \over 3}\,g^{-1} + 
\left({3 \over 4}\,j^2-{19 \over 12}-{17 \over 4} \, E^2 \right) \, g}
\nonumber\\
& & + \left( \frac{227}{8} \, E^3 - \frac{77}{8}\,j^2\,E + 
\frac{187}{8} \, E\right) \, g^2 
\nonumber\\
& & + \left(-\frac{47431}{192}\,E^4 -\frac{34121}{96}\,E^2
+ \frac{3717}{32}\,j^2\,E^2 
- \frac{341}{64}\,j^4 
\right.
\nonumber\\
& & \qquad \left.
+ \frac{1281}{32}\,j^2 - \frac{28829}{576} \right)\,g^3
\nonumber\\
& & + \left(\frac{317629}{128}\,E^5 
+ \frac{264725}{48}\,E^3
+ \frac{842909}{384}\,E \right.
+ \frac{19215}{128}\,j^4\,E
\nonumber\\
& & \qquad \left.
- \frac{4445}{3}\,j^2\,E^3
- \frac{253045}{192}\,j^2\,E\right)\,g^4
+  {\mathcal O}(g^5)\,.
\end{eqnarray}
In (\ref{eBEgj}) and (\ref{eAEgj}), the functions 
$B_\nu$ and $A_\nu$ are given up to terms of the order 
$g^2$, respectively.

% If $j$ is not an integer, it is easy to calculate the imaginary part
% of the one-instanton contribution. For $E_{j,N}=1+j+2N + {\mathcal O}(g)$, 
% one finds
% %
% \begin{equation} 
% \label{einstImEselfadj}
% {\rm Im}\, E_{j,N}
% \mathop{\sim}_{g\to 0_+}
% {1\over N!\,\Gamma (N+j+1)} \, \left(2\over g\right)^{2N+j+1} \,
% \e^{-1/3g}\,
% \end{equation}
% %
% in agreement with the expression (\ref{einstImEj}). 

%
% Resonances
%
\section{Resonances}
\label{sAOresonances}

%
% The ${\mathcal O}(2)$ anharmonic oscillator
%
\subsection{The ${\mathcal O}(2)$--Anharmonic Oscillator}
\label{sO2}

We first discuss resonances in the example case $\nu=2$
(see also chapter~\ref{ssAOresonances}), 
the generalization then being simple.
Note that for $\nu=2$, we have $j = l + \nu/2-1 = l$
[see equations (\ref{edefj}) and (\ref{edefjii})].
As in chapter~\ref{ssinstdegb}, we extensively use here the ideas 
outlined in chapter~\ref{ssContribPartition}.
For $g > 0$, as discussed in chapter~\ref{ssOnuOrientation},
there are no saddle points of the Euclidean action beyond 
the trivial one, and thus no instantons to consider.

For $g < 0$, the instanton solution exists and has the form 
\begin{equation} 
{\bf q} (t)= {\bf u} \, f (t)\,, 
\end{equation}
in which $ {\bf u} $ is a fixed unit vector. 
The leading one-instanton
contribution to the ground state energy is 
\begin{equation} 
\label{eE1g}
{\rm Im} \, E^{(1)}(g) = {4 \over g} \, \e^{1/3g} \,
\left(1 + {\mathcal O}\left(g \right) \right) 
\quad {\rm for\ } g \rightarrow 0_{-}\,. 
\end{equation}
It is easy to calculate the instanton interaction, 
and thus $n$-instanton action  
\begin{equation} 
A(\theta_{i} ) = -{1 \over 3} \, n - 4 \sum_i \e^{-\theta_{i}}
\cos \varphi_{i}\,, 
\end{equation}
in which $ \theta_{i} $ is the distance between two successive
instantons and $ \varphi_{i} $ the angle between them:
\begin{equation} 
\cos \varphi_{i} = 
{\bf u}_{i}\cdot {\bf u}_{i+1}\,. 
\end{equation}
It is convenient to consider the quantity 
\begin{equation}
{\mathcal Z}(\beta ,\alpha ) = 
\tr\left[R (\alpha ) \e^{-\beta H}\right] = 
\int \left[ \d  {\bf q} (t) \right] 
\exp\left[-{\mathcal S}(q )/g \right]\,,
\label{etrRalph} 
\end{equation}
where $ R (\alpha ) $ is a rotation matrix which
rotates vectors by an angle $ \alpha $. 
In the path integral, it leads to the boundary condition
that $ {\bf q} (t) $ at initial and final times differ 
by an angle $ \alpha $: 
${\bf \hat q} (-\ud \beta ) \cdot {\bf \hat q} (\ud\beta )
= \cos \alpha$. 

We recall that in two dimensions with a radial coordinate $\rho$
and an angle $\phi$, we have
$\bm{\nabla}^2 = \rho^{-1} (\partial/\partial\rho)
(\rho \, \partial/\partial\rho) + \rho^{-2}\, (\partial^2/\partial\phi^2)$.
Eigenfunctions of the angular part are of the form 
$\e^{-{\rm i}\,l\,\phi}$ with $l = -\infty, \dots, \infty$.
The r.h.s.\ of equation (\ref{etrRalph}) can be rewritten as
\begin{equation} 
\tr\left[R (\alpha ) \, \e^{-\beta H}\right] = 
\sum_{l,N} \e^{-{\rm i}\,l\,\alpha -\beta \, E_{l, N}}, 
\label{eangmom}
\end{equation}
where $ l $ is the angular momentum. 
The boundary
condition in the path integral (\ref{etrRalph}) 
implies for the multi-instanton configuration the constraint 
\begin{equation} 
\sum^{n}_{i=1} \varphi_{i} = \alpha \, , 
\end{equation}
which can be implemented through the identity 
\begin{equation} 
\delta \left( \sum^{n}_{i=1} \varphi_{i} -
\alpha \right)={1 \over 2\pi} \,
\sum^{+\infty}_{l=-\infty} 
\exp  \left[ {\rm i}\, l \, \left( \sum ^{n}_{i=1}\varphi_{i} \right)-
{\rm i}\, \alpha \, l \right]\,. 
\end{equation}
As before, we introduce the Laplace transform of the 
twisted partition function ${\mathcal Z}(\beta ,\alpha )$
defined in (\ref{etrRalph}),
\begin{equation}
G(E, \alpha) = 
\int_0^\infty \d\beta \,\e^{\beta E} \,
\tr\left[R (\alpha ) \, \e^{-\beta H}\right] = 
\sum_{l,N} \e^{-{\rm i}\, l \,\alpha} {1\over E_{l,N}-E}\,. 
\end{equation}
The $ n $-instanton contribution to expression 
(\ref{eangmom}) then takes the form
\begin{equation} 
G^{(n)}(E, \alpha) \sim{ \lambda^{n} \over 2 \,{\rm i}\,\pi n} \,
\sum^{+\infty}_{l=-\infty} 
\e^{-{\rm i}\, l \, \alpha} \left[ I(E-1, l, \mu) \right]^{n}, 
\end{equation}
where we have again redefined $\lambda$ and $\mu$
[cf. equations (\ref{einstmulampar}) and (\ref{einstcoslampar})],
\begin{eqnarray}
\label{eredeflambda}
\lambda & = & - {\rm i} \, \e^{1/3g}, \qquad \mu =-{4 \over g}\,,
\\
I(s, l, \mu ) & = & {\mu \over 2\pi} \,
\int^{2\pi}_{0} \d \varphi \int^{+\infty}_{0} \d  \theta \, 
\exp \left( s \, \theta + {\rm i}\, l \, \varphi -
\mu \, \e^{-\theta} \, \cos  \varphi \right)\,.
\end{eqnarray}
We introduce the generating
function of {$n$-instanton} contributions at fixed angular 
momentum $l$:
\begin{equation} 
{\mathcal G}_{l} (E ,g ) = -{\partial \over \partial E} \,
\ln\left[1 + \lambda \, I(E-1, l, \mu) \right]\,. 
\label{elinstsum} 
\end{equation}
To evaluate $ I(s, l, \mu )$, we first 
integrate over $ \theta $, where we employ the 
approximations $\mu \to \infty$ and thus $g \to 0_-$, 
as in going from 
(\ref{ederivation1}) to (\ref{ederivation2}). We find 
\begin{equation} 
I(s, l, \mu ) = \mu^{s+1} \, \Gamma( -s ) \, 
\int^{2\pi}_{0}{ \d  \varphi \over 2\pi} \, \e^{{\rm i}\,l\,\varphi} \,
\left( \cos \varphi \right)^s. 
\end{equation}
The last integration yields
\begin{equation}
\int^{2\pi}_{0}{ \d  \varphi \over 2\pi} \, \e^{{\rm i}\,l\varphi} \,
\left( \cos \varphi \right)^s =
{2^{-s-1} \, \left(1+(-1)^{l } \, \e^{{\rm i}\,\pi s}\right) \, 
\Gamma (1+s) \over
\Gamma\bigl(\ud (s - l) + 1\bigr) \, \Gamma\bigl(\ud (s+l )+1\bigr)}, 
\end{equation}
and thus 
\begin{eqnarray}
\label{everif1}
I(s, l, \mu) & = &
- \left({\mu\over2}\right)^{s+1} \, {\pi \over \sin(\pi s)} \,
{1 + (-1)^{l}\, \e^{{\rm i}\,\pi s}\over
\Gamma \bigl(\ud(s-l )+1\bigr)\,
\Gamma \bigl(\ud(s+l )+1\bigr)} 
\nonumber\\
& = & 
\left({\mu \over 2}\right)^{s+1} \, e^{{\rm i}\,\pi(s+l)/2} \,
{\Gamma \bigl(\ud(l - s)\bigr) \over \Gamma\bigl(\ud\,(s + l) + 1 \bigr)}\,,
\end{eqnarray}
the first expression being explicitly even in $l$. 

The sum of leading order contributions to the Fredholm determinant 
at fixed angular momentum $l$ thus is
\begin{equation} 
\Delta_l(E) = 
{1\over\Gamma \left({ 1 \over 2} (l + 1 - E) \right)} - 
{\rm i}\, \left(- {2\over g } \right)^E \, 
{\e^{{\rm i}\,\pi (E + l + 1)/2} \, \e^{1/ 3g} \over 
\Gamma \left({1 \over 2} (l+1+E ) \right)} \,.
\label{eOdeinsti}
\end{equation} 
The eigenvalues $E_{l,N}$ are solutions of  
the equation $\Delta_l(E)=0$ that satisfy
\begin{equation} 
E_{l,N } = l + 2N + 1 + {\mathcal O}(g)\,, \quad N\geq 0\,. 
\end{equation}

%
% The ${\mathcal O}(\nu)$ anharmonic oscillator
%
\subsection{The ${\mathcal O}(\nu)$--Symmetric Hamiltonian}
\label{sOnu}

One can extend this result to the general $ {\mathcal O}(\nu) $ case since, 
at fixed angular momentum $l$, 
the Hamiltonian depends only on the combination $ j = l + \nu/2 - 1$,
and thus the generalization of the discussion in the 
previous chapter simply involves the replacement $l \to j$.
Hence, making in equation (\ref{eOdeinsti}) the corresponding substitution,
one obtains
\begin{equation} 
\label{einstResonance}
\Delta_j(E) = 
{1\over\Gamma \left(  { 1 \over 2} (j + 1 - E) \right)} 
- {\rm i}\, \left(- {2\over g } \right)^E \,
{\e^{ {\rm i}\, \pi (E + j + 1)/2} \, \e^{1/3g} \over 
\Gamma \left( {1 \over 2} (j+1+E ) \right)}\,. 
\end{equation}
This equation is consistent with (\ref{eOdeinst})
at leading order in $g$. The imaginary parts of
the energy eigenvalues 
\begin{equation} 
E_{j, N} = j + 2N + 1 + {\mathcal O}(g)\,, \quad N\geq 0\,, 
\end{equation}
for $ g\to0_- $  follow:
\begin{equation} 
\label{einstImEj}
{\rm Im}\, E_{j, N} \mathop{=}_{ g \rightarrow 0_{-}}  
- {2 \over N! \,\, \Gamma ( j+1+N )} \,
\left(-{2 \over g} \right)^{ j+1+2N} \,
\e^{1/3g} \, \bigl(1 + {\mathcal O} \left(g \right) \bigr)\,,
\end{equation}
in full agreement with the ground-state result (\ref{eE1g}).
Using the Cauchy formula, one can then derive from this expression
large-order estimates for perturbative 
expansions~\cite{BrLGZJ1977long}.
At next order in $ \lambda $, 
one obtains the two-instanton contribution, 
which is related by the same dispersion relation to the large-order 
behaviour of the perturbative expansion around one instanton. 

Finally, note that  checks about these expressions
are provided by the  perturbative relation between the
${\mathcal O}(\nu)$ anharmonic oscillator with negative coupling
and the double-well potential with linear symmetry breaking
derived in chapter \ref{ssinstdegb}.

%
% Instanton Calculations in the Double--Well Problem
%
\chapter{Instanton Calculations in the Double--Well Problem}
\label{Explicit}

%
% Matchings, Higher--Order Coefficients, and the like
%
% \section{Matchings, Higher--Order Coefficients, and the like}
% \label{ssMatchings}

%
% Orientation
%
\section{Orientation}

The purpose of the current chapter consists in the illustration
of the general discussion of instantons by concrete calculational 
example. We choose the quantum-mechanical double-well oscillator 
with a potential of the form $V(q) = q^2\,(1-q)^2/2$ as in 
equation (\ref{ehamdw}). The resurgent expansion for the energy
of a level with principal quantum number $N$ and parity $\epsilon$
is then given by equation (\ref{ecomexp}). We remember the 
nonperturbative factor (\ref{defxi})
\begin{equation}
\xi(g) = \frac{1}{\sqrt{\pi g}} \, \exp\left[ -{1 \over 6\,g} \right]\,,
\end{equation} 
which characterizes the instanton contributions and the 
logarithm (\ref{defchi})
\begin{equation}
\chi(g) = \ln\left( - \frac{2}{g} \right)\,,
\end{equation} 
which generates, for $g$ positive, imaginary contributions that
cancel among the resummed perturbative expansion and the instanton
contributions.

The calculations, and even the results which are discussed in the 
current chapter, have a somewhat involved structure and are rather 
lengthy. The formulas should be taken {\em cum grano salis}, exemplifying
the application of general concepts discussed in previous chapters
to a case of special interest. The lengthy formulas and numerical
results (see also Table~\ref{tableeee} below)
are not displayed for their own sake;
they receive a meaning and an interpretation 
in the context of the general 
conjectures introduced and motivated in 
chapters~\ref{BSqf},~\ref{BSWKB} and~\ref{ssninstdw}.

%
% Corrections to Asymptotics
%
\section{Corrections to Asymptotics}
\label{ssCorrAsymp}

According to (\ref{ePert}), the perturbation series for the level 
$N$ is independent of the parity $\epsilon$ and can be written as
\begin{equation}
E^{(0)}_N(g) = \sum^\infty_{K=0} E^{(0)}_{N,K} \, g^{K}\,.
\end{equation}
We have determined the first 300 terms of the perturbative
expansion of the ground state(s) with $N=0$ ($\epsilon = \pm$)
in closed analytic form (a complete list is available for
internet download at~\cite{JeHome}). The coefficients are expressed in terms
of rational numbers. An example is given in equation (\ref{eE017}).
The first terms read [see also equation (\ref{esimground})]:
\begin{equation}
E^{(0)}_0(g) = {1 \over 2} - g - {9 \over 2} \, g^2 - {89 \over 2} \, g^3
- {5013 \over 8} \, g^4 - \frac{88251}{8} \, g^5 -\frac{3662169}{16} \, g^6
+ {\mathcal O}(g^7) \,.
\end{equation}
The magnitude of the terms grows factorially, and the series
is nonalternating: all coefficients except the first are negative. 
We had stressed in chapter~\ref{ssResurgent} that the 
imaginary part incurred by analytic continuation from $g$ negative 
is compensated by the explicit imaginary part of the two-instanton
contribution to the energy eigenvalues. For the ground state(s)
with $N = 0$, the formal expansion of the two-instanton effect
as given by equation (\ref{eInstn}) reads  
\begin{equation}
\label{eInst2formal}
E^{(2)}_{\epsilon,0}(g) = 
\frac{\e^{-1/3g}}{\sqrt{\pi g}} \,
\left[ \ln\left(-\frac{2}{g}\right) \,
\sum^{\infty}_{K=0} e_{0,21K} \, g^{K} +
\sum^{\infty}_{K=0} e_{0,20K} \, g^{K} \right]\,.
\end{equation}
The imaginary part (generated by the logarithm for positive $g$)
reads
\begin{equation}
\label{eIm2inst}
{\rm Im} E^{(2)}_{\epsilon,0}(g) =  \pm \,
{\e^{-1/3g}\over g} \, 
\left[ e_{0,210} + e_{0,211} \, g + e_{0,212} \, g^2 +
{\mathcal O}(g^3) \right]\,.
\end{equation}
The leading factorial growth of the perturbative coefficients 
$E^{(0)}_{0,K}$ for the ground state is known~\cite{BrPaZJ1977} to be
of the form $E^{0}_{0,K} \sim -3^{K+1} \, K!/\pi$.
By analytic continuation of the perturbation series for positive
$g$, an imaginary part of the order of $\exp(-1/3 g)$ is obtained.
The large-order behaviour of the 
perturbative coefficients and the imaginary part of the 
two-instanton energy shift are connected by the 
(dispersion-type) relations (\ref{einstImii})---(\ref{eEzerok}).
Indeed, as discussed in chapter~\ref{ssLargeOrder}, the power corrections in
$g$ [equation (\ref{eIm2inst})] are connected with the corrections
of order $K^{-1}$ to the perturbative coefficients $E^{(0)}_{0,K}$
(see also appendix~\ref{ssDispersion}).

It is therefore interesting to numerically determine the
corrections to the leading factorial growth of the 
perturbative coefficients. 
We start with the available results for the first 300 perturbative
coefficients and subtract the known leading asymptotics of the 
form $E^{0}_{0,K} \sim -3^{K+1} \, K!/\pi$, as well as divide the results
of the subtraction by the 
leading asymptotics. An approximation
for the coefficient of the 
correction term of relative order $1/K$ can be found via 
multiplication of the results of the previous operation by $K$.
This approximation can be improved by elimination of 
power corrections of higher order in $1/K$ via 
``extrapolation to $K=\infty$.'' For this latter step,
various algorithms may be used. One possibility consists in the
Neville algorithm~\cite{PrFlTeVe1993}, which is a variant
of Richardson extrapolation~\cite{Ri1927}. The result
is again a {\em numerical} estimate for the coefficient of the 
$1/K$-term. The exact, rational form can then be found easily by 
routines built into modern-day computer algebra systems,
for example the {\em Rationalize} function of~\cite{Wo1988}. 
These operations may be repeated, and conjectures may be found
for the coefficients multiplying the corrections of relative
order $1/K$ to the factorial growth of the perturbative
coefficients. We find 
\begin{eqnarray}
\label{Corrections}
E^{0}_{0,K} &\sim&
-\frac{3^{K+1} \, K!}{\pi} \,
\left[ 1 - 
\frac{53}{18} \, \frac{1}{K} -
\frac{1277}{648} \, \frac{1}{K^2} -
\frac{405\,395}{34\,992} \, \frac{1}{K^3} \right.
\nonumber\\
& & \left. \qquad -\frac{218\,793\,923}{2\,519\,424} \, \frac{1}{K^4} -
\frac{35\,929\,260\,709}{45\,349\,632} \, \frac{1}{K^5} +
{\mathcal{O}}\left(\frac{1}{K^6}\right)
\right]\,.
\end{eqnarray}
If we rewrite the corrections to the leading factorial
growth of the perturbative coefficients in terms of a 
factorial series,
\begin{eqnarray}
E^{0}_{0,K} &\sim&
-\frac{3^{K+1} \, K!}{\pi} \,
\left[ 1 -
\sum_{j=1}^\infty \frac{a_{j}}{(K-j+1)_{j}} \right] = 
\nonumber\\
&=&  -\frac{3^{K+1} \, K!}{\pi} 
\left[ 1 - \frac{a_1}{K} -  \frac{a_2}{K\,(K-1)} \right.
\nonumber\\
& & \left. - \frac{a_3}{K\,(K-1)\,(K-2)} -
\frac{a_4}{K\,(K-1)\,(K-2)\,(K-3)} - \dots \right]\,,
\end{eqnarray}
then it is easy to identify the coefficients
entering into (\ref{eIm2inst}) 
with the $a_j$ coefficients. 
Based on the discussion in chapter~\ref{ssLargeOrder},
it is easy to show that
\begin{equation}
3^j \, (-a_j) = \epsilon_{0,21j}\,.
\end{equation}
The results listed in equation (\ref{Corrections})  
therefore lead to the following conjectures for the 
two-instanton coefficients:
\begin{eqnarray}
\label{eeconj}
e_{0,211} &=& -\frac{53}{6}\,, \qquad
e_{0,212} = -\frac{1277}{72}\,, \qquad
e_{0,213} = -\frac{336\,437}{1296} \,, \nonumber\\
e_{0,214} &=& -\frac{141\,158\,555}{31104}\,, \qquad
e_{0,215} = -\frac{17\,542\,610\,737}{186\,624}\,.
\end{eqnarray}
The numbers $53$, $1277$ and $336\,437$ are prime.

%
% Reference Values
%
\section{Perturbative Expansion and Instanton Theory}
\label{ssFunctions}

We will now use a completely different and more direct route to
the calculation of instanton coefficients: It is based on the 
explicit solution of the quantization condition (\ref{equantization})
by an ansatz as given by the resurgent expansion (\ref{ecomexp}).
In order to carry out a calculation, we first have to 
determine the coefficients in the expansions of $B_{\rm dw}(E,g)$ and 
$A_{\rm dw}(E,g)$ defined in (\ref{defB}) and (\ref{defA}).
An accurate calculation requires more terms
than those given in (\ref{eBdble}) and (\ref{eAdble}).

The calculation of the perturbative 
function $B_{\rm dw}(E,g)$ can be carried out easily based on
equation (\ref{equantCii}). Indeed, recursive algorithms are
known~\cite{ZJ1981jmplong}.
The generalization of (\ref{eBdble}) to higher orders is as
follows:
\begin{scriptsize}
\begin{eqnarray}
\label{BEg8}
\lefteqn{B_{\rm dw}(E,g) = E + g \, \left( 3 E^2 + \frac{1}{4} \right) }
\nonumber\\
&& + g^2 \, \left( 35\, E^3 + \frac{25}{4} \, E\right) 
\nonumber\\
&& + g^3 \, \left( \frac{1155}{2} \, E^4 + \frac{735}{4} \, E^2 + 
  \frac{175}{32} \right) 
\nonumber\\
&& + g^4 \, \left( \frac{45045}{4} \, E^5 + \frac{45045}{8} \, E^3 +
  \frac{31185}{64} \, E \right) 
\nonumber\\
&& + g^5 \, \left( \frac{969969}{4} \, E^6 +
\frac{2807805}{16} \, E^4 + 
\frac{1924923}{64} \, E^2 +
\frac{159159}{256} \right) +
\nonumber\\
&& + g^6 \, \left( \frac{22309287}{4} \, E^7 +
\frac{88267179}{16} \, E^5 +
\frac{100553453}{64} \, E^3 +
\frac{25746721}{256} \, E \right)
\nonumber\\
&& + g^7 \, \left( \frac{2151252675}{16} \, E^8 +
\frac{2788660875}{16} \, E^6 +
\frac{9526065549}{128} \, E^4 \right.
\nonumber\\
&& \qquad \left. + \frac{2506538463}{256} \, E^2 +
\frac{692049787}{4096} \right)
\nonumber\\
&& + g^8 \, \left( \frac{214886239425}{64} \, E^9 +
\frac{353522522925}{64} \, E^7 +
\frac{1691601686775}{512} \, E^5 
\right.
\nonumber\\
&& \qquad \left. + \frac{7583829646255}{1024} \, E^3 +
\frac{663834081625}{16384} \, E \right) + {\mathcal O}(g^9)\,.
\end{eqnarray}
\end{scriptsize}
The calculation of the ``instanton function''
$A_{\rm dw}(E,g)$ is a little more difficult
(see the appendices~\ref{sExplicit}
and~\ref{sAlternative}):
The calculation is based on
the evaluation of successively higher orders of the 
contour integral of the WKB expansion as given 
in (\ref{econj}).
The generalization of (\ref{eAdble}) to higher orders
is given by:
\begin{scriptsize}
\begin{eqnarray}
\label{AEg8}
\lefteqn{A_{\rm dw}(E,g) = \frac{1}{3\,g} +
g \, \left( 17 E^2 + \frac{19}{2} \right)} 
\nonumber\\
&& +g^2 \, \left( 227\, E^3 + \frac{187}{4} \, E\right) +
\nonumber\\
&& +g^3 \, \left( \frac{47431}{12} \, E^4 + \frac{34121}{24} \, E^2 +
  \frac{28829}{576} \right) +
\nonumber\\
&& +g^4 \, \left( \frac{317629}{4} \, E^5 + \frac{264725}{6} \, E^3 +
  \frac{842909}{192} \, E \right) +
\nonumber\\
&& + g^5 \, \left( \frac{26145967}{15} \, E^6 +
 \frac{16601579}{12} \, E^4 +
 \frac{63996919}{240} \, E^2 +
 \frac{6167719}{960} \right) +
\nonumber\\
&& + g^6 \, \left( \frac{812725953}{20} \, E^7 +
 \frac{3490889111}{80} \, E^5 
\right.
\nonumber\\
&& \qquad \left.+
 \frac{4398906487}{320} \, E^3 +
 \frac{1280980929}{1280} \, E \right)
\nonumber\\
&& + g^7 \, \left( \frac{443323117271}{448} \, E^8 +
 \frac{265222473925}{192} \, E^6 +
 \frac{4948336000477}{7680} \, E^4 \right.
\nonumber\\
&& \qquad \left. + \frac{10166658134543}{107520} \, E^2 +
 \frac{3228992367577}{172032} \right) +
\nonumber\\
&& + g^8 \, \left( \frac{22315986340103}{896} \, E^9 +
 \frac{4909541135621}{112} \, E^7 +
 \frac{29042282605297}{1024} \, E^5 \right.
\nonumber\\
&& \qquad \left. + \frac{15025619362858}{21504} \, E^3 +
 \frac{29626122220400}{688128} \, E \right) + {\mathcal O}(g^9)\,.
\end{eqnarray}
\end{scriptsize}

%
% Instanton Coefficients
%
\section{Instanton Coefficients}
\label{ssCoefficients}

All coefficients, up to eight-instanton order, up 
to seventh order in $g$, have been calculated
(see Table~\ref{tableeee}). This requires the WKB expansion up to 
$g^8 S_8(g, E, q)$. Due to the prefactor $1/g$, we obtain relevant terms with 
positive powers of $E$ from $S_{10}$ of order $g^9$. 
This is irrelevant if we want to calculate all terms up to the order of
$g^8$, and we may therefore neglect $S_{10}$
(see also appendix~\ref{sExplicit}).

The perturbation theory function $B_{\rm dw}(E,g)$, together with the
instanton function $A_{\rm dw}(E,g)$, entirely determine the 
perturbative expansion about the $n$-instanton effect 
($n$ arbitrary) to order $g^8$, i.e. all coefficients 
$e_{0,nkl}$ with $l \leq 8$
in the resurgent expansion (\ref{ecomexp}),
for arbitrary $n$ and $k$. 
In order to illustrate the effect of the instanton contributions
on the energy levels, we give here formulas for the 
complete instanton effects through the order 
of $g^8$ rather than lists of coefficients.
This is inspired by the notation (\ref{eInstn}).
For the states with parity $\epsilon=\pm$ and $N=0$, the one-instanton
effect reads [see also (\ref{defxi}) and (\ref{defchi})]:
\begin{subequations}
%\begin{scriptsize}
\label{eE1N0}
\begin{eqnarray}
E^{(1)}_{\epsilon,0} &=& -\epsilon \, \xi(g) \,
\left(1
-\frac{71}{12}\, g
-\frac{6299}{288}\, g^2
-\frac{2691107}{10368}\, g^3
\right.
\nonumber\\
& & \left.
-\frac{2125346615}{497664}\, g^4
-\frac{509978166739}{5971968}\, g^5
-\frac{846134074443319}{429981696}\, g^6
\right.
\nonumber\\
& & 
-\frac{262352915454007369}{5159780352}\, g^7
-\frac{717976540715437267525}{495338913792}\, g^8 \nonumber\\
& & \left.
+ {\mathcal O}(g^9)\right)\,.
\end{eqnarray}
%\end{scriptsize}
%
Note that according to~\cite{ZJ1981jmplong}, many more terms in the 
perturbative expansion about one instanton may be calculated.
There are even recursive formulae available.
The first coefficients in the perturbative expansion about one
instanton therefore read as follows:
\begin{eqnarray}
\label{eE1N0coeff}
e_{0,100} &=& 1 \,, \qquad
e_{0,101} = -\frac{71}{12} \,, \qquad
e_{0,102} = -\frac{6299}{288}\,,
\nonumber\\
e_{0,103} &=& -\frac{2691107}{10368}\,,\qquad
e_{0,104} = -\frac{2125346615}{497664}\,.
\end{eqnarray}
\end{subequations}
The coefficients $e_{0,10l}$ ($l = 0,\dots,300$) are available
in electronic form~\cite{JeHome}.

Now we consider the
two-instanton energy shift.
Equation (\ref{eInst2formal}) clarifies that the two-instanton
shift is given by the sum of two infinite series in $g$,
which differ by the presence (or absence) of the logarithmic
prefactor $\chi(g)$.
The two-instanton effect is parity independent,
and the explicit formula for $N=0$ reads
\begin{subequations}
%\begin{scriptsize}
\label{eE2N0}
\begin{eqnarray}
E^{(2)}_{\epsilon,0}(g) & = & \xi^2(g) \, \left[ \chi(g) \,
\left(1
-\frac{53}{6}\, g
-\frac{1277}{72}\, g^2
-\frac{336437}{1296}\, g^3
\right. \right.
\nonumber\\
& & 
-\frac{141158555}{31104}\, g^4
-\frac{17542610737}{186624}\, g^5
\nonumber\\
& & 
-\frac{14922996684685}{6718464}\, g^6
-\frac{2359159111315567}{40310784}\, g^7
\nonumber\\
& & \left.
-\frac{3279840218627988925}{1934917632}\, g^8 + {\mathcal O}(g^9)\right)
\nonumber\\
& &  
+ \left(\gamma
+\left(-\frac{23}{2}-\frac{53}{6}\,\gamma\right)\, g
+\left(\frac{13}{12}-\frac{1277}{72}\,\gamma\right)\, g^2
\right.
\nonumber\\
& & 
+\left(-\frac{45941}{144}-\frac{336437}{1296}\,\gamma\right)\, g^3
+\left(-\frac{20772221}{2592}-\frac{141158555}{31104}\,\gamma\right)\, g^4
\nonumber\\
& & 
+\left(-\frac{12783531515}{62208}-
  \frac{17542610737}{186624}\,\gamma\right)\, g^5
\nonumber\\
& &
+\left(-\frac{2110670726275}{373248}-
  \frac{14922996684685}{6718464}\,\gamma\right)\, g^6
\nonumber\\
& & 
+\left(-\frac{2249987155449805}{13436928}-
  \frac{2359159111315567}{40310784}\,\gamma\right)\, g^7
\nonumber\\
& & 
+\left(-\frac{429051019455941467}{80621568}-
  \frac{3279840218627988925}{1934917632}\,\gamma\right)\, g^8 + 
\nonumber\\
& & \left. \left. 
{\mathcal O}(g^9)\right) \right]\,.
\end{eqnarray}
%\end{scriptsize}
\end{subequations}
The logarithmic coefficients $e_{0,21l}$ ($l=0,\dots,8$), 
determined here by direct 
analytic calculation, verify the conjectures ($l=0,\dots,5$)
which were previously
derived based on the corrections to the large-order growth of the 
perturbative coefficients (\ref{eeconj}) and the dispersion relations
discussed in chapter~\ref{ssLargeOrder}.

The three-instanton correction to energy eigenvalues,
for states with $N=0$, involves three powers of 
the nonperturbative factor $\xi(g)$.
The analytic expressions become very involved, as there
are three infinite perturbative series in
$g$, multiplying the logarithmic 
terms $\chi^2(g)$, $\chi(g)$, and a series
multiplying the terms which lack the logarithm.
The coefficients of the instanton expansion can still be 
calculated in closed analytic form, by direct reference
to the quantization condition (\ref{egenisum}).
Up to terms of order $g^8$, the three-instanton
shift is given by
\begin{scriptsize}
\begin{eqnarray}
\label{eE3N0}
E^{(3)}_{\epsilon,0}(g) & = & -\epsilon \, \xi^3(g) \, \left[ \chi^2(g) \,
\left\{\frac{3}{2}
-\frac{141}{8}\, g
-\frac{489}{64}\, g^2
-\frac{89635}{256}\, g^3
-\frac{27325159}{4096}\, g^4
-\frac{2362765483}{16384}\, g^5 \right.\right.
\nonumber\\
& & \left.
-\frac{1378627835215}{393216}\, g^6
-\frac{148426925165305}{1572864}\, g^7
-\frac{139916898211047653}{50331648}\, g^8 + {\mathcal O}(g^9)\right\}
\nonumber\\
& & + \chi(g) \,
\left\{3\,\gamma
+\left(-\frac{63}{2}-\frac{141}{4}\,\gamma\right)\, g
+\left(\frac{825}{8}-\frac{489}{32}\,\gamma\right)\, g^2
+\left(-\frac{24483}{64}-\frac{89635}{128}\,\gamma\right)\, g^3
\right.
\nonumber\\
& &
+\left(-\frac{3369081}{256}-\frac{27325159}{2048}\,\gamma\right)\, g^4
+\left(-\frac{1534609037}{4096}-\frac{2362765483}{8192}\,\gamma\right)\, g^5
\nonumber\\
& &
+\left(-\frac{179089589633}{16384}-
  \frac{1378627835215}{196608}\,\gamma\right)\, g^6
\nonumber\\
& & 
+\left(-\frac{44105311958423}{131072}-
  \frac{148426925165305}{786432}\,\gamma\right)\, g^7
\nonumber\\
& & \left.
+\left(-\frac{17311037863755515}{1572864}-
  \frac{139916898211047653}{25165824}\,\gamma\right)\, g^8 + 
{\mathcal O}(g^9)\right\}
\nonumber\\
& & + 
\left\{\left(\frac{3}{2}\,{\gamma}^2+\frac{1}{2}\,\zeta(2)\right)
+\left(-\frac{17}{2}-\frac{63}{2}\,\gamma-
  \frac{141}{8}\,{\gamma}^2-\frac{47}{8}\,\zeta(2)\right)\, g
\right.
\nonumber\\
& & 
+\left(\frac{297}{2}+\frac{825}{8}\,\gamma-\frac{489}{64}\,{\gamma}^2-
   \frac{163}{64}\,\zeta(2)\right)\, g^2
\nonumber\\
& & 
+\left(\frac{21199}{192}-\frac{24483}{64}\,\gamma-
  \frac{89635}{256}\,{\gamma}^2-\frac{89635}{768}\,\zeta(2)\right)\, g^3
\nonumber\\
& & 
+\left(-\frac{271647}{128}-\frac{3369081}{256}\,\gamma-
   \frac{27325159}{4096}\,{\gamma}^2-
   \frac{27325159}{12288}\,\zeta(2)\right)\, g^4
\nonumber\\
& & 
+\left(-\frac{9180880297}{61440}-\frac{1534609037}{4096}\,\gamma-
   \frac{2362765483}{16384}\,{\gamma}^2-
    \frac{2362765483}{49152}\,\zeta(2)\right)\, g^5
\nonumber\\
& & 
+\left(-\frac{572797894871}{92160}-
   \frac{179089589633}{16384}\,\gamma-
   \frac{1378627835215}{393216}\,{\gamma}^2-
   \frac{1378627835215}{1179648}\,\zeta(2)\right)\, g^6
\nonumber\\
& & +\left(-\frac{9833533558034329}{41287680}-
   \frac{44105311958423}{131072}\,\gamma \right.
\nonumber\\
& & \qquad \left. - \frac{148426925165305}{1572864}\,{\gamma}^2-
   \frac{148426925165305}{4718592}\,\zeta(2)\right)\, g^7
\nonumber\\
& & + \left(-\frac{150070370854927511}{16515072}-
   \frac{17311037863755515}{1572864}\,\gamma\right.
\nonumber\\
& & \qquad \left. \left. \left. 
  - \frac{139916898211047653}{50331648}\,{\gamma}^2
  - \frac{139916898211047653}{150994944}\,\zeta(2)\right)\, g^8 + 
  {\mathcal O}(g^9) \right\} \right]\,.
\end{eqnarray}
\end{scriptsize}

%
% Reference Values
%
\section{Reference Values}
\label{ssReference}

All accurate verifications of the theory of instantons, 
at small coupling $g$, necessitate a very precise 
determination of energy eigenvalues in the small-$g$ region.
A numerically accurate determination of eigenvalues can be a rather 
hard problem, even in a one-dimensional case.
However, calculations are simplified when using 
multiprecision libraries~\cite{Ba1990tech,Ba1993,Ba1994tech}.
In the current chapter, we intend to give reference values 
for specific small-$g$ example cases. The results for 
$E_{\pm,0}$ in all cases are rather close together, and 
the tiny energy differences then point toward the instanton-mediated
energy shifts. In~\cite{JeZJ2001long},
we indicated two such values, valid up to 180 decimals,
for the case $g=0.001$. Here, we restrict the discussion to 
slightly fewer decimals while stressing that essentially
arbitrary accuracy is available when using appropriate
numerical algorithms~\cite{JeZJ2001long}.
For $g=0.002$, we obtain for the ground state
to 100 decimals
\begin{equation}
\label{Result0plus} 
\begin{array}{r@{}l}
E_{+,0}(0.002) = 0.&
\underline{49798~16336~05614~52785~33444~97756~93929}~30135~47830\\
&41905~18141~65406~30388~85981~28620~52208~12662~00662\,,
\end{array}
\end{equation}
whereas the first excited state has the energy
\begin{equation}
\label{Result0minus}
\begin{array}{r@{}l}
E_{-,0}(0.002) = 0.&
\underline{49798~16336~05614~52785~33444~97756~93930}~90653~58949\\
& 74478~49607~06416~37435~43472~00173~52993~86232~45517\,.
\end{array}
\end{equation}
The two energies differ on the level of the one-instanton
contribution which is given by equation (\ref{eE1N0}) and has
opposite sign for states with opposite parity.
Indeed, the energy difference $E_{-,0}(0.002) - E_{+,0}(0.002)$
can be calculated to high accuracy by simply evaluating the
partial sums of the perturbative expansion about one
instanton, in analogy to the 
case $g=0.001$ considered in~\cite{JeZJ2001long}.
As mentioned previously, the coefficients $e_{0,10l}$ 
($l = 0,\dots, 300$) are available electronically~\cite{JeHome}
to the interested reader.
For $g=0.005$, we obtain as reference values, to 45 decimals,
\begin{subequations}
\label{example0005}
\begin{equation}
\label{Result0005plus}
E_{+,0}(0.005) = 
\underline{0.49488~15073~206}47~62721~54849~56033~00402~87323~00532\,,
\end{equation}
and
\begin{equation}
\label{Result0005minus}
E_{-,0}(0.005) = 
\underline{0.49488~15073~206}99~29084~80981~89004~43542~33907~22906\,.
\end{equation}
The mean energy is
\begin{equation}
\frac{E_{+,0}(0.005) + E_{-,0}(0.005)}{2} = 
\underline{0.49488~15073~20673~45903~17915}~72518\,.
\end{equation}
Again, the real part of the Borel sum of the perturbation series
(\ref{defcalBE0})
is slightly different, due to the two-instanton correction,
\begin{equation}
{\rm Re} \, {\mathcal B}\left\{E^{(0)}_0 (0.005)\right\} = 
\underline{0.49488~15073~20673~45903~17915}~68107\,.
\end{equation}
The energy difference is
\begin{equation}
E_{-,0}(0.005) - E_{+,0}(0.005) = 
\underline{5.16636~32613~23297~14313~9465}8~42237 \times 10^{-14}\,.
\end{equation}
Due to the three-instanton shift,
the quantity $E_{-,0}(0.005) - E_{+,0}(0.005)$
is not quite equal to (twice) the real part of the 
Borel sum of the perturbative series (\ref{defcalBE1})
about one-instanton, 
\begin{equation}
2 \, {\mathcal B}\left\{E^{(1)}_0 (0.005)\right\} = 
\underline{5.16636~32613~23297~14313~9465}6~65773 \times 10^{-14}\,.
\end{equation}
\end{subequations}
The case $g=0.007$ yields the following energies
\begin{subequations}
\label{example0007}
\begin{equation}
\label{Result0007plus}
E_{+,0}(0.007) = 
\underline{0.49276~251}38~34552~88807~85274~35274~12515~77733~55043\,,
\end{equation}
and
\begin{equation}
\label{Result0007minus}
E_{-,0}(0.007) = 
\underline{0.49276~251}44~24291~38099~89370~58280~41759~92080~41287\,.
\end{equation}
The mean energy is
\begin{equation}
\frac{E_{+,0}(0.007) + E_{-,0}(0.007)}{2} = 
\underline{0.49276~25141~29422~13}453~87322\,.
\end{equation}
The real part of the Borel sum of the perturbation series
(\ref{defcalBE0})
is slightly different, due to the two-instanton correction,
\begin{equation}
{\rm Re} \, {\mathcal B}\left\{E^{(0)}_0 (0.007)\right\} = 
\underline{0.49276~25141~29422~13}399~23438\,.
\end{equation}
The energy difference is
\begin{equation}
E_{-,0}(0.007) - E_{+,0}(0.007) = 
\underline{5.89738~49292~04096~2}3006~29244~14346 \times 10^{-10}\,.
\end{equation}
Again, due to the three-instanton shift,
the quantity $E_{-,0}(0.007) - E_{+,0}(0.007)$
is slightly different from the real part of twice the 
Borel sum of the perturbative series (\ref{defcalBE1})
about one-instanton, 
\begin{equation}
2\,{\mathcal B}\left\{E^{(1)}_0 (0.007)\right\} = 
\underline{5.89738~49292~04096~2}0714~40266~74322 \times 10^{-10}\,.
\end{equation}
\end{subequations}
Finally, for the slightly less problematic 
case of $g = 0.01$, we only list the 
energies
\begin{equation}
E_{+,0}(0.01) = 0.489\,497\,520\,976\,030\,,\qquad
E_{-,1}(0.01) = 0.489\,498\,132\,721\,197\,.
\end{equation}

%
% Instanton Function Delta to g^8
%
\begin{figure}[htb!]
\begin{center}
\begin{minipage}{12.0cm}
\begin{center}
\epsfxsize=121.mm
\epsfysize=85.mm
\centerline{\epsfbox{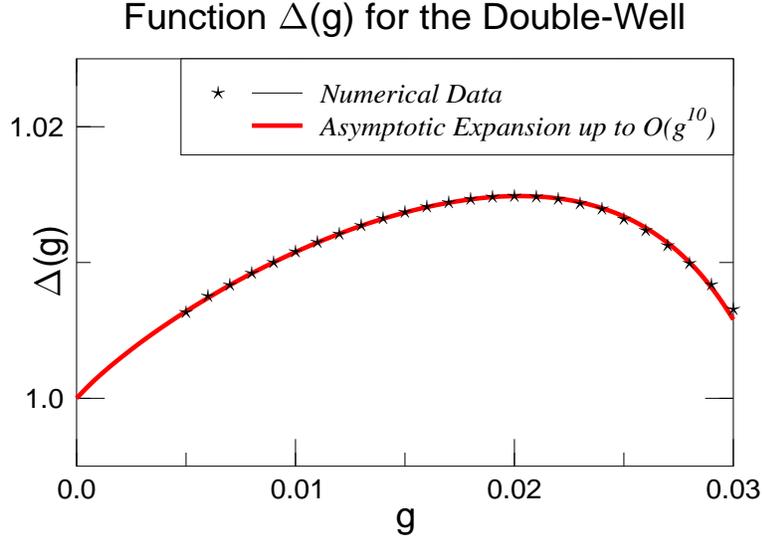}}
\caption{\label{figdelta2}
Double-well potential: Comparison of numerical
data obtained for the function $\Delta(g)$ with the sum of
the terms of order $g^j$ (where
$j \leq 10$) of its asymptotic expansion
for $g$ small, where we express both the 
numerator as well as the denominator 
of (\ref{defDelta}) as a power series in $g$. 
[When the resulting fraction is further 
expanded in $g$, then the 
leading terms of order $g$ and $g^2$ are given
in equation (\ref{asympDelta}) and represented
graphically in figure~\ref{figdelta}].
The higher-order terms employed here are easily derived
using equations (\ref{eE1N0}) and (\ref{eE2N0}). 
There is good agreement between 
numerically determined (``exact'') values (data points) and 
the smooth curve given by the analytic asymptotics.}
\end{center}
\end{minipage}
\end{center}
\end{figure}

%
% A Closer Look at the Function $\Delta(g)$
%
\section{The Function $\Delta(g)$ for States with $N=0$}
\label{ssDelta}

We recall the definition of the function $\Delta(g)$,
given in (\ref{defDelta}),
\begin{equation} 
\Delta (g) = 4 { \left\{ { 1 \over 2} \,
\left(E_{+,0}+E_{-,0} \right) - 
{\mathcal B}\left\{E^{(0)}_0 (g)\right\} \right\} \over 
\left(E_{+,0}-E_{-,0} \right)^2
\left[ \ln(2g^{-1}) + \gamma \right]}. 
\end{equation}
We provide here, based in part on the data given in
chapter~\ref{ssReference}, some reference values of the for 
the function $\Delta(g)$, for a somewhat larger coupling $g$ 
as compared to those listed in Table~\ref{tabcirmi}:
\begin{scriptsize}
\begin{eqnarray}
\Delta(0.011) &=& 1.01147,\quad
\Delta(0.012) \;=\; 1.01211, \quad
\Delta(0.014) \;=\; 1.01322, \quad
\Delta(0.015) \;=\; 1.01368, \nonumber\\
\Delta(0.016) &=& 1.01408, \quad
\Delta(0.017) \;=\; 1.01439, \quad
\Delta(0.018) \;=\; 1.01465, \quad
\Delta(0.019) \;=\; 1.01482, \nonumber\\
\Delta(0.020) &=& 1.01489, \quad
\Delta(0.021) \;=\; 1.01486, \quad
\Delta(0.022) \;=\; 1.01464, \quad
\Delta(0.023) \;=\; 1.01444, \nonumber\\
\Delta(0.024) &=& 1.01392, \quad
\Delta(0.025) \;=\; 1.01339, \quad
\Delta(0.026) \;=\; 1.01233, \quad
\Delta(0.027) \;=\; 1.01144, \nonumber\\
\Delta(0.028) &=& 1.00991, \quad
\Delta(0.029) \;=\; 1.00816, \quad
\Delta(0.030) \;=\; 1.00653\,.
\end{eqnarray}
\end{scriptsize}
The available higher-order corrections (in $g$) to the 
two-instanton energy shift [see equation (\ref{eE2N0})] 
allow for an accurate comparison of the numerically
determined energy eigenvalues with the asymptotic expansion of the 
function $\Delta(g)$ (see figure~\ref{figdelta2}).
We recall that 
the calculation of the instanton coefficients
(\ref{eE2N0}) relies on higher-order WKB expansions
are described in appendices~\ref{sExplicit} and~\ref{sAlternative}.

%
% Leading Instanton Effects
%
\section{Leading Instanton Effects for States with $N=0$}
\label{ssLeading}

Based on the simplified quantization condition (\ref{epoles}),
derived using the path-integral formalism, it is relatively easy 
to determine the leading (in $g$) coefficients of the 
instanton expansion (\ref{ecomexp}) to any order in 
the nonperturbative factor $\xi(g) \equiv \exp(-1/6g)/\sqrt{\pi g}$,
and zeroth order in $g$ (this is equivalent to the coefficients
$e_{N,nkl}$ with $l=0$). In the current chapter, we intend to 
provide results for the states with $N=0$, up to eight-instanton order,
in closed analytic form. The coefficients display an interesting 
analytic structure; a rapid growth of their absolute magnitude 
is observed in higher instanton orders 
for constant (zeroth) order in $g$ (see also Table~\ref{tableeee}).
We briefly recall here that the leading
two- and three-instanton
coefficients have been given in
(\ref{ePthree}),  (\ref{eE2N0}) and (\ref{eE3N0}).
The four-instanton correction is discussed
in appendix~\ref{sFourInstanton}. The four-instanton
coefficients determine 
the polynomial $P^0_4$ implicitly defined
by~(\ref{einstEnn}):
\begin{subequations}
\begin{eqnarray}
e_{0,430} &=& \frac{8}{3} \,, \quad
e_{0,420} = 8\,\gamma \,, \quad
e_{0,410} = 8\,\gamma^2 + 2 \,\zeta(2) \,,
\nonumber\\
e_{0,400} &=& \frac{8}{3}\,\gamma^3 + 2\,\gamma\,\zeta(2) +
\frac{1}{3} \, \zeta(3)\,.
\end{eqnarray}
\end{subequations}
The five-instanton coefficients which enter into
$P^0_5$, read:
\begin{subequations}
\begin{eqnarray}
e_{0,540} &=& \frac{125}{24} \,,
\\
e_{0,530} &=& \frac{125}{6}\,\gamma\,,
\\
e_{0,520} &=& \frac{125}{4}\,\gamma^2 + \frac{25}{4}\,\zeta(2)\,,
\\
e_{0,510} &=& \frac{125}{6}\,\gamma^3 + \frac{25}{2}\,\gamma\,\zeta(2) +
\frac{5}{3} \, \zeta(3)\,,
\\
e_{0,500} &=& \frac{125}{24}\,\gamma^4 + \frac{25}{4}\,\gamma^2\,\zeta(2) +
\frac{5}{3} \, \gamma \, \zeta(3) + \frac{29}{16}\, \zeta(4)\,.
\end{eqnarray}
\end{subequations}
The result for the six-instanton coefficients are a little
more complex:
\begin{subequations}
\begin{eqnarray}
e_{0,650} &=& \frac{54}{5}\,,
\\
e_{0,640} &=& 54 \, \gamma\,,
\\
e_{0,630} &=& 108\,\gamma^2 + 18 \,\zeta(2)\,,
\\
e_{0,620} &=& 108\,\gamma^3 + 54 \,\gamma\,\zeta(2) +
6\,\zeta(3)\,,
\\
e_{0,610} &=& 54\,\gamma^4 + 54 \,\gamma^2\,\zeta(2) +
12\,\gamma\,\zeta(3) + \frac{51}{4}\,\zeta(4)\,,
\\
e_{0,600} &=& \frac{54}{5}\,\gamma^5 + 18\,\gamma^3\,\zeta(2) +
6\,\gamma^2 \,\zeta(3) \nonumber\\
& & + \zeta(2)\,\zeta(3) + 
\frac{51}{4} \,\gamma\,\zeta(4) + \frac{1}{5}\,\zeta(5)\,.
\end{eqnarray}
\end{subequations}

The seven-instanton coefficients, to lowest order in $g$, read:
\begin{scriptsize}
\begin{subequations}
\begin{eqnarray}
e_{0,760} &=& \frac{7^5}{720} \,,
\\
e_{0,750} &=& \frac{7^5}{120} \,\gamma \,,
\\
e_{0,740} &=& \frac{7^5}{48} \,\gamma^2 +
\frac{2401}{48}\,\zeta(2)\,,
\\
e_{0,730} &=& \frac{7^5}{36} \,\gamma^3 +
\frac{2401}{12}\,\gamma\,\zeta(2) +
\frac{343}{18}\,\zeta(3)\,,
\\
e_{0,720} &=& \frac{7^5}{48} \,\gamma^4 +
\frac{2401}{8}\,\gamma^2\,\zeta(2) +
\frac{343}{6}\,\gamma\,\zeta(3) +
\frac{1911}{32}\,\zeta(4) \,,
\\
e_{0,710} &=& \frac{7^5}{120} \,\gamma^5 +
\frac{2401}{8}\,\gamma^3\,\zeta(2) +
\frac{343}{6}\,\gamma^2\,\zeta(3) +
\frac{49}{6}\,\zeta(2)\,\zeta(3)
\nonumber\\
& & + \frac{1911}{16}\,\gamma\,\zeta(4) +
\frac{7}{5}\,\zeta(5)\,,
\\
e_{0,700} &=& \frac{7^5}{720} \,\gamma^6 +
\frac{2401}{48}\,\gamma^4\,\zeta(2) +
\frac{343}{18}\,\gamma^3\,\zeta(3) +
\frac{49}{6}\,\gamma\,\zeta(2)\,\zeta(3) 
\nonumber\\
& &  + \frac{7}{18}\,\zeta^2(3)
+ \frac{1911}{32}\,\gamma^2\,\zeta(4) +
\frac{7}{5}\,\gamma\,\zeta(5) +
\frac{789}{128}\,\zeta(6)\,.
\end{eqnarray}
\end{subequations}
\end{scriptsize}
The eight-instanton effect features the following coefficients:
\begin{scriptsize}
\begin{subequations}
\begin{eqnarray}
e_{0,870} &=& \frac{2^{14}}{315}\,,
\\
e_{0,860} &=& \frac{2^{14}}{45}\, \gamma\,,
\\
e_{0,850} &=& \frac{2^{14}}{15}\, \gamma^2
+ \frac{2^{11}}{15}\,\zeta(2)\,,
\\
e_{0,840} &=& \frac{2^{14}}{9}\,\gamma^3 +
\frac{2^{11}}{3}\,\gamma\,\zeta(2) +
\frac{2^{9}}{9}\,\zeta(3)\,,
\\
e_{0,830} &=& \frac{2^{14}}{9}\,\gamma^4 +
\frac{2^{12}}{3}\,\gamma^2\,\zeta(2) +
\frac{2^{11}}{9}\,\gamma\,\zeta(3) +
\frac{704}{3}\,\zeta(4) \,,
\\
e_{0,820} &=& \frac{2^{14}}{15}\,\gamma^5+
\frac{2^{12}}{3}\,\gamma^3\,\zeta(2) +
\frac{2^{10}}{3}\,\gamma^2\,\zeta(3) +
\frac{2^7}{3}\,\zeta(2)\,\zeta(3) \nonumber\\
& & + 704\,\gamma\,\zeta(4) + \frac{2^5}{5}\,\zeta(5) \,,
\\
e_{0,810} &=& \frac{2^{14}}{45}\,\gamma^6+
\frac{2^{11}}{3}\,\gamma^4\,\zeta(2) +
\frac{2^{11}}{9}\,\gamma^3\,\zeta(3) +
\frac{2^8}{3}\,\gamma\,\zeta(2)\,\zeta(3) +
\frac{2^5}{9}\,\zeta^2(3)
\nonumber\\
& & + 704\,\gamma^2\,\zeta(4) +
\frac{2^6}{5}\,\gamma\,\zeta(5) +
62\,\zeta(6)\,,
\\
e_{0,800} &=& \frac{2^{14}}{315}\,\gamma^7 +
\frac{2^{11}}{15}\,\gamma^5\,\zeta(2) +
\frac{2^9}{9}\,\gamma^4\,\zeta(3) +
\frac{2^7}{3}\,\gamma^2\,\zeta(2)\,\zeta(3) 
\nonumber\\
& & + \frac{2^5}{9}\,\gamma\,\zeta^2(3) +
\frac{704}{3}\,\gamma^3\,\zeta(4) +
\frac{22}{3}\,\zeta(3)\,\zeta(4) +
\frac{2^5}{5}\,\gamma^2\,\zeta(5) 
\nonumber\\
& & + \frac{4}{5}\,\zeta(2)\,\zeta(5) 
+ 62\,\gamma\,\zeta(6) 
+ \frac{1}{7}\,\zeta(7)\,.
\end{eqnarray}
\end{subequations}
\end{scriptsize}
It is intriguing to observe that in all individual terms
contributing to a particular coefficient,
the sum of the power of $\gamma$ and of all 
(integer) arguments of $\zeta$ functions equals
a constant. An analogous pattern may be observed 
in the context of specific sums that enter into
the evaluation of higher-order corrections to 
the vacuum-polarization charge density around a 
nucleus, as used in the derivation of 
equations~(66) and (67) of~\cite{WiKr1956}
(relevant formulas are given in appendix~IV
{\em ibid.}). 

%
% States with $N=1$
%
\section{The Function $\Delta_1(g)$ for States with $N=1$}
\label{sDeltaN1}

In this chapter, we will be concerned with the 
generalization of the function $\Delta(g)$ 
[equation~(\ref{defDelta})] to states
with $N=1$. The perturbation series for $N=1$,
up to order $g^{70}$, has been determined 
analytically based on the recursive techniques 
outlined in \cite{ZJ1981jmplong}. Coefficients are
made available at~\cite{JeHome}.
The instanton coefficients for 
states with $N=1$ are
different from those listed in equations~(\ref{epsiloncoeff}),
(\ref{ecoeff}), (\ref{eE1N0}), (\ref{eE1N0coeff}) and
(\ref{eE2N0}), and in the
Table~\ref{tableeee}, where the states with $N=0$ are considered.
However, it is relatively easy to calculate the coefficients
$e_{1,nkl}$ with $n \leq 2$ and $l \leq 6$ based on the
quantization condition (\ref{egenisum}) and the
explicit formulas for the function $B_{\rm dw}(E,g)$ and $A_{\rm dw}(E,g)$
given in (\ref{BEg8}) and (\ref{AEg8}).
For the states with parity $\epsilon$ and $N=1$, the one-instanton
effect is enhanced by a factor $g^{-1}$ in comparison to the 
$N=0$-states [see also (\ref{defxi}) and (\ref{defchi})]:
\begin{eqnarray}
\label{eE1N1}
E^{(1)}_{\epsilon,1} &=& -\frac{2}{g} \, \epsilon \, \xi(g)\,
\left(1
-\frac{347}{12}\, g
+\frac{5317}{288}\, g^2
-\frac{15995159}{10368}\, g^3
-\frac{28062012119}{497664}\, g^4
\right.
\nonumber\\
& & \left. -\frac{12918255230839}{5971968}\, g^5
-\frac{38332497252543415}{429981696}\, g^6 + {\mathcal O}(g^7)\right)\,.
\end{eqnarray}
The coefficient $e_{1,100}$ is governed by the general 
formula (\ref{eN100}).
The two-instanton energy shift, for states with $N=1$, reads
\begin{scriptsize}
\begin{eqnarray}
\label{eE2N1}
E^{(2)}_{\epsilon,1} & = & \frac{4}{g^2} \, \xi^2(g) \, \left[ \chi(g) \,
\left(1
-\frac{293}{6}\, g
+\frac{39823}{72}\, g^2
-\frac{2767481}{1296}\, g^3
-\frac{902381531}{31104}\, g^4
\right.
\right.
\nonumber\\
& &
\left.
-\frac{233318679457}{186624}\, g^5
-\frac{377478189958993}{6718464}\, g^6 + {\mathcal O}(g^7)\right)
\nonumber\\
& &  
\left(\left(\gamma - 1\right)
+\left(\frac{61}{3}-\frac{293}{6}\,\gamma\right)\, g
+\left(\frac{21179}{72}+\frac{39823}{72}\,\gamma\right)\, g^2
+\left(-\frac{1592699}{648}-\frac{2767481}{1296}\,\gamma\right)\, g^3
\right. 
\nonumber\\
& &  
+\left(\frac{548736791}{31104}-\frac{902381531}{31104}\,\gamma\right)\, g^4
+\left(\frac{33161353697}{93312}-
  \frac{233318679457}{186624}\,\gamma\right)\, g^5
\nonumber\\
& &  
\left.
\left.
+\left(\frac{6148524786991}{6718464}-
  \frac{377478189958993}{6718464}\,\gamma\right)\, g^6 + {\mathcal O}(g^7)\right)
\right]\,.
\end{eqnarray}
\end{scriptsize}
Again, the coefficients $e_{1,210}$ and $e_{1,200}$ are governed by the 
general formulas~(\ref{eN210}) and~(\ref{eN200}).

%
% Instanton Function Delta_1 to g^6
%
\begin{figure}[htb!]
\begin{center}
\begin{minipage}{12.0cm}
\begin{center}
\epsfxsize=121.mm
\epsfysize=85.mm
\centerline{\epsfbox{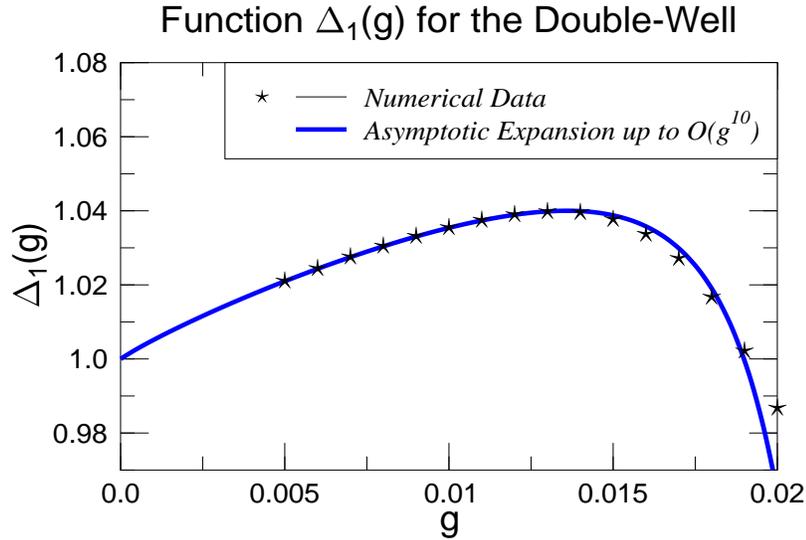}}
\caption{\label{figdelta3}
Double-well potential: Comparison of numerical
data obtained for the function $\Delta_1(g)$ 
defined in (\ref{defDelta1}) with the sum of
the terms up to the order of $g^{10}$
of its asymptotic expansion
for $g$ small, where we express both the 
numerator as well as the denominator of (\ref{defDelta1})
as a power series in $g$. There is good agreement between
numerically determined (``exact'') values (data points) and
the smooth curve given by the analytic asymptotics.
See also Table~\ref{tabcirmi1}.}
\end{center}
\end{minipage}
\end{center}
\end{figure}

We consider the example $g = 0.005$, both in order
to illustrate the calculation and in order to provide reference
values for independent verification.
The energy of the state $(+,1)$ as determined numerically reads:
\begin{equation}
E_{+,1}(0.005) = 
1.46321\,33515\,77109\,24646\,17593\,40427\,75209\,25202\,.
\end{equation}
The energy of the state $(-,1)$ is slightly higher:
\begin{equation}
E_{-,1}(0.005) = 
1.46321\,33515\,95341\,45149\,69757\,27023\,63549\,18578\,.
\end{equation}
The energy difference
\begin{equation}
\Delta E(0.005) = E_{-,1}(0.005) - E_{+,1}(0.005)
\approx 1.823 \times 10^{-11}\,.
\end{equation}
The mean energy is
\begin{equation}
\frac{E_{+,1}(0.005) + E_{-,1}(0.005)}{2} = 
\underline{1.46321\,33515\,86225\,34897}\,93675\,33725\,.
\end{equation} 
The real part of the Borel sum of the perturbation series is
\begin{equation}
{\mathcal B}\left\{E^{(0)}_1 (0.005)\right\} = 
\underline{1.46321\,33515\,86225\,34897}\,88950\,50((1)\,.
\end{equation}
Those decimal figures which agree are underlined. The difference
is due to the two-instanton energy shift.
The definition (\ref{defDelta}) of the 
function $\Delta$ which relates the one- and 
two-instanton effects to the Borel sum of the perturbation series, should 
be modified in order to accommodate for the changed structure
of the leading two-instanton coefficients. We define the 
function $\Delta_1(g)$ as
\begin{equation} 
\label{defDelta1}
\Delta_1 (g) = 4 { \left\{ { 1 \over 2} \,
\left(E_{+,1}+E_{-,1} \right) - 
{\mathcal B}\left\{E^{(0)}_1 (g)\right\} \right\} \over 
\left(E_{+,1}-E_{-,1} \right)^2 
\left[ \ln(2\,g^{-1}) - 1 + \gamma \right]}. 
\end{equation}
In order to determine the leading asymptotics of the function
$\Delta_1 (g)$, we have used the perturbative expansions
about one- and two-instantons given in equations~(\ref{eE1N1})
and~(\ref{eE2N1}). If we additionally perform an 
expansion in inverse powers of the logarithm $\ln(g)$,
we arrive at the result 
\begin{equation} 
\label{asympDelta1}
\Delta_1 (g) = 1 + 9\,g - 
\frac{57}{2} \, \frac{g}{\ln(2/g)} \,
\left( 1 +
\frac{1 - \gamma}{\ln(2/g)} +
{\mathcal O}\left(\frac{1}{\ln^2(2/g)}\right) \right) 
+ {\mathcal O}\left( g^2 \right)\,,
\end{equation}
which generalizes the formula (\ref{asympDelta}) to the 
case $N=1$. Of course, many more terms in the asymptotic
expansion of $\Delta_1(g)$ can be 
determined based on equations~(\ref{eE1N1})
and~(\ref{eE2N1}). Sample values for the function $\Delta_1(g)$ 
are given in Table~\ref{tabcirmi1}, and the function
is plotted against its small-$g$ asymptotic expansion 
in figure~\ref{figdelta3}.

%
% table1
%
\begin{table}[tbh]
\begin{center}
\begin{minipage}{13cm}
\begin{center}
\caption{\label{tabcirmi1} The ratio $ \Delta_1(g) $ as a function
of $ g $. The values obtained from the asymptotic
expansion for $g$ small (up to the order $g^{10}$,
denoted ``asymp.'') are compared
to numerically determined values (denoted as ``num.'').
Useful numerical data for an independent verification
is given in equations~(\ref{example0005}) and~(\ref{example0007}).}
\vspace*{0.3cm}
\begin{tabular}{cr@{.}lr@{.}lr@{.}lr@{.}lr@{.}lr@{.}l%
r@{.}lr@{.}lr@{.}lr@{.}lr@{.}l}
\hline
\hline
\rule[-3mm]{0mm}{8mm} coupling $g$ &
 $0$ & $005$ &
 $0$ & $006$ &
 $0$ & $007$ &
 $0$ & $008$ &
 $0$ & $009$ &
 $0$ & $010$ \\
\hline
$\Delta_1(g)$ num. &
\rule[-3mm]{0mm}{8mm}
 $1$ & $02097$ &
 $1$ & $02434$ &
 $1$ & $02748$ &
 $1$ & $03039$ &
 $1$ & $03304$ &
 $1$ & $03540$ \\
$\Delta_1(g)$ asymp. &
\rule[-3mm]{0mm}{8mm}
 $1$ & $02098$ &
 $1$ & $02434$ &
 $1$ & $02748$ &
 $1$ & $03039$ &
 $1$ & $03305$ &
 $1$ & $03542$ \\
\hline
\hline
\end{tabular}
\end{center}
\end{minipage}
\end{center}
\end{table}

%
% Instanton Function D to g^6
%
\begin{figure}[htb!]
\begin{center}
\begin{minipage}{12.0cm}
\begin{center}
\epsfxsize=121.mm
\epsfysize=85.mm
\centerline{\epsfbox{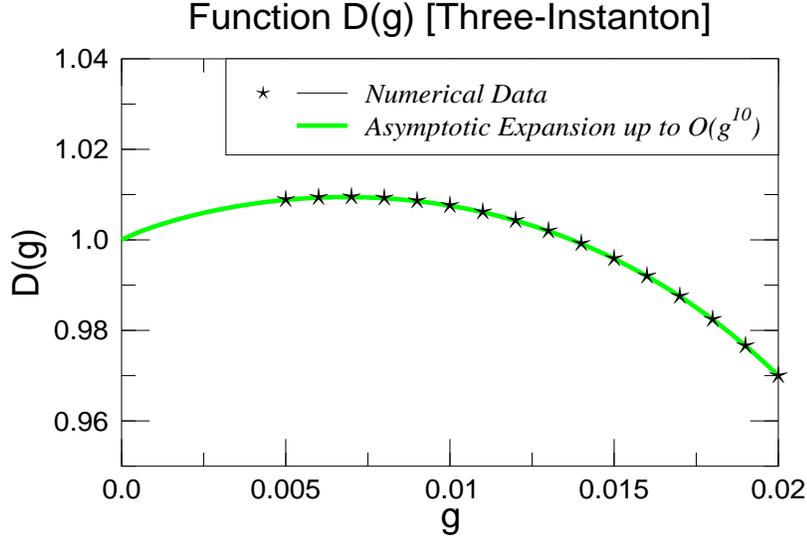}}
\caption{\label{figd}
Double-well potential: Comparison of numerical
data obtained for the function $D(g)$
defined in (\ref{defD}) with the sum of
the terms up to the order of $g^{10}$ 
of its asymptotic expansion for $g$ small.
Again, we express both the numerator as well as the denominator
of the expression defining $D(g)$ [see equation (\ref{defD})]
as a power series in $g$. }
\end{center}
\end{minipage}
\end{center}
\end{figure}

%
% Three--Instanton Effect 
%
\section{Three--Instanton Effects for $N=0$}
\label{sThreeInstanton}

According to the equations~(\ref{defxi}), 
(\ref{einstEnn}) and (\ref{ePthree}),  
the three-instanton shift of the states with $N=0$,
at leading order in $g$,  is given by
\begin{equation}
E^{(3)}_0(g) = -\varepsilon \, \xi^3(g) \,
\left\{ P^0_3[\ln(-2/g)] + {\mathcal O}(g \ln^2(g)) \right\}\,,
\end{equation}
where the expression $P^0_3[\ln(-2/g)]$ reads
\begin{equation}
P^0_3(g) = \frac{3}{2}
\left[\ln\left(-\frac{2}{g}\right) + \gamma\right]^2 +
\frac{\pi^2}{12}\,.
\end{equation}
In order to evaluate the physically relevant part 
of the three-instanton shift, we have to give an interpretation 
to the squared imaginary part generated by the logarithm,
for positive $g$, under the 
analytic continuation of the logarithm to negative argument,
\begin{equation}
\label{eanalytic}
\ln\left(-\frac{2}{g}\right) \to
\ln\left(\frac{2}{g}\right) + {\rm i}\,\pi\,.
\end{equation}
There is a potential problem. The 
explicit imaginary part of the three-instanton shift, generated by the 
analytic continuation of the logarithm, has to cancel
the imaginary part of the (generalized) Borel sum of
the one-instanton shift. In chapter~\ref{ssLargeOrder}, 
this fact has been used 
in order to derive the leading large-order asymptotics of the 
perturbative expansion about one instanton. The 
imaginary part therefore cancels when the resurgent
expansion (\ref{ecomexp}) is summed, and we need not consider
it when evaluating real energy shifts.
On squaring the right-hand side
of (\ref{eanalytic}), we isolate the physically relevant part of 
$P^0_3(g)$ which for $g > 0$ reads:
\begin{equation}
\label{ReP03}
g > 0: \qquad {\rm Re}\,P^0_3(g) = \frac{3}{2}
\left[\ln\left(\frac{2}{g}\right) + \gamma\right]^2 -
\frac{17}{12}\,\pi^2\,.
\end{equation}

%
% table1
%
\begin{table}[tbh]
\begin{center}
\begin{minipage}{13cm}
\begin{center}
\caption{\label{tabfuncd} The ratio $ D(g) $ as a function
of $ g $. The values obtained from the asymptotic
expansion for $g$ small (up to the order $g^8$,
denoted ``asymp.'') are compared
to numerically determined values (see chapter~\ref{ssReference},
values are denoted as ``num.''). The interested reader may verify 
the numerical values based on the data given in 
equations~(\ref{example0005}) and~(\ref{example0007}).}
\vspace*{0.3cm}
\begin{tabular}{cr@{.}lr@{.}lr@{.}lr@{.}lr@{.}lr@{.}l%
r@{.}lr@{.}lr@{.}lr@{.}lr@{.}l}
\hline
\hline
\rule[-3mm]{0mm}{8mm} coupling $g$ &
 $0$ & $005$ &
 $0$ & $006$ &
 $0$ & $007$ &
 $0$ & $008$ &
 $0$ & $009$ &
 $0$ & $010$ \\
\hline
$D(g)$ num. &
\rule[-3mm]{0mm}{8mm}
 $1$ & $00883$ &
 $1$ & $00931$ &
 $1$ & $00944$ &
 $1$ & $00919$ &
 $1$ & $00856$ &
 $1$ & $00754$ \\
$D(g)$ asymp. &
\rule[-3mm]{0mm}{8mm}
 $1$ & $00877$ &
 $1$ & $00920$ &
 $1$ & $00925$ &
 $1$ & $00891$ &
 $1$ & $00816$ &
 $1$ & $00698$ \\
\hline
\hline
\end{tabular}
\end{center}
\end{minipage}
\end{center}
\end{table}

According to the conjecture (\ref{ecomexp}), the 
energy difference $E_{-,0}(g) - E_{+,0}(g)$,
for $g > 0$, is approximately
equal to twice the value of the (positive) one-instanton
energy shift of the state $(-,0)$.
The difference of the quantity $E_{-,0}(g) - E_{+,0}(g)$
and the real part of the Borel sum of the perturbative 
expansion about one instanton is approximately 
given by the three-instanton shift,  
\begin{equation}
E_{-,0}(g) - E_{+,0}(g) - 2\, {\mathcal B}\left\{ E^{(0)}_1(g) \right\}
\approx
2\,E^{(3)}_{-,0}(g)\,,
\end{equation}
where we neglect the five-instanton contribution and recall
that $E^{(3)}_{-,0}(g) > 0$ for $g > 0$. In order
to define a function $D(g)$ in distant analogy to 
(\ref{defDelta}) and (\ref{defDelta1}), we should normalize
to unity in the limit $g \to 0$. 
This can be achieved as follows: According to 
(\ref{eN100}), the quantity $E_{-,0}(g) - E_{+,0}(g)$
is approximately equal to $2 \xi(g)$ in the small-$g$ limit,
where $\xi(g)$ is defined in (\ref{defxi}). The third power
of $E_{-,0}(g) - E_{+,0}(g)$ then compensates the three powers
of $\xi(g)$ comprised by the three-instanton effect (\ref{eE3N0}). 
In order to normalize to unity, the only remaining missing
terms are a prefactor $4$ and an inverse normalization factor 
${\rm Re} P^0_3(g)$ which is given in (\ref{ReP03}). 
We then define the function $D(g)$ as
\begin{equation}
\label{defD}
D(g) = \frac{4}{\displaystyle \frac{3}{2} 
\left[\ln\left(\frac{2}{g}\right) + \gamma\right]^2 - 
\frac{17}{12}\,\pi^2}\,
\frac{E_{-,0} - E_{+,0} - 2\, {\mathcal B}\left\{ E^{(0)}_1(g) \right\}}
{(E_{-,0} - E_{+,0})^3}
\end{equation}
The leading asymptotics for small $g$ read
[we also expand in inverse powers of the logarithm,
as in (\ref{asympDelta}) and~(\ref{asympDelta1})]:
\begin{equation}
\label{asympD}
D (g) = 1 + 6\,g + \frac{g}{\ln(2/g)} \,
\left( -\frac{21}{2} + \frac{63 \, \gamma - 17}{3 \, \ln(2/g)} +
{\mathcal O}\left(\frac{1}{\ln^2(2/g)}\right) \right)
+ {\mathcal O}\left( g^2 \right)\,.
\end{equation}
Of course, more terms in the asymptotic expansion 
of $D(g)$ for small $g$ can be calculated in
a straightforward manner on the basis
of equations (\ref{eE1N0}) and (\ref{eE3N0}).
A table of numerically calculated values of $D(g)$,
and of the values suggested by the expansion up to the order
$g^8$, is given in table~\ref{tabfuncd}, and a graphical
representation is obtained in figure~\ref{figd}.
In the obtaining the table~\ref{tabfuncd} and figure~\ref{figd},
the expansion of $D(g)$ is carried out through the order
of $g^8$, and no additional expansion in inverse powers
of the logarithm $\ln(2 \, g^{-1})$ is performed.

%
% Conclusions
%
\chapter{Conclusions}
\label{conclusions}

The central theme of the current article is the 
discussion of (multi-)instanton effects, which manifest themselves in 
nonperturbative, nonanalytic (in the coupling) contributions
to the path integral which in turn defines the partition 
function (see chapters~\ref{intro},~\ref{BSqf} and~\ref{BSWKB}). 
The expansion about nontrivial saddle
points of the Euclidean action leads naturally 
to energy shifts which involve nonanalytic 
factors of the form $\exp(-a/g)$.  The instanton 
interaction and multi-instanton effects find a natural 
interpretation in terms of classical trajectories along which 
the particle may oscillate between degenerate minima.
A particular detailed
discussion of instantons in the (symmetric) double-well problem  
is provided in chapter~\ref{ssninstdw},
and the considerations are generalized to 
asymmetric wells in chapter~\ref{ssninstg},
to the periodic cosine potential in 
chapter~\ref{ssninstcos}, to radially symmetric 
oscillators (self-adjoint extension and resonances),
and potentials of the 
Fokker--Planck type (chapter~\ref{ssOnuanhar}).
In chapter~\ref{Explicit}, we complement the preceding chapters
by higher-order analytic as well as numerical
calculations related to the double-well
problem, up to the order of eight instantons
(see also appendix~\ref{appWKB}).

The conjectured quantization conditions are
presented in chapter~\ref{sSummary}.
For completeness, we give a list
of the classes of potentials for which 
conjectures on quantization conditions are discussed 
in the current article,
\begin{itemize}
\item the double-well potential (section~\ref{sDouble} and
chapter~\ref{ssninstdw}),
\item more general symmetric potentials with degenerate
minima (section~\ref{sssym} and appendix~\ref{ssSchrodsplit}),
\item a potential with two equal minima but asymmetric wells
(section~\ref{ssasym} and chapter~\ref{ssninstg}),
\item a periodic-cosine potential
(section~\ref{ssPeriodicCosine} and chapter~\ref{ssninstcos}),
\item resonances of the ${\mathcal O}(\nu)$-symmetric anharmonic oscillator,
for negative coupling (sections~\ref{ssAOFP} and~\ref{sselfadjoint}),
\item a special potential which has the property
that the perturbative expansion of the ground-state energy
vanishes to all orders of the coupling constant
(see chapters~\ref{ssAOFP} and~\ref{sfp}),
\item and eigenvalues of the ${\mathcal O}(\nu)$-symmetric 
anharmonic oscillator,
for negative coupling but with the Hamiltonian
endowed with nonstandard boundary conditions
(section~\ref{ssAOresonances} and~\ref{sAOresonances}).
\end{itemize}
Indeed, exact results may be derived for large classes
of analytic potentials, as discussed in chapter~\ref{BSqf}.
The considerations presented in chapters~\ref{sGeneralConsiderations}
and~\ref{ssresoldeg} are important for more general investigations
in later chapters of the current work, and indeed might 
be useful for any further conceivable generalizations.
Specifically, the rather general quantization condition summarized in 
equations (\ref{equant}) and (\ref{econj}) may be 
adapted to large classes of analytic potentials.
The perturbative $B$-function and the instanton $A$-function
are known to higher orders in the coupling $g$, for a 
many of the potentials discussed here.
For the periodic cosine potential [Eqs.~(\ref{eBEgpcg4})
and~(\ref{eAEgpcg4})] and the Fokker--Planck potential
[Eqs.~(\ref{eBEgFPO4}) and~(\ref{eAEgFPO4})],
as well as the ${\mathcal O}(\nu)$ anharmonic oscillator
[Eqs.~(\ref{eBEgjg4}) and (\ref{eAEgjg4})],
we perform higher-order calculations of the $B$- and $A$-functions
up to the order $g^4$.
For the double-well potential, we perform calculations
up to ${\mathcal O}(g^8)$ [see Eqs.~(\ref{BEg8}) and (\ref{AEg8})],
i.e.~up to eight-instanton order.

The investigation of nontrivial saddle points of the 
Euclidean action allows for the identification of finite-action 
contributions to the path integral, the instantons, which may 
be calculated exactly in the limit of a large separation 
of the instanton tunneling configurations
(see the discussion in chapter~\ref{ssInstantonContribution}). 
Approximate quantization conditions may be derived on the basis of related
considerations that evaluate the sum of the 
leading instanton effects, with an arbitrary number
of oscillations on the instanton trajectory but 
considering only the leading term
in the instanton interaction [see equations~(\ref{epoles}),
(\ref{egenpole}), (\ref{ecospole}), (\ref{egenbpole}),
(\ref{eninstFP}), (\ref{eOdeinsti}), and~(\ref{einstResonance})].

We now briefly recall general consequences 
of the resurgent expansion for the 
energy levels of the double-well potential
(\ref{ecomexp}), which reads
\begin{equation} 
\label{ecomexpii}
E_{\varepsilon,N}(g)= 
\sum^{\infty}_{l = 0} E^{(0)}_{N,l} \, g^{l} + 
\sum^{\infty}_{n=1} \left(2\over g\right)^{Nn}\, 
\left( - \varepsilon {\e^{-1/6g}\over\sqrt{\pi g}} \right)^{n} \,
\sum^{n-1}_{k=0} \left\{ \ln\left(-\frac{2}{g}\right) \right\}^k \,
\sum^{\infty}_{l=0} e_{N,nkl} \, g^{l}\,.
\end{equation} 
It is thus clear that 
ordinary perturbation theory, which relies on an expansion
in powers of the coupling constant, is not even qualitatively
sufficient for a description of the energy levels as a function
of the coupling constant $g$. In other words,
whenever there are nontrivial saddle points of the 
classical action, it is not sufficient to consider
perturbation theory about the trivial saddle point 
of the action. Generalized expansions are required which
typically include nonanalytic factors of the form $\exp(-a/g)$
where [see equation (\ref{edefA})]
\begin{equation}
a = 2 \, \int_0^{q_0}\d q\,\sqrt{2V(q)}\,,
\end{equation}
$q$ and $q_0$ are the positions of the degenerate 
minima, and $V$ is the potential.

Rather important differences prevail among cases where there is 
parity symmetry, and asymmetric cases. Indeed, for the 
(symmetric) double-well
potential, the splitting of opposite-parity states is 
given by a one-instanton effect of the 
order of $2 \, \exp[-a/(2\,g)]$.
By contrast, there is no such degeneracy of states 
in the case of asymmetric wells, 
and for the ground-state energy (\ref{efp}) 
of the Fokker--Planck Hamiltonian, we have 
an energy of the order
of $\exp(-a/g)$. For the asymmetric case, 
the classical trajectories have to return
to the original minimum (this corresponds to an even number of
``tunnelings'' between the degenerate minima),
or else the trajectories do not contribute to the path integral.
This is discussed in chapters~\ref{sPathIntegrals}
and~\ref{sOrientationGeneral}.

Let us outline a few unsolved questions related to the 
discussed problems, which may be investigated in the 
future.
\begin{itemize}
\item
First and foremost, the exact WKB- and instanton-inspired
methods as well as the 
numerical investigations discussed here may be extended to 
more general potentials [see the general 
equation (\ref{econj})]. One example is provided 
by potentials with three or more degenerate minima
which interpolate between the cases of two degenerate 
minima on the one hand and the periodic cosine potential 
on the other hand. A further, particularly intriguing case
is the Fokker--Planck potential (see chapter~\ref{sfp}) 
for which the perturbation series vanishes to all orders
(the manifestly positive ground-state energy is determined 
in this case by instanton effects).
\item
Second, there is a certain unsolved issue related
to the complete (re-)summation of the instanton expansion.
Indeed, the nonperturbative factor $\exp(-a/g)$ may assume 
rather large values for moderate values of $g$.
In the specific case (\ref{ecomexp}), while the power series
\begin{equation}
\sum^{\infty}_{l=0} e_{N,nkl} \, g^{l} 
\end{equation}
find a natural summation procedure in terms of the Borel 
method in complex directions, the large-order 
properties of the expansion in powers of 
\begin{equation}
{\e^{-1/6g}\over\sqrt{\pi g}}
\end{equation}
are largely unknown. The explicit calculations in chapter~\ref{Explicit}
might be helpful for a verification of large-order estimates
[regarding the expansion in $\exp(-1/6g)$]. The same applies to 
the entries of Table~\ref{tableeee}.
\item
Third, as outlined in chapter~\ref{ssPerturbative}, 
there is an interesting connection
to fundamental properties 
of Borel summability. Note that even in situations where
the perturbative expansion of eigenvalues is Borel summable,
it is not clear whether the functions $B_i(E,g)$ and 
$A(E,g)$ [see equation (\ref{econj})] are Borel summable
in $g$ at $E$ fixed. Indeed, the wave function $\psi(q)$ is
unambiguously defined only when the quantization condition 
(\ref{equant}) is satisfied
with $N$ a non-negative integer. When $E$ is not an eigenvalue,
the solution of the Schr\"odinger
equation is an undefined linear combination of two particular solutions.
Thus, it may be interesting to investigate the behaviour of 
the functions $B_i(E,g)$ and $A(E,g)$ when $E$ approaches an eigenvalue,
in closer detail. It is conceivable 
that this approach might reveal connections
between the $A$- and the $B$-functions and eventually 
lead to a unified understanding of the structure of these
functions. Again, the explicit calculations in chapter~\ref{Explicit}
might be useful in the context of a verification of such relations,
if they indeed exist.
\item
Fourth, there might be some possibilities for an extension of the 
exact methods discussed in the current article to 
higher-dimensional scenarios where our current understanding is more
limited. There is a well-known analogy between a one-dimensional field 
theory and one-dimensional quantum mechanics, the one-dimensional
field configurations being associated with the 
classical trajectory of the particle. Therefore, model 
problems derived from field-theoretic models via a reduction 
of the dimensionality may find a natural and exact treatment via 
the semi-classical methods discussed in the current article.
Indeed, the loop expansion in field theory corresponds 
to the semi-classical expansion~\cite[chapter~6]{ItZu1980}.
\end{itemize}

We conclude with a few remarks 
on conceivable applications to field theories, 
and in particular non-Abelian gauge field 
theories. These questions follow naturally after the 
famous examples found by Lipatov~\cite{Li1976,Li1977Lett,Li1977}
for scalar theories.
For a study of the non-Abelian case,
one could be tempted to assume that the periodic potential discussed in
chapter~\ref{ssninstcos} might be a good starting point.
However, one of the pre-eminent problems is related
to the scale invariance of the classical 
equations of motion (classical chromodynamics)
of the fields which are invariant
under the transformation $A_\mu(x) \to \lambda \, 
A_\mu(\lambda x)$. In a multi-dimensional 
field theory, instantons are relevant at all possible length
scales (small and large separations), and it is not 
{\em a priori} clear
how to perform the integration  
of the small- and large-scale instantons. 
An explicit infrared cutoff
destroys the gauge invariance of the theory.
For a survey of field-theoretic 
aspects related to  instantons, 
one may consult the following
review articles~\cite{Sh1994,St2002,Fo2001,Fo2002,Fo2000primer}
(the list is necessarily incomplete). 
The nonlinear $\sigma$ model
(see e.g.~\cite[chapter~14]{ZJ1996}),
or the $\phi^4$ theory~\cite[chapter~34]{ZJ1996},
or even manifestly nonrelativistic field theories
might provide better defined scenarios for a systematic,
desirably exact study of instanton effects.
In general, a better and
more systematic understanding of situations which have by now
precluded a thorough analysis, might be gained 
by suitable generalizations of the methods discussed here.
As demonstrated, these have been successfully applied to 
solvable model problems, in one-dimensional quantum mechanics,
which is equivalent to a field theory (point-like in
space, with one time dimension).

%
% Acknowledgements 
%
\section*{Acknowledgements}

The authors would like to acknowledge the Institute of Physics,
University of Heidelberg, for the stimulating atmosphere during
a visit in January 2004, on the occasion of which part of 
this paper was completed, and the Alexander--von--Humboldt Foundation
for support. Roland Rosenfelder is acknowledged for helpful conversations.
The stimulating atmosphere at the National Institute
of Standards and Technology has contributed to
the completion of this project.

\appendix

\renewcommand{\chaptermark}[1]%
{\markboth{\MakeUppercase{APPENDIX~\thechapter: #1\\[2ex]}}{}}

\setcounter{chapter}{5}

%
% WKB calculation and instanton contributions
%
\chapter{WKB Calculation and Instanton Contributions}
\label{appWKB}

%
% An important asymptotic expansion
%
\section{An Important Asymptotic Expansion}

To obtain the perturbative expansion of the function $A(E,g)$ 
(equation (\ref{econj})) we need the WKB expansion 
(\ref{eSWKB})~of $S(q)$. 
To extract $A$ from equation (\ref{econj})~we have first to expand 
the term $ -\ln \Gamma[\frac12-B(E,g)] + B(E,g) \, \ln(-g/2)$ for $B$ large. 
This necessity arises because $B(E,g) \sim E = {\mathcal O}(g^{-1})$
in the context of the WKB expansion.

For the derivative $\psi'$ of 
the $\psi$ function [$\psi(z) = \partial/\partial z 
\ln\Gamma(z)$], the following integral representation 
is known: 
\begin{equation}
\psi'\left(\frac12+z\right)={1\over2}\int_0^\infty 
\frac{t}{\sinh(t/2)}\, e^{-tz} \, \d t =
\frac{\partial^2}{\partial z^2}\,\Gamma\left(\frac12 + z\right)\,.
\end{equation}
For large $z$, the integral is dominated by the region of 
small $t$, and one may therefore derive an asymptotic 
expansion of $\psi'(\frac12+z)$ by expanding the integrand
[excluding the exponential factor $\exp(-t\,z)$] in $t$, and subsequent
integration of the terms resulting from this expansion.
Since $\psi'(z) = \partial^2/\partial z^2
\ln\Gamma(z)$, the integration constants have to 
be adjusted properly.
One obtains the following asymptotic expansion of the
$\Gamma$-function,
\begin{eqnarray}
\label{eGammaAsymp}
\ln\Gamma\left(\frac12+z\right) &=&
z \ln z - z + \frac12 \, \ln(2\pi) - \frac{1}{24 \, z} + \frac{7}{2880 \, z^3}
\nonumber\\
& & - \frac{31}{40320 \, z^5} + \frac{127}{215040 \, z^7} 
- \frac{511}{608256 \, z^7} + {\mathcal O}(z^{-9}).
\end{eqnarray}

%
% Mellin Transform
%
\section{Properties of the Mellin Transform}

The Mellin transform of a function $F$ is commonly defined as
\begin{equation}
M(s) = \int^1_0 \d E \, F(E) \, E^{-s-1}\,.
\end{equation}
A rather useful property of the Mellin transform is this:
If the function $F$ can be  expanded into 
a power series in its argument, then the transform 
$M$ develops poles at positive integer argument,
and the corresponding residues yield the 
coefficients of the power series for $F$. Moreover, for 
functions $F(E)$ which contain logarithmic terms
of the form $E^n \,\ln E$, the Mellin
transform has double poles whose residues, again, give the 
coefficients of the logarithmic terms.

Let us consider a slight generalization of the Mellin
transform with an arbitrary upper limit of integration
$L$. Then,
\begin{equation}
\int^L_0 \d E \, E^n \, E^{-s-1} = \frac{L^{n-s}}{n-s} =
\frac{1}{n - s} + \ln L + {\mathcal O}(n-s)\,.
\end{equation}
The residue at $s = n$ is $-a$.
For a term of the form
\begin{eqnarray}
\int^L_0 \d E \,\, E^n \, \ln E \, E^{-s-1} &=&
- \frac{L^{n-s}}{(n-s)^2} + \frac{L^{n-s} \ln L}{n-s}
\nonumber\\
&=& - \frac{1}{(n - s)^2} + \frac{\ln^2 L}{2} + {\mathcal O}(n-s)\,.
\end{eqnarray}
The coefficient of the leading term in the Laurent series 
about $n = s$ is independent of $L$. This is important
for the considerations presented in the following chapters.

%
% Mellin Transform of the WKB Expansion
%
\section{Mellin Transform of the WKB Expansion}

To calculate the successive terms of $A(E,g)$ in an expansion in powers of $g$
we have then to replace in equation (\ref{econj})~$S_{+}$ by its WKB expansion,
and expand each term for $Eg$, that is $E$ small. A standard method to
obtain this expansion which contains only powers of $E$ and powers of $E$
multiplied by $\ln E$ is to calculate the Mellin transform of integral 
(\ref{econj}).
We thus consider the function
\begin{equation}
M(s) = {1\over g} \, \oint \d E\,\d q\, S_+(q,E,g) \, E^{-s-1}\,.
\end{equation}
One verifies, replacing $S_+$ by its WKB expansion, that the function
$M(s)$ has double poles at integer values of $s$. The residue of the double
pole at $s=n$ yields the coefficient of $-E^n \, \ln E$ in the expansion of
the integral (\ref{econj}) for $g$ small and the residue of the simple pole the
coefficient of $-E^n$: 
\begin{subequations}
\label{eMellin}
\begin{eqnarray}
\label{eMellinPow}
{1 \over n-s} &\mapsto& E^n \,, \\
\label{eMellinLog}
- \left({1 \over n-s}\right)^2 &\mapsto&
E^n\, \ln E. 
\end{eqnarray} 
\end{subequations}
The formulas lead to a proper identification of the terms in the
Mellin transform of the contour integrals of successive orders 
of the WKB expansion, with the expansion in $g$ of the contour integrals
themselves of the successive orders of the WKB expansion. Therefore,
the equation (\ref{eMellin}), for each order in the 
expansion in $g$, is in fact a Mellin backtransformation.

We now investigate the contour integrals of successive 
orders of the perturbative expansion of the WKB expansion (sic!).
We recall that the WKB expansion is an expansion in powers of $g$
at $gE$ fixed [equation (\ref{eSWKB})], and that a general
result for $S_0$ has been given in (\ref{eWKBlead}).
[A general  
second-order result for $g^2\,S_2$ has been given in (\ref{eWKBii}).]
Expansions in $g$ of $S_0$ and $S_2$ are given in (\ref{eS0powerii}) 
and (\ref{eS2powerii}), respectively.
Formulas relevant for the contour integrals are given in 
(\ref{eperti}) and (\ref{epertii}).

We call $I_0(s)$ the leading contribution to $M(s)$ coming from $S_0$,
and $I_2(s)$ the contribution coming from  $g^2\,S_2$:
\begin{equation}
M(s) = I_0(s) + I_2(s) +
{\mathcal O}\left(g^3\right).
\end{equation}
The function $I_0(s)$ is then given by
[see equations (\ref{eS0powerii}) and (\ref{eperti})]:
\begin{eqnarray}
I_0(s) &=& 
-{1 \over ig} \, \oint 
\d E \, \d q \, \left[2gE-U^2(q)\right]^{1/2}\, E^{-s-1}
\nonumber\\
 &=& 
-{g^s \over i} \, \oint 
\d E \, \d q \, \left[2gE-U^2(q)\right]^{1/2}\, (g E)^{-s-1}
\nonumber\\
&=& 2^{s+1}g^{s-1} \, {\Gamma(-s) \, \Gamma(3/2) \over \Gamma(3/2-s)} \,
\int_0^{q_0} \d q\, U^{1-2s}(q)\,,
\label{einstIsi}
\end{eqnarray}
where $q = 0$ and $q = q_0$ are the two minima of the potential. 
The function $I_0(s)$
has a simple pole for $s=0$ and double poles for $s$ positive integer,
indicating the presence of logarithms in higher powers of $g$.

The contribution from $g^2\,S_2$ is [see equation (\ref{econtourii})]:
\begin{equation} 
I_2(s) = {1\over8} \, (2g)^{s+1} \, \Gamma(-s) \,
{\Gamma(-3/2)\over \Gamma(-3/2-s)}\,
\int_0^{q_0}\d q\,U^{-3-2s} \, U'{}^2\,. 
\label{ebarii}
\end{equation} 
The function $I_2(s)$ has a simple pole for $s=-1$ and double poles for
$s$ integer, $s\ge 0$. 

%
% General potentials at leading WKB order
%
\section{General Potentials at Leading WKB Order}
\label{sGenLeadWKB}
 
The residue at $s=0$ of $I_0(s)$ 
as defined in (\ref{einstIsi}) 
is proportional to the instanton action 
(\ref{einstAdegen}): 
\begin{equation}
{2\over g}\int_{0}^{q_0}\d q \,U (q) \equiv {a\over g}\,.
\end{equation}

The residue and the double pole at $s=1$ are also of crucial importance
for the quantization condition (\ref{egeneral}). Indeed,
by calculating the
residues of the double and simple poles of $I_0(s)$ at $s=1$, we obtain 
the coefficients of $E\ln E$ and $E$ and thus terms which are generated only
by the expansion of the $\Gamma$-functions and and the factor $(-2C_i/g)^B$ of
equation (\ref{econj}). We now combine (\ref{eS0powerii}), (\ref{eperti}),
and (\ref{einstIsi}) in order
to evaluate $I_0(s)$. If we denote by $1$ and $\omega^2$
the values of the second derivatives of the potential at the minima
$q = 0$ and $q = q_0$, respectively [see (\ref{eobda})], then
\begin{eqnarray}
2 \int_{0}^{q_0}\d q\,U^{1-2s}(q) &=& 
2\int_{0}^{q_0}\d q\left[U^{1-2s}(q) - q^{1-2s} -
\bigl(\omega(q_0-q)\bigr)^{1-2s} \right]
\nonumber\\
&  & + q_0^{2-2s} \, {1+\omega^{1-2s} \over (1-s)} 
\nonumber\\
& = & 2\int_{0}^{q_0}
\d q \left[ U^{-1}(q) - {1\over q} -
{1 \over \omega(q_0-q)}\right] 
\nonumber\\
& & + {1 \over (1-s)}\, \left(1+{1 \over \omega}\right)+
2 \, \left(1+{1 \over \omega}\right) \, \ln q_0+
2{\ln \omega \over \omega}\,.
\end{eqnarray}
It follows, according to the correspondence 
(\ref{eMellin}), that the contributions 
to the integral on the left-hand side of (\ref{econj})~are
\begin{eqnarray}
\label{ecorr1}
\lefteqn{{E \over \omega} \,
\bigl(\ln(-E/\omega)-1\bigr)+
E \, \bigl(\ln(-E)-1\bigr) }
\nonumber\\
& & + 
E\left[\left(1+{1\over\omega}\right)\, \ln(-g/2)-
2 \, {\ln\omega \over \omega} - 2\ln\widetilde{C}
-2 \, \left(1+{1\over \omega}\right) \ln q_0  \right]\,,
\end{eqnarray}
with
\begin{equation}
\label{etildeCi}
\ln\widetilde{C} =
\int_{0}^{q_0} \d q \left[{1\over U(q)} -
{1\over q}-{1 \over \omega(q_0-q)}\right]\,.
\end{equation}
This result is fully consistent with equations 
(\ref{edefCgen}) and (\ref{edefCgenii}) as well as
(\ref{edefCgeniii}) below,
\begin{equation}
B_1(E,g) = E + {\mathcal O}(g),
\quad 
B_2(E,g) = E/\omega + {\mathcal O}(g),
\quad
C_\omega = 
q_0^2\, \left(\omega^2 \, \widetilde{C}^{2\omega}\right)^{1/(1+\omega)}\,.
\label{etildeCii}
\end{equation} 
We have used here the relation $B(E, g) = E \, (1 + 1/\omega)$
which is valid for asymmetric wells in the convention (\ref{eobda}).
The result implied by this 
calculation, if combined with equations (\ref{equant}) and~(\ref{econj}), 
is fully consistent with the 
quantization condition (\ref{egenpole}), which in turn is the expansion
of the quantization condition(\ref{egeneral})~at
leading order in $g$.

%
% Symmetric potentials: next to leading order
%
\section{Symmetric Potentials: Next--to--Leading WKB Order}
\label{sssymAEg}

We now specialize the treatment of the 
previous chapter to symmetric potentials.
In the case of symmetric potentials with 
degenerate minima, which 
have $\omega_1 = \omega_2 = 1$ in the sense 
of~(\ref{eomega1omega2}) and~(\ref{eobda}), 
one can introduce the parameterization (\ref{einstUpU}).
Then,
\begin{subequations}
\label{einstIIii}
\begin{eqnarray}
\int_{0}^{q_0}
\d q\,U^{1-2s}(q) 
& = & 
2 \, \int_0 ^{u_0} \d u\, {u^{1-2s} \over \sqrt{\rho (u)}}\,,
\label{einstIIiia}
\\
\int_0^{q_0}\d q \, U^{-3-2s}(q) \, U'{}^2(q)
& = & 
2 \, \int_0^{u_0} \d u\, u^{-3-2s} \, \sqrt{\rho(u)}\,,
\label{einstIIiib}
\end{eqnarray}
\end{subequations}
where $u_0$ is the zero of the function $\rho $:
\begin{equation}
\label{edefu0}
\rho(u_0) = 0\,.
\end{equation}

We now consider the residues of $I_0(s)$.
\begin{itemize}
\item The residue at $s=0$ of $I_0(s)$ yields the leading
contribution to $A(E,g)$ [equation (\ref{einstIsi})]:
\begin{equation}
{a\over g}={2\over g}\int_0^{q_0}\d q\, 
U(q)={4\over g}\int_0 ^{u_0}\d u\, {u \over \sqrt{\rho (u)}} ,
\end{equation}
\item The residues of the poles at $s=1$ of $I_0(s)$ 
require slightly more work:
\begin{equation} 
2\int_0 ^{u_0}\d u\, {u^{1-2s}
\over \sqrt{\rho (u)}}=
{1\over 1-s}+2\int_{0}^{u_0}{\d u  \over u}\,
\left({1\over \sqrt{\rho (u)}}-1\right)
+2\ln u_0 +{\mathcal O}(s-1). 
\end{equation}
We set, in accordance with (\ref{edefC}),
\begin{equation}
\label{edefCagain}
\ln C = 2\int_{0}^{u_0} {\d u \over u}\,
\left({1\over \sqrt{\rho (u)}}-1\right)
+ 2\ln u_0\,.
\end{equation}
Combining with the pole of the factor $\Gamma(-s)$, one obtains  the
coefficients of $E\ln(E)$ and $E$, respectively,
\begin{equation} 
2E \, \ln E-2E+2E\ln(g/2C)\,.
\end{equation}

\item We consider the residue at $s=2$ of $I_0(s)$. 
For the term of order $g$ we need the small-$u$ expansion 
(\ref{einstUpU}) of the function $\rho$. Then,  
\begin{eqnarray}
& & 2\,\int_0 ^{u_0}\d u\, {u^{1-2s} \over \sqrt{\rho (u)}}
\nonumber\\
& & = 2\int_0 ^{u_0}\d u\,
\left[u^{1-2s} - \frac12 \, \alpha_1 \, u^{2-2s} +
\left(- \frac12 \alpha_2 + \frac{3}{8} \alpha_1^2\right) \, u^{3-2s}\right]
\nonumber\\
& & \qquad + 2\int_0 ^{u_0}{\d u \over u^3}
\left[ {1\over \sqrt{\rho (u)}} - 1 + \frac12 \alpha_1 u
- \left(-\frac12 \alpha_2+\frac{3}{8} \alpha_1^2\right)u^2\right]
\nonumber\\
& & = \left( -\frac12 \alpha_2+\frac{3}{8} \alpha_1^2\right)
  {1\over 2-s}+\tilde a_{2,2} + {\mathcal O}(s-2)
\end{eqnarray}
with
\begin{eqnarray}
\label{etildea22}
\tilde a_{2,2} &=&
2 \, \int_0^{u_0} {\d u \over u^3}\,
\left[ {1\over \sqrt{\rho (u)}} - 1 + \frac12 \alpha_1 u
- \left(-\frac12 \alpha_2 + \frac{3}{8} \alpha_1^2 \right) \, u^2\right]
\nonumber\\
& & - {1\over u_0^2} + {\alpha_1 \over u_0} +
2 \, \left( - \frac12 \, \alpha_2 + \frac{3}{8} \, \alpha_1^2\right)\,
\ln u_0 \,.
\end{eqnarray}
The factor then yields 
\begin{equation}
2^{s+1} \, g^{s-1} \, 
{\Gamma(-s) \Gamma(3/2) \over \Gamma(3/2-s)} =
-{g\over 2-s} + {g\over 2} + g \ln(g/2) + {\mathcal O}(s-2) . 
\end{equation}
Combining factors, one obtains a contribution to $g E^2 \ln E$:
\begin{equation}
\left( -\frac12 \alpha_2+\frac{3}{8} \alpha_1^2\right)g E^2 \ln E 
\end{equation}
and to $g E^2$:
\begin{equation}
\left[\left(\frac12+\ln\left(\frac{g}{2}\right) \right)
\left( -\frac12 \alpha_2+\frac{3}{8} \alpha_1^2\right)-
\tilde a_{2,2}\right] g E^2\,.
\end{equation}
\end{itemize}

We now consider the residues of $I_2(s)$.
We first rewrite the expression (\ref{ebarii}) in terms of 
the function (\ref{einstUpU}) as
\begin{equation}
I_2(s) = {1\over4} \, (2g)^{s+1} \, \Gamma(-s) \,
{\Gamma(-3/2)\over \Gamma(-3/2-s)} \,
\int_0^{u_0}\d u \, u^{-3-2s} \, \sqrt{\rho(u)}\,.
\end{equation} 
The first residues of $I_2(s)$ may now be evaluated as follows:
\begin{itemize}
\item From $s \to -1$, one obtains the contribution $-1/12E$. 
\item For $s \to 0$, the factor in front of the integral 
in (\ref{ebarii}) has the expansion 
\begin{equation}
\label{ePrefactorSpecial}
-{g\over2 s}-{g\over2}\ln(g/2)-{4g\over3} \,.
\end{equation}
In combining (\ref{einstIIiib}) with the parameterization 
(\ref{einstUpU}), we conclude that the integral itself yields
\begin{eqnarray}
& & \int_0^{u_0}\d u\,u^{-3-2s}\, \sqrt{\rho(u)} 
\nonumber\\
& & = \int_0^{u_0}\d u\,u^{-3-2s}\,
\left[ 1 + \frac12 \, \alpha_1 u + 
\left(\frac12 \, \alpha_2 - \frac{1}{8} \, \alpha_1^2 \right) u^2\right]
\nonumber\\
& & \qquad + \int_0^{u_0}{\d u \over u^3} \,
\left[\sqrt{\rho(u)} - 1 - \frac12 \, \alpha_1 u - 
\left(\frac12 \, \alpha_2 - \frac{1}{8} \alpha_1^2 \right) u^2\right]
\nonumber\\
& & = \left( -{1\over2s} + \ln u_0\right) \,
\left(\frac12 \, \alpha_2 - \frac{1}{8} \, \alpha_1^2 \right) -
{1\over 2 \, u_0^2} - {\alpha_1\over 2\, u_0} 
\nonumber\\
& & \qquad + \int_0^{u_0}{\d u \over u^3} \,
\left[\sqrt{\rho(u)} - 1 - \frac12 \, \alpha_1 u - 
\left(\frac12 \, \alpha_2 - \frac{1}{8} \alpha_1^2 \right) u^2\right]\,.
\end{eqnarray}
Using (\ref{ePrefactorSpecial}) and (\ref{eMellinLog}), 
we infer that the contribution proportional to $g\ln E$ thus is
\begin{equation}
-g\ln E \, \left(\frac{1}{8} \, \alpha_2 - \frac{1}{32} \, \alpha_1^2\right)\,. 
\end{equation}
The coefficient of $g$ (without logarithms) may be deduced from
(\ref{ePrefactorSpecial}) and (\ref{eMellinPow}),
\begin{eqnarray}
& & {1\over2} \, \int_0^{u_0}{\d u \over u^3} \,
\left[\sqrt{\rho(u)} -1 - \frac12 \, \alpha_1 u - 
\left(\frac12 \, \alpha_2 - \frac{1}{8} \alpha_1^2 \right) \, u^2 \right] 
\nonumber\\
& & - {1\over 4 \, u_0^2} - {\alpha_1 \over 4\,u_0} 
- \left(\frac{1}{8} \, \alpha_2 - \frac{1}{32} \, \alpha_1^2 \right) \,
\left[ \ln\left(\frac{g}{2}\right) + \frac{8}{3} - 2\ln u_0 \right]\,.
\end{eqnarray}
\end{itemize}
Meanwhile, we have evaluated certain contributions
to the Mellin transform of the contour integrals 
of the leading orders of the WKB expansion.
These Mellin transforms were denoted by $I_0$ and $I_2$,
and after a Mellin backtransformation (\ref{eMellin}),
these terms enter on the left-hand side of equation (\ref{econj}).
We would now like to regroup these terms 
into a form which is amenable to the 
identification of the functions $A(E, g)$ and $B(E, g)$.
These latter functions are found on the right-hand side of (\ref{econj}). 
Using the  expansion of the $\Gamma$ function and
setting $B(E, g) = E + g \, b_2(E) + {\mathcal O}(g^2)$, one finds 
\begin{eqnarray}
\lefteqn{\frac12\ln(2\pi)-\ln\Gamma\left(\frac12-B(E,g)\right)+
B(E,g) \, \ln\left(-{g\over2C}\right)}
\nonumber\\
& & \sim \bigl[ E + g \, b_2(E) \bigr]\,
\ln\left( {g E \over 2 C} \right) - E - {1 \over 24 E}
+ \cdots\,.
\label{expGam}
\end{eqnarray}
Here, advantage has been taken of the fact that 
the WKB-expansion is an expansion in $g$ at fixed $E g$,
wherefore $E = {\mathcal O}(1/g)$.
In order to obtain $A(E,g)$, one has to subtract this 
contribution twice, because we have a symmetric potential,
with two equal contributions from each of the two wells. 
One sees immediately
that the terms of order $g^0$ cancel and as well as the term
$2 b_2(E) \ln(g E/2)$ 
[see equation (\ref{einstBgengii})], as expected 
from general arguments.

We are now in the position to write down 
explicit expressions for the function $A(E,g)$ at order $g$,
for symmetric potentials which follow the parameterization 
(\ref{einstUpU}).
The contribution at order $g$ is then
\begin{equation}
A(E,g) = {a\over g} +
g \, \left(a_{2,2} \, E^2 + a_{2,0}\right) + {\mathcal O}(g^2)\,,
\end{equation}
where $a_{2,2}$ and $a_{2,0}$ are given by
\begin{subequations}
\begin{eqnarray}
a_{2,2} &=& \left(\frac12+\ln C\right)\,
\left( -\frac12\, \alpha_2+\frac{3}{8} 
\alpha_1^2\right)-\tilde a_{2,2} \,,
\\
a_{2,0} &=& {1\over2}\int_0^{u_0}{\d u \over u^3}\,
\left[\sqrt{\rho(u)} - 1 - \frac12  \alpha_1 u - 
\left(\frac12 \, \alpha_2 - \frac{1}{8} \alpha_1^2 \right) \, u^2 \right] 
\nonumber\\
& & - {1\over 4u_0^2} - {\alpha_1\over4u_0}
+ \left(\frac{1}{8}  \alpha_2 - 
\frac{1}{32} \alpha_1^2 \right) 
\left(\ln C-\frac{8}{3}+2\ln u_0 \right)\,.
\end{eqnarray}
\end{subequations}
The explicit result for $\tilde a_{2,2}$ can be found in 
(\ref{etildea22}).

%
% Explicit evaluation for a Special Family of Potentials
%
\section{Explicit evaluation for a Special Family of Potentials}
\label{sExplicit}

Let us remember at this stage that the basic paradigm for the 
evaluation of the ``instanton function'' $A(E,g)$ that enters into
(\ref{eAdble}) is the following: start from the 
conjectured structure of the WKB expansion (\ref{econj}),
\begin{eqnarray}
& & {1\over g}\oint_{C'}\d z\, S_+(z) =
A(E,g) + \ln(2\pi) \nonumber\\
& & \qquad - \sum_{i=1}^2
\left\{ \ln \Gamma\bigl(\frac12-B_i(E,g)\bigr) +
B_i(E,g)\ln(-g/2C_i) \right\}\,.
\label{econjii}
\end{eqnarray}
from which using (\ref{equantCiii}) 
\begin{equation}
\exp\left[-{1\over g}\,\oint_{C'}\d z\, S_+(z) \right]+1=0\,,
\end{equation}
the quantization condition (\ref{egeneral})
follows immediately. The perturbative expansions $B_i(E,g)$ can in general 
be evaluated easily using the techniques described in
chapter~\ref{ssPerturbative}. The goal is the calculation
of $A(E,g)$.

One may calculate successive orders in the
WKB expansion using the algorithm (\ref{eWKBCalc}),
resulting in an approximation for the left-hand
side of (\ref{econjii}). Because the 
$B_i(E,g)$ on the right-hand side of (\ref{econjii})
may easily be calculated and expanded according 
to (\ref{eGammaAsymp}), it is then possible to 
calculate $A(E,g)$ by subtracting those
terms that are generated by the $B_i(E,g)$ on the 
right-hand side of (\ref{econjii}), from the 
result obtained for the WKB expansion on
the left-hand side, and obtain a result for $A(E,g)$.

For classes of potentials for which the function $\rho $ has a simple form,
the expressions simplify. We thus consider now the class 
(\ref{eclassm}) of potentials which satisfy
$\rho (u)=1-4u^m$. Then, the two integrals (\ref{einstIIiia}) 
and (\ref{einstIIiib}) can be calculated 
explicitly. One finds
\begin{subequations}
\begin{eqnarray}
\int_0 ^{u_0}\d u\, {u^{1-2s}\over \sqrt{1-4u^m}}
&=&
{1\over m} 4^{2(s-1)/m}\,
{\Gamma \bigl({2\over m}(1-s)\bigr)\,
\Gamma\bigl(\frac12\bigr)\over 
\Gamma \bigl(\frac12+{2\over m}(1-s)\bigr)}\,,
\\
\int_0^{u_0}\d u\,u^{-3-2s}\sqrt{1-4u^m}
&=&
{1\over m} 4^{2(s+1)/m}
{\Gamma \bigl(-{2\over m}(1+s)\bigr)\,
\Gamma \bigl(\frac{3}{2}\bigr) \over 
\Gamma \bigl(\frac{3}{2}-{2\over m}(1+s)\bigr)}
\,.
\end{eqnarray}
\end{subequations}
The leading WKB order then yields
\begin{equation}
\label{eI0Res}
I_0(s)= {1\over m}g^{s-1}2^{s(1+4/m)+2-4/m}
  {\Gamma(-s) \Gamma\bigl(\frac{3}{2}\bigr) \over
\Gamma(\frac{3}{2}-s)}{\Gamma\bigl(\frac12\bigr)
  \Gamma\bigl({2\over m}(1-s)\bigr) \over
\Gamma\left[\frac12+{2\over m}(1-s)\right]}\,. 
\end{equation}
At next order one finds
\begin{equation} 
\label{eI2Res}
I_2(s) ={1\over m} \, g^{s+1} \, 2^{s(1+4/m)+4/m-1} \,
{\Gamma(-s) \, \Gamma(-{3\over2}) \over \Gamma(-{3\over2}-s)} \,
{\Gamma\bigl(\frac{3}{2}\bigr) \, \Gamma\bigl(-{2\over m}(1+s)\bigr) \over
\Gamma\bigl({3\over2}-{2\over m}(1+s)\bigr)}\,.   
\end{equation} 
One verifies that the residue of the double pole in $s$ coincide, as expected,
with the terms appearing in the expansion of $B(E,g)$.

({\em The double-well potential.}) This is a symmetric case
in which $B_1 = B_2 = B$ and $C = 1$ 
[see equations (\ref{equantization}),
(\ref{egenisum}), (\ref{eBdble}) and (\ref{edefC})]. 
Note that according to
(\ref{edefu0}),
$u_0 = 1/2$ for the double-well potential.
We recall the relations $U(q) = q\,(1-q)$, as
well as $\rho = \sqrt{{U'}^2}$
and $\rho(u) = \sqrt{1 - 4 u}$ valid for the double-well.
At leading order, one finds (the duplication formula of 
the $\Gamma $ function has been used)
\begin{equation}
I_0(s) = -g^{s-1} \, 2^{3s-1} \,
{\Gamma^2(1-s) \, \Gamma(3/2) \over s \, \Gamma(5/2-2s)}.
\end{equation}
For $s\to0$, one recovers the contribution $1/3g$.\par
Expanding for $s\to 1$ and again using the correspondence 
(\ref{eMellin}), 
one obtains the contributions to (\ref{econj})
\begin{equation}
2 E \ln(-E) - 2E + 2E \, \ln (- g/2)
\sim - 2 \, \ln \Gamma(\frac12 - E) - 2 E \ln(- 2/g ).
\end{equation}
Similarly, for $s\to 2$,
\begin{equation}
6 \, g E^2 \ln\left(\frac{Eg}{2}\right) + 17 \, g E^2
\end{equation} 
and for $s\to 3$:
\begin{equation} 
70 \, g^2 E^3 \ln\left(\frac{E g}{2}\right) + 236 \, g^2 E^3. 
\end{equation}
To obtain the contributions to $A(E,g)$, 
one has to subtract various contributions
of order $g^2 \, E^3$.
One of these originates from the 
expansion of the term 
\begin{eqnarray}
B(E,g) \, \ln B(E,g) &\sim&
g \, b_2(E) \, \ln[E \, (1 + (g/E) \, b_2(E)] 
\nonumber\\
&\sim&
g^2 \, b^2_2(E)/E \sim 3^2 g^2 E^4/E =
9 g^2 E^3 
\end{eqnarray}
[see equations (\ref{defB}) and (\ref{eBdble})].
The term $B(E,g) \, \ln B(E,g)$, in turn,
comes from the expansion (\ref{eGammaAsymp}) of the $\Gamma$-function in 
(\ref{expGam}). The other terms of order
$g^2 E^3$ cancel against each other.
The result for the term of order
$g^2 E^3$, which is $236 - 9 = 227$, 
is in agreement with (\ref{eAdble}).  
In order to obtain the results presented in chapter~\ref{ssCoefficients},
the contributions of order $g^3$ and $g^4$ are useful.
We have from the contribution of $s\to 4$:
\begin{equation}
1155 \, g^3 E^4 \ln\left(\frac{E g}{2}\right) + 
\frac{49843}{12} \, g^3 E^4
\end{equation}
and from $s\to 5$:
\begin{equation}
\frac{45045}{2} \, g^4 E^5 \ln\left(\frac{E g}{2}\right) + 
\frac{335183}{4} \, g^4 E^5\,.
\end{equation}

The complete result for the contour integral of the 
leading WKB term, through the order of $g^6$, is
\begin{eqnarray}
\label{eS0Dbleg6}
\lefteqn{\frac{1}{g} \, \oint \d z \, S_0(q, g, E) =
\frac{1}{3\,g} + 
\left\{ 2 E \ln\left(\frac{E g}{2}\right) - 2E \right\}} 
\nonumber\\
& & +
g \, \left( 6 E^2 \ln\left(\frac{E g}{2}\right) + 17 E^2\right) 
\nonumber\\
& & +
g^2 \, \left( 70 \, E^3 \ln\left(\frac{E g}{2}\right) + 236 E^3 \right) 
\nonumber\\
& & +
g^3 \, \left( 1155 \, E^4 
\ln\left(\frac{E g}{2}\right) + \frac{49843}{12} E^4 \right) 
\nonumber\\
& & +
g^4 \, \left( \frac{45045}{2} \,
E^5 \ln\left(\frac{E g}{2}\right) + 
\frac{335183}{4} E^5 \right)\,.
\nonumber\\
& & +
g^5 \, \left( \frac{11\,056\,741}{6} \,
E^6 \ln\left(\frac{E g}{2}\right) +
\frac{969\,969}{2} E^6 \right)
\nonumber\\
& & +
g^6 \, \left( \frac{515\,954\,137}{12} \,
E^7 \ln\left(\frac{E g}{2}\right) +
\frac{22\,309\,287}{2} E^7 \right) + 
\nonumber\\
& & +
g^7 \, \left( \frac{2\,151\,252\,675}{8} \,
E^8 \ln\left(\frac{E g}{2}\right) +
\frac{469\,212\,586\,743}{448} E^8 \right)
\nonumber\\
& & +
g^8 \, \left( \frac{214\,886\,239\,425}{32} \,
E^9 \ln\left(\frac{E g}{2}\right) +
\frac{70\,860\,581\,490\,397}{2688} E^9 \right)\,, 
\end{eqnarray}
where we neglect terms of order $g^8$ and higher.
At next WKB order 
\begin{equation}
I_2(s)= g^{s+1} 2^{5s+1} \,
{\Gamma(-1-s)\, \Gamma(-{3\over2}) \over
\Gamma(-{3\over2}-s)} \,
{\Gamma(\frac12) \, \Gamma(-1-2s) \over \Gamma(-\frac12-2s)}.
\end{equation}
$I_2(s)$ can be rewritten as
\begin{equation}
I_2(s)=-g^{s+1}2^{3s-1}{\Gamma^2(-s)\Gamma(-\frac12)\over \Gamma(-\frac12-2s)}
{1+\frac{2}{3}s \over 1+s}.
\end{equation}
For $s\to -1$, one obtains a singular term contributing to  the asymptotic
expansion of the $\Gamma$-function:
\begin{equation}
I_2\sim{1 \over12(s+1)}\ \mapsto -{1 \over 12 E}.
\end{equation}
From the expansion for $s\to 0$, one infers
\begin{equation} 
-{g \over 2s^2}\,\left(1-s\ln\left(\frac{2}{g}\right) + {11 \over 3}s\right).
\end{equation}
This yields a contribution to (\ref{econj}):
\begin{equation}
{1\over 2}\, g \ln\left(\frac{Eg}{2}\right) + {11 \over 6}g.
\end{equation}
To obtain a contribution to $A$, one must now take into account the
correction coming from replacing $E$ by $B$ in the expansion 
(\ref{expGam}) of
$\Gamma(\frac12-B)$. At this order only $-1 / (12B)$ contributes:
\begin{equation}
-{1\over 12B} = -{1\over12 E} + {g \over 4}+\cdots.
\end{equation}
It follows that the contribution to $A(E,g)$ is $(11/6)-1/4=19/12$.
Thus, the expansion (\ref{eAdble}) has been verified.

For $s\to 1$, one infers the contribution to (\ref{econj}):
\begin{equation}
{25\over2} g^2 \, E \ln\left(\frac{E g}{2}\right)+ {605\over12} \, g E.
\end{equation}
Now three terms in (\ref{expGam}) 
contribute, involving $B$ up to order $g^2$ and
(equation (\ref{eBdble})) 
one finds $605/12 - 3/2 - 35/12 + 9/12 = 187/4$, in agreement with
(\ref{eAdble}).  

The complete result for the contour integral of the 
WKB term $S_2$, up to the order $g^8$, is
\begin{eqnarray}
\label{eS2Dbleg6}
\lefteqn{\frac{1}{g} \oint \d z \, [g^2 \, S_2(q, g, E)] =
- \frac{1}{12 E}
+ g \, \left( \frac12 \, \ln\left(\frac{E g}{2}\right) + 
\frac{11}{6} \right) }
\nonumber\\
& & +
g^2 \, \left( \frac{25}{2} \, E \ln\left(\frac{E g}{2}\right) + 
\frac{605}{12} E \right) 
\nonumber\\
& & +
g^3 \, \left(  \frac{735}{2} \, E^2 \ln\left(\frac{E g}{2}\right) + 
\frac{4\,522}{3} \, E^2 \right) 
\nonumber\\
& & +
g^4 \, \left( \frac{45\,045}{4} \, E^3 \ln\left(\frac{E g}{2}\right) + 
\frac{743\,439}{16} \, E^3 \right)
\nonumber\\
& & +
g^5 \, \left( \frac{2\,807\,805}{8} \, E^4 \ln\left(\frac{E g}{2}\right) +
\frac{69\,706\,241}{48} \, E^4 \right)
\nonumber\\
& & +
g^6 \, \left( \frac{88\,267\,179}{8} \, E^5 \ln\left(\frac{E g}{2}\right) +
\frac{1\,097\,349\,517}{24} \, E^5 \right) 
\nonumber\\
& & +
g^7 \, \left( \frac{2\,788\,660\,875}{8} \, E^6 \ln\left(\frac{E g}{2}\right) +
\frac{69\,402\,310\,265}{48} \, E^6 \right) 
\nonumber\\
& & +
g^8 \, \left( \frac{353\,522\,522\,925}{32} \, 
E^7 \ln\left(\frac{E g}{2}\right) +
\frac{82\,167\,014\,713\,033}{1792} \, E^7 \right)\,, 
\end{eqnarray}
where again terms of order $g^9$ and higher are neglected.
This result is required for the calculations presented 
in chapter~\ref{ssCoefficients}.

Without further calculational details, we also give here 
the contour integral of $g^4 \, S_4$ up to the
order $g^8$:
\begin{eqnarray}
\label{eS4Dbleg6}
\lefteqn{\frac{1}{g} 
\oint \d z \, [g^4 \, S_4(q, g, E)] =
\frac{7}{1440 E^3} - \frac{11 g}{480 E^2} + 
\frac{101 g^2}{480 E}}
\nonumber\\
& & +
g^3 \, \left( \frac{175}{16} \, \ln\left(\frac{E g}{2}\right) +  
\frac{17\,473}{288} \right)
\nonumber\\
& & +
g^4 \, \left( \frac{31\,185}{32} \, E\, \ln\left(\frac{E g}{2}\right) + 
\frac{616\,601}{128} \, E \right)
\nonumber\\
& & +
g^5 \, \left( \frac{1\,924\,923}{32} \, 
E^2 \, \ln\left(\frac{E g}{2}\right) + 
\frac{544\,644\,431}{1920} \, E^2 \right)
\nonumber\\
& & +
g^6 \, \left(  \frac{100\,553\,453}{32} \,
E^3 \ln\left(\frac{E g}{2}\right) +
\frac{83\,125\,560\,313}{5760} \, E^3 \right)
\nonumber\\
& & +
g^7 \, \left(  \frac{9\,526\,065\,549}{64} \,
E^4 \ln\left(\frac{E g}{2}\right) +
\frac{645\,115\,861\,327}{960} \, E^3 \right)
\nonumber\\
& & +
g^8 \, \left(  \frac{1\,691\,601\,686\,775}{256} \,
E^5 \ln\left(\frac{E g}{2}\right) +
\frac{60\,366\,482\,211\,337}{2048} \, E^5 \right)\,.
\end{eqnarray}
The contour integral of $g^6 \, S_6$,
up to ${\mathcal O}(g^8)$, reads:
\begin{eqnarray}
\label{eS6Dbleg6}
\frac{1}{g} 
& & \oint \d z \, [g^6 \, S_6(q, g, E)] \nonumber\\
& & =
-\frac{31}{20160\,E^5} + 
\frac{87\,g}{4480\, E^4} +
\frac{359\,g^2}{40320\, E^3} -
\frac{15\,g^3}{128\, E^2} + 
\frac{2515\, g^4}{480 E}
\nonumber\\
& & +
g^5 \, \left( \frac{159\,159}{128} \, 
\ln\left(\frac{E g}{2}\right) +  
\frac{59\,665\,081}{7680} \right)
\nonumber\\
& & +
g^6 \, \left( \frac{25\,746\,721}{128} \, 
E\, \ln\left(\frac{E g}{2}\right) +
\frac{25\,285\,094\,891}{23040} \, E \right)
\nonumber\\
& & +
g^7 \, \left( \frac{2\,506\,538\,463}{128} \, 
E^2 \, \ln\left(\frac{E g}{2}\right) +
\frac{5\,392\,814\,329\,807}{53760} \, E^2 \right)
\nonumber\\
& & +
g^8 \, \left(  \frac{758\,382\,964\,625}{512} \,
E^3 \ln\left(\frac{E g}{2}\right) +
\frac{1\,884\,561\,008\,165\,335}{258048} \, E^3 \right)\,.
\end{eqnarray}
The term $g^8 \, S_8$, upon contour integration, yields
to the order $g^8$,
\begin{eqnarray}
\label{eS8Dbleg8}
\lefteqn{\frac{1}{g}
\oint \d z \, [g^8 \, S_8(q, g, E)] =
\frac{127}{107520\,E^7} -
\frac{7381\,g}{322560\, E^6} +
\frac{217\,g^2}{9216\, E^5}}
\nonumber\\
& & -
\frac{3377\,g^3}{215040\, E^4} +
\frac{2195\, g^4}{12288\,E^3} -
\frac{593\,329\, g^5}{12288\,E^2} +
\frac{3344883\, g^6}{20480\,E} 
\nonumber\\
& & +
g^7 \, \left( \frac{692\,049\,787}{2048} \,
\ln\left(\frac{E g}{2}\right) +
\frac{5\,827\,886\,716\,943}{2580480} \right)
\nonumber\\
& & +
g^8 \, \left(  \frac{663\,834\,081\,625}{8192} \,
E \ln\left(\frac{E g}{2}\right) +
\frac{9\,724\,807\,577\,177\,167}{20643840} \, E \right) \,.
\end{eqnarray}
The leading terms in $1/E$ reproduce the coefficients 
of the asymptotic expansion of the $\Gamma$-function
in equation (\ref{eGammaAsymp}).

A similar calculation can immediately be repeated for 
the cosine potential.
Note that for the radial Schr\"odinger equation, similar calculations
are possible with slight technical modifications, 
because the WKB expansion
is valid for both $E$ and $\nu+2l$ large.

%
% Alternative Methods for the Contour Integrals
%
\section{Alternative Methods for the Contour Integrals}
\label{sAlternative}

In chapter~\ref{sExplicit},
we have considered the evaluation of contour integrals
of successive terms in the WKB expansion by the method
of the Mellin transform. In the current chapter,
two alternative methods will be discussed: 
(i) a ``subtraction procedure'' and (ii) an explicit direct evaluation
of the contour integral which is inspired by Lamb-shift
calculations~\cite{JePa2002}.

({\em Method (i).}) We consider the 
double-well problem and use the convention $U(q) = q \, (1-q)$.
Starting from the expansion (\ref{eS0powerii}),
\begin{equation}
S_0(q, g, E) = q \, (1-q) \, \sum_{n=0} 
\left(\frac{2\,g\,E}{q^2\,(1-q)^2} \right)^n\,
{\Gamma(n - \frac12) \over \Gamma(n+1) \, \Gamma(-\frac12)}\,,
\end{equation}
the contour integral of $S_0(q, g, E)$ can be written as
\begin{eqnarray}
\label{edivergence}
& & \frac1g \, \oint \d q\, S_0(q, g, E) =
\frac2g \, \int^1_0 \d q\, S_0(q, g, E) =
\frac2g \, \int^1_0 \d q\, \sqrt{U^2(q) - 2 g E} 
\nonumber\\
& & \qquad = \frac2g \, \int^1_0 \d q\, \sqrt{q^2 \, (1 - q)^2 - 2 g E} 
\nonumber\\
& & \qquad = \frac2g \, 
\sum_{n=0} {\Gamma(n - \frac12) \over \Gamma(n+1) \, \Gamma(-\frac12)}\,
(2\,g\,E)^n \,
\int^1_0 \d q \, q^{1 - 2n} \, (1 - q)^{1-2n}\,.
\end{eqnarray}
Here, we have interchanged the infinite sum with the 
integration, which is not a mathematically rigorous procedure
in the current situation. Moreover, 
formally, the resulting integrals
do not converge for general positive integer $n$,
because of the asymptotic behaviour of the integrand 
near $q \to 0$ and $q \to 1$.
However, it is easy to evaluate the integral in terms of the 
Beta function for arbitrary $n$, and to perform 
an expansion about integer $n$, by setting 
$n \to n + \eta$. The following series results,
\begin{eqnarray}
\lefteqn{\frac{1}{g} \, \oint \d z \, S_0(q, g, E) =
\frac{1}{3\,g} +
\left\{ 2 E \ln\left(\frac{E g}{2}\right) - 2E \right\}} 
\nonumber\\
& & +
g \, \left( 
6 E^2 \left\{ \ln\left(\frac{E g}{2}\right) + 
\frac{1}{\eta} + {\mathcal O}(\eta) \right\} + 
17 E^2\right)
\nonumber\\
& & +
g^2 \, \left(
70 E^3 \left\{ \ln\left(\frac{E g}{2}\right) + 
\frac{1}{\eta} + {\mathcal O}(\eta) \right\} + 
236 E^3 \right)
\nonumber\\
& & +
g^3 \, \left( 1155 E^4 \left\{ \ln\left(\frac{E g}{2}\right) + 
\frac{1}{\eta} + {\mathcal O}(\eta) \right\} + 
\frac{49843}{12} E^4 \right)
\nonumber\\
& & +
g^4 \, \left( 
\frac{45045}{2} E^5 \left\{ \ln\left(\frac{E g}{2}\right) +
\frac{1}{\eta} + {\mathcal O}(\eta) \right\} + 
\frac{335183}{4} E^5 \right) + {\mathcal O}(g^5)\,.
\end{eqnarray}
When employing the prescription of neglecting the 
divergent terms in $\eta$, the result (\ref{eS0Dbleg6}) is recovered.
This strategy 
is in formal analogy to the ``minimal subtraction'' procedure---however,
we here do not change the dimensionality of the integration.

({\em Method (ii).}) We now discuss the applicability of the 
so-called $\epsilon$-method for the evaluation of the 
contour integrals.  This method has been used
for the evaluation of integrals which give rise to 
logarithmic terms in addition to the usual power terms 
in an asymptotic expansion
(for a discussion, including a number of 
illustrative examples, and further references,
the reader may consider~\cite{JePa2002,JeKePa2002}).
We observe that the logarithms originate because of the 
integration regions near $q=0$ and near $q=1$ in~(\ref{edivergence}).
The divergences are cut off 
naturally at the scale $q \approx g\,E$, but actual
divergences result when the expansion in powers
of $g\,E$ is performed.
We may therefore consider to introduce an arbitrary
separation parameter $\epsilon$ which separates
the regions $(0,\epsilon)$ and $(1-\epsilon,1)$ on the 
one side from the region  $(\epsilon, 1-\epsilon)$ on the 
other side. In the latter region, we may safely expand in powers
of $g\,E$, whereas in the first region, we may expand in $q$ 
(or $1-q$) which is effectively an expansion in $g\,E$
(this is in analogy to the example considered in appendix
A of~\cite{JePa2002}). The arbitrary parameter $\epsilon$
cancels if the results of the integrations are expanded 
first in $g\,E$, then in $\epsilon$.

The integration regions $(\epsilon, 1-\epsilon)$  gives the 
following contribution:
\begin{eqnarray}
\lefteqn{\frac{2}{g} \,
\int^{1-\epsilon}_\epsilon {\rm d}q\, \sqrt{q^2 \, (1-q)^2 - 2 g E} }
\nonumber\\
&=&
\frac{1}{3\,g} + 4 E \,\ln \epsilon 
\nonumber\\
&& + g \, E^2 \left[ 7 + 12 \,\ln\epsilon -  
\frac{6}{\epsilon} - \frac{1}{\epsilon^2} \right]
\nonumber\\
&& + g^2 \, E^3 \left[ \frac{533}{6} +
140 \, \ln\epsilon - \frac{70}{\epsilon} - \frac{15}{\epsilon^2} -
\frac{10}{3\,\epsilon^3} - \frac{1}{2\,\epsilon^4} 
\right]
\nonumber\\
&& + g^3 \, E^4 \left[
\frac{18107}{12} - 2310 \, \ln\epsilon - 
\frac{1155}{\epsilon} - \frac{525}{2\,\epsilon^2} -
\frac{70}{\epsilon^3} - \frac{35}{2\,\epsilon^4} -
\frac{7}{2\,\epsilon^5} - \frac{5}{12\,\epsilon^6} 
\right]
\nonumber\\
&& + g^4 \, E^5 \left[ \frac{477745}{16} + 45045 \, \ln\epsilon -
\frac{45045}{2\,\epsilon} - \frac{21021}{4\,\epsilon^2} \right.
\nonumber\\
& & \qquad \qquad 
- \left. \frac{3003}{2\,\epsilon^3} - \frac{3465}{8\,\epsilon^4} -
\frac{231}{2\,\epsilon^5} - \frac{105}{4\,\epsilon^6} -
\frac{9}{2\,\epsilon^7} - \frac{7}{16\,\epsilon^8} 
\right]
\end{eqnarray}
We should add the result from the regions $(0,\epsilon)$ and $(1-\epsilon,1)$:
\begin{scriptsize}
\begin{eqnarray}
\lefteqn{\frac{2}{g} \,
\left( \int^\epsilon_0 + 
\int^1_{1 - \epsilon} \right) {\rm d}q\, \sqrt{q^2 \, (1-q)^2 - 2 g E} }
\nonumber\\
&=& 2 E \, 
\left[ \ln\left( \frac{E g}{2} \right) - 1 - 2\,\ln \epsilon \right]
\nonumber\\
&& + g \, E^2 \left[ 6 \ln\left( \frac{E g}{2} \right) +
10 - 12 \,\ln\epsilon + 
\frac{6}{\epsilon} + \frac{1}{\epsilon^2} \right]
\nonumber\\
&& + g^2 \, E^3 \left[ 70 \, \ln\left( \frac{E g}{2} \right) +
\frac{883}{6} - 140 \, \ln\epsilon + 
\frac{70}{\epsilon} + \frac{15}{\epsilon^2} +
\frac{10}{3\,\epsilon^3} + \frac{1}{2\,\epsilon^4} 
\right]
\nonumber\\
&& + g^3 \, E^4 \left[ 1155 \, \ln\left( \frac{E g}{2} \right) +
\frac{7934}{3} - 2310 \, \ln\epsilon + 
\frac{1155}{\epsilon} + \frac{525}{2\,\epsilon^2} +
\frac{70}{\epsilon^3} + \frac{35}{2\,\epsilon^4} +
\frac{7}{2\,\epsilon^5} + \frac{5}{12\,\epsilon^6} 
\right]
\nonumber\\
&& + g^4 \, E^5 \left[ 45045 \, \ln\left( \frac{E g}{2} \right) +
\frac{862987}{16} - 45045 \, \ln\epsilon +
\frac{45045}{2\,\epsilon} + \frac{21021}{4\,\epsilon^2} \right.
\nonumber\\
& & \qquad \qquad 
+ \left. \frac{3003}{2\,\epsilon^3} + \frac{3465}{8\,\epsilon^4} +
\frac{231}{2\,\epsilon^5} + \frac{105}{4\,\epsilon^6} +
\frac{9}{2\,\epsilon^7} + \frac{7}{16\,\epsilon^8} +
\right]\,.
\end{eqnarray}
\end{scriptsize}
Adding the contributions, the dependence on 
$\epsilon$ cancels, and the result (\ref{eS0Dbleg6})
is recovered.

%
% Four-Instanton Coefficients 
%
\section{Four--Instanton Coefficients}
\label{sFourInstanton}

\begin{scriptsize}
We give here the analytic expression for the four-instanton 
shift of the energy eigenvalues of the double-well oscillator
up to eighth order in $g$. There are four infinite series in $g$,
each multiplying a specific 
power of the logarithmic factor $\lambda(g) \equiv \ln(-2/g)$.
The shift $E^{(4)}_0(g)$ is the sum 
\begin{equation}
E^{(4)}_0(g) = \sum_{k=0}^3 L_k(g) \,,
\end{equation}
with the $L_k(g)$ given below. The leading term, for small $g$,
is of the order $\xi^4(g) \lambda^3(g)$,
\begin{eqnarray}
L_3(g) & = & \xi^4(g) \lambda^3(g) \,
\left\{\frac{8}{3}
-\frac{352}{9}\, g
+\frac{1156}{27}\, g^2
-\frac{133394}{243}\, g^3
-\frac{16276129}{1458}\, g^4
-\frac{550613176}{2187}\, g^5 \right. 
\nonumber\\
& & 
\left.
-\frac{123969635693}{19683}\, g^6
-\frac{40904182753031}{236196}\, g^7
-\frac{14705773688939005}{2834352}\, g^8 + {\mathcal O}(g^9)\right\}\,.
\end{eqnarray}
$L_2(g)$ is given by
\begin{eqnarray}
L_2(g) &=& \xi^4(g) \, \lambda^2(g)\,
\left\{8\,\gamma
+\left(-80-\frac{352}{3}\,\gamma\right)\, g
+\left(\frac{1522}{3}+\frac{1156}{9}\,\gamma\right)\, g^2
\right.
\nonumber\\
& & 
+\left(-\frac{5350}{9}-\frac{133394}{81}\,\gamma\right)\, g^3
+\left(-\frac{3332681}{162}-\frac{16276129}{486}\,\gamma\right)\, g^4
\nonumber\\
& & 
+\left(-\frac{158239171}{243}-\frac{550613176}{729}\,\gamma\right)\, g^5
+\left(-\frac{117850396597}{5832}
- \frac{123969635693}{6561}\,\gamma\right)\, g^6
\nonumber\\
& & 
+\left(-\frac{4249565076178}{6561}
-\frac{40904182753031}{78732}\,\gamma\right)\, g^7
\nonumber\\
& & 
\left.
+\left(-\frac{1717917660985273}{78732}-
\frac{14705773688939005}{944784}\,\gamma\right)\, g^8 + 
{\mathcal O}(g^9)\right\}\,.
\end{eqnarray}
The coefficients entering into $L_1(g)$ are more complex,
\begin{eqnarray}
L_1(g) &=& \xi^4(g) \, \lambda(g) 
\left\{\left(8\,{\gamma}^2+2\,\zeta(2)\right)
+\left(-34-160\,\gamma-\frac{352}{3}\,{\gamma}^2-
  \frac{88}{3}\,\zeta(2)\right)\, g \right.
\nonumber\\
& & 
+\left(\frac{2564}{3}+\frac{3044}{3}\,\gamma+
  \frac{1156}{9}\,{\gamma}^2+\frac{289}{9}\,\zeta(2)\right)\, g^2
\nonumber\\
& & 
+\left(-\frac{9781}{18}-\frac{10700}{9}\,\gamma-
   \frac{133394}{81}\,{\gamma}^2-\frac{66697}{162}\,\zeta(2)\right)\, g^3
\nonumber\\
& & 
+\left(-\frac{153739}{648}-\frac{3332681}{81}\,\gamma-
   \frac{16276129}{486}\,{\gamma}^2-
   \frac{16276129}{1944}\,\zeta(2)\right)\, g^4
\nonumber\\
& & 
+\left(-\frac{2623137449}{9720}-\frac{316478342}{243}\,\gamma-
   \frac{550613176}{729}\,{\gamma}^2-
   \frac{137653294}{729}\,\zeta(2)\right)\, g^5
\nonumber\\
& & 
+\left(-\frac{824874353753}{58320}-\frac{117850396597}{2916}\,\gamma-
   \frac{123969635693}{6561}\,{\gamma}^2-
   \frac{123969635693}{26244}\,\zeta(2)\right)\, g^6 
\nonumber\\
& & 
+\left(-\frac{2215465381303411}{3674160}-
      \frac{8499130152356}{6561}\,\gamma \right.
\nonumber\\
& & 
\left. - \frac{40904182753031}{78732}\,{\gamma}^2-
      \frac{40904182753031}{314928}\,\zeta(2)\right)\, g^7
\nonumber\\
& & 
+\left(-\frac{1079940312204816343}{44089920}-
      \frac{1717917660985273}{39366}\,\gamma\right.
\nonumber\\
& & \left. \left. - \frac{14705773688939005}{944784}\,{\gamma}^2-
      \frac{14705773688939005}{3779136}\,\zeta(2)\right)\, g^8 + 
{\mathcal O}(g^9)\right\}\,.
\end{eqnarray}
The series $L_0(g)$ is free of any logarithms,
\begin{eqnarray}
L_0(g) &=& \xi^4(g) \, 
\left\{\left(\frac{8}{3}\,{\gamma}^3+2\,\gamma\,\zeta(2)+
  \frac{1}{3}\,\zeta(3)\right)
 \right.
\nonumber\\
& & + \left(-34\,\gamma-80\,{\gamma}^2-\frac{352}{9}\,{\gamma}^3-
  20\,\zeta(2)-\frac{88}{3}\,\gamma\,\zeta(2)-
  \frac{44}{9}\,\zeta(3)\right)\, g 
\nonumber\\
& & 
+ \left(\frac{555}{2}+\frac{2564}{3}\,\gamma+
  \frac{1522}{3}\,{\gamma}^2+\frac{1156}{27}\,{\gamma}^3+
  \frac{761}{6}\,\zeta(2)+\frac{289}{9}\,\gamma\,\zeta(2)+
  \frac{289}{54}\,\zeta(3)\right)\, g^2
\nonumber\\
& & 
+ \left(-\frac{4931}{4}-\frac{9781}{18}\,\gamma-
  \frac{5350}{9}\,{\gamma}^2-\frac{133394}{243}\,{\gamma}^3-
  \frac{2675}{18}\,\zeta(2)-\frac{66697}{162}\,\gamma\,\zeta(2)-
  \frac{66697}{972}\,\zeta(3)\right)\, g^3
\nonumber\\
& & 
+ \left(\frac{98395}{48}-\frac{153739}{648}\,\gamma-
  \frac{3332681}{162}\,{\gamma}^2-\frac{16276129}{1458}\,{\gamma}^3-
\right.
\nonumber\\
& & \left. \qquad
  \frac{3332681}{648}\,\zeta(2)-\frac{16276129}{1944}\,\gamma\,\zeta(2)-
  \frac{16276129}{11664}\,\zeta(3)\right)\, g^4 
\nonumber\\
& & 
+ \left(\frac{5322143}{108}-\frac{2623137449}{9720}\,\gamma-
  \frac{158239171}{243}\,{\gamma}^2-\frac{550613176}{2187}\,{\gamma}^3
\right.
\nonumber\\
& & \left. \qquad -
  \frac{158239171}{972}\,\zeta(2)-\frac{137653294}{729}\,\gamma\,\zeta(2)-
  \frac{68826647}{2187}\,\zeta(3)\right)\, g^5
\nonumber\\
& & + \left(-\frac{1803744497}{2592}-\frac{824874353753}{58320}\,\gamma-
  \frac{117850396597}{5832}\,{\gamma}^2-
  \frac{123969635693}{19683}\,{\gamma}^3\right.
\nonumber\\
& & \left. \qquad - 
  \frac{117850396597}{23328}\,\zeta(2)-
  \frac{123969635693}{26244}\,\gamma\,\zeta(2)-
  \frac{123969635693}{157464}\,\zeta(3)\right)\, g^6
\nonumber\\
& &  + \left(-\frac{4001442382153}{38880}-
  \frac{2215465381303411}{3674160}\,\gamma-
  \frac{4249565076178}{6561}\,{\gamma}^2-
  \frac{40904182753031}{236196}\,{\gamma}^3\right.
\nonumber\\
& & \left. \qquad - \frac{2124782538089}{13122}\,\zeta(2)-
  \frac{40904182753031}{314928}\,\gamma\,\zeta(2)-
  \frac{40904182753031}{1889568}\,\zeta(3)\right)\, g^7
\nonumber\\
& & 
+ \left(-\frac{6253193288225413}{979776}-
  \frac{1079940312204816343}{44089920}\,\gamma-
  \frac{1717917660985273}{78732}\,{\gamma}^2
\right.
\nonumber\\
& & \left. \qquad - \frac{14705773688939005}{2834352}\,{\gamma}^3-
  \frac{1717917660985273}{314928}\,\zeta(2)\right.
\nonumber\\
& & \left. \left. \qquad - 
  \frac{14705773688939005}{3779136}\,\gamma\,\zeta(2)-
  \frac{14705773688939005}{22674816}\,\zeta(3)\right)\, g^8 + 
  {\mathcal O}(g^9)\right\}\,.
\end{eqnarray}
\end{scriptsize}

%
% Four-Instanton Coefficients 
%
\section{Higher--Order Coefficients}
\label{sHigherInstanton}

Some instanton coefficients have already been presented in
chapters~\ref{ssCoefficients},~\ref{ssLeading} 
as well as in appendix~\ref{sFourInstanton}.
Here, we present numerical data for all coefficients up to eight-instanton
order, and up to seventh order in $g$. The analytic expressions become rather
involved; numerical data in table~\ref{tableeee} exhibits the 
rapid (factorial) growth of the coefficients in higher orders
in $g$, as well as the rapid growth for fixed order in $g$
in higher instanton-order.

\begin{table}[htb]
\begin{center}
\begin{minipage}{13.5cm}
\begin{center}
\begin{tiny}
\begin{tabular}{cc|rlrlrlrlrlrlrlrl}
\hline
\hline 
\rule[-2mm]{0mm}{3.0mm}
  $n$ & $k$ & \multicolumn{2}{c}{$e_{0,nk0}$} & 
\multicolumn{2}{c}{$e_{0,nk1}$} & 
\multicolumn{2}{c}{$e_{0,nk2}$} & 
\multicolumn{2}{c}{$e_{0,nk3}$} &
\multicolumn{2}{c}{$e_{0,nk4}$} &
\multicolumn{2}{c}{$e_{0,nk5}$} & 
\multicolumn{2}{c}{$e_{0,nk6}$} & 
\multicolumn{2}{c}{$e_{0,nk7}$} \\
\hline
\hline
\rule[-1.5mm]{0mm}{3.0mm} 1 & 0 & $ 1.0 \times $ & \hspace{-0.55cm} $10^{0}$ & $- 5.9 \times $ & \hspace{-0.55cm} $10^{0}$ & $- 2.2 \times $ & \hspace{-0.55cm} $10^{1}$ & $- 2.6 \times $ & \hspace{-0.55cm} $10^{2}$ & $- 4.3 \times $ & \hspace{-0.55cm} $10^{3}$ & $- 8.5 \times $ & \hspace{-0.55cm} $10^{4}$ & $- 2.0 \times $ & \hspace{-0.55cm} $10^{6}$ & $- 5.1 \times $ & \hspace{-0.55cm} $10^{7}$ \\ 
\hline 
\rule[-1.5mm]{0mm}{3.0mm} 2 & 0 & $ 5.8 \times $ & \hspace{-0.55cm} $10^{-1}$ & $- 1.7 \times $ & \hspace{-0.55cm} $10^{1}$ & $- 9.2 \times $ & \hspace{-0.55cm} $10^{0}$ & $- 4.7 \times $ & \hspace{-0.55cm} $10^{2}$ & $- 1.1 \times $ & \hspace{-0.55cm} $10^{4}$ & $- 2.6 \times $ & \hspace{-0.55cm} $10^{5}$ & $- 6.9 \times $ & \hspace{-0.55cm} $10^{6}$ & $- 2.0 \times $ & \hspace{-0.55cm} $10^{8}$  \\ 
\rule[-1.5mm]{0mm}{3.0mm} 2 & 1 & $ 1.0 \times $ & \hspace{-0.55cm} $10^{0}$ & $- 8.8 \times $ & \hspace{-0.55cm} $10^{0}$ & $- 1.8 \times $ & \hspace{-0.55cm} $10^{1}$ & $- 2.6 \times $ & \hspace{-0.55cm} $10^{2}$ & $- 4.5 \times $ & \hspace{-0.55cm} $10^{3}$ & $- 9.4 \times $ & \hspace{-0.55cm} $10^{4}$ & $- 2.2 \times $ & \hspace{-0.55cm} $10^{6}$ & $- 5.9 \times $ & \hspace{-0.55cm} $10^{7}$\\ 
\hline 
\rule[-1.5mm]{0mm}{3.0mm} 3 & 0 & $ 1.3 \times $ & \hspace{-0.55cm} $10^{0}$ & $- 4.2 \times $ & \hspace{-0.55cm} $10^{1}$ & $ 2.0 \times $ & \hspace{-0.55cm} $10^{2}$ & $- 4.2 \times $ & \hspace{-0.55cm} $10^{2}$ & $- 1.6 \times $ & \hspace{-0.55cm} $10^{4}$ & $- 4.9 \times $ & \hspace{-0.55cm} $10^{5}$ & $- 1.6 \times $ & \hspace{-0.55cm} $10^{7}$ & $- 5.2 \times $ & \hspace{-0.55cm} $10^{8}$  \\ 
\rule[-1.5mm]{0mm}{3.0mm} 3 & 1 & $ 1.7 \times $ & \hspace{-0.55cm} $10^{0}$ & $- 5.2 \times $ & \hspace{-0.55cm} $10^{1}$ & $ 9.4 \times $ & \hspace{-0.55cm} $10^{1}$ & $- 7.9 \times $ & \hspace{-0.55cm} $10^{2}$ & $- 2.1 \times $ & \hspace{-0.55cm} $10^{4}$ & $- 5.4 \times $ & \hspace{-0.55cm} $10^{5}$ & $- 1.5 \times $ & \hspace{-0.55cm} $10^{7}$ & $- 4.5 \times $ & \hspace{-0.55cm} $10^{8}$  \\ 
\rule[-1.5mm]{0mm}{3.0mm} 3 & 2 & $ 1.5 \times $ & \hspace{-0.55cm} $10^{0}$ & $- 1.8 \times $ & \hspace{-0.55cm} $10^{1}$ & $- 7.6 \times $ & \hspace{-0.55cm} $10^{0}$ & $- 3.5 \times $ & \hspace{-0.55cm} $10^{2}$ & $- 6.7 \times $ & \hspace{-0.55cm} $10^{3}$ & $- 1.4 \times $ & \hspace{-0.55cm} $10^{5}$ & $- 3.5 \times $ & \hspace{-0.55cm} $10^{6}$ & $- 9.4 \times $ & \hspace{-0.55cm} $10^{7}$ \\ 
\hline 
\rule[-1.5mm]{0mm}{3.0mm} 4 & 0 & $ 2.8 \times $ & \hspace{-0.55cm} $10^{0}$ & $- 1.2 \times $ & \hspace{-0.55cm} $10^{2}$ & $ 1.2 \times $ & \hspace{-0.55cm} $10^{3}$ & $- 2.6 \times $ & \hspace{-0.55cm} $10^{3}$ & $- 2.5 \times $ & \hspace{-0.55cm} $10^{4}$ & $- 8.6 \times $ & \hspace{-0.55cm} $10^{5}$ & $- 3.1 \times $ & \hspace{-0.55cm} $10^{7}$ & $- 1.1 \times $ & \hspace{-0.55cm} $10^{9}$  \\ 
\rule[-1.5mm]{0mm}{3.0mm} 4 & 1 & $ 6.0 \times $ & \hspace{-0.55cm} $10^{0}$ & $- 2.1 \times $ & \hspace{-0.55cm} $10^{2}$ & $ 1.5 \times $ & \hspace{-0.55cm} $10^{3}$ & $- 2.5 \times $ & \hspace{-0.55cm} $10^{3}$ & $- 4.9 \times $ & \hspace{-0.55cm} $10^{4}$ & $- 1.6 \times $ & \hspace{-0.55cm} $10^{6}$ & $- 5.2 \times $ & \hspace{-0.55cm} $10^{7}$ & $- 1.7 \times $ & \hspace{-0.55cm} $10^{9}$  \\ 
\rule[-1.5mm]{0mm}{3.0mm} 4 & 2 & $ 4.6 \times $ & \hspace{-0.55cm} $10^{0}$ & $- 1.5 \times $ & \hspace{-0.55cm} $10^{2}$ & $ 5.8 \times $ & \hspace{-0.55cm} $10^{2}$ & $- 1.5 \times $ & \hspace{-0.55cm} $10^{3}$ & $- 4.0 \times $ & \hspace{-0.55cm} $10^{4}$ & $- 1.1 \times $ & \hspace{-0.55cm} $10^{6}$ & $- 3.1 \times $ & \hspace{-0.55cm} $10^{7}$ & $- 9.5 \times $ & \hspace{-0.55cm} $10^{8}$  \\ 
\rule[-1.5mm]{0mm}{3.0mm} 4 & 3 & $ 2.7 \times $ & \hspace{-0.55cm} $10^{0}$ & $- 3.9 \times $ & \hspace{-0.55cm} $10^{1}$ & $ 4.3 \times $ & \hspace{-0.55cm} $10^{1}$ & $- 5.5 \times $ & \hspace{-0.55cm} $10^{2}$ & $- 1.1 \times $ & \hspace{-0.55cm} $10^{4}$ & $- 2.5 \times $ & \hspace{-0.55cm} $10^{5}$ & $- 6.3 \times $ & \hspace{-0.55cm} $10^{6}$ & $- 1.7 \times $ & \hspace{-0.55cm} $10^{8}$  \\ 
\hline 
\rule[-1.5mm]{0mm}{3.0mm} 5 & 0 & $ 7.1 \times $ & \hspace{-0.55cm} $10^{0}$ & $- 3.7 \times $ & \hspace{-0.55cm} $10^{2}$ & $ 5.4 \times $ & \hspace{-0.55cm} $10^{3}$ & $- 2.5 \times $ & \hspace{-0.55cm} $10^{4}$ & $- 3.8 \times $ & \hspace{-0.55cm} $10^{4}$ & $- 1.7 \times $ & \hspace{-0.55cm} $10^{6}$ & $- 6.2 \times $ & \hspace{-0.55cm} $10^{7}$ & $- 2.4 \times $ & \hspace{-0.55cm} $10^{9}$  \\ 
\rule[-1.5mm]{0mm}{3.0mm} 5 & 1 & $ 1.8 \times $ & \hspace{-0.55cm} $10^{1}$ & $- 8.4 \times $ & \hspace{-0.55cm} $10^{2}$ & $ 1.0 \times $ & \hspace{-0.55cm} $10^{4}$ & $- 3.4 \times $ & \hspace{-0.55cm} $10^{4}$ & $- 1.2 \times $ & \hspace{-0.55cm} $10^{5}$ & $- 4.0 \times $ & \hspace{-0.55cm} $10^{6}$ & $- 1.4 \times $ & \hspace{-0.55cm} $10^{8}$ & $- 5.3 \times $ & \hspace{-0.55cm} $10^{9}$  \\ 
\rule[-1.5mm]{0mm}{3.0mm} 5 & 2 & $ 2.1 \times $ & \hspace{-0.55cm} $10^{1}$ & $- 8.2 \times $ & \hspace{-0.55cm} $10^{2}$ & $ 7.7 \times $ & \hspace{-0.55cm} $10^{3}$ & $- 1.7 \times $ & \hspace{-0.55cm} $10^{4}$ & $- 1.3 \times $ & \hspace{-0.55cm} $10^{5}$ & $- 4.2 \times $ & \hspace{-0.55cm} $10^{6}$ & $- 1.4 \times $ & \hspace{-0.55cm} $10^{8}$ & $- 4.8 \times $ & \hspace{-0.55cm} $10^{9}$  \\ 
\rule[-1.5mm]{0mm}{3.0mm} 5 & 3 & $ 1.2 \times $ & \hspace{-0.55cm} $10^{1}$ & $- 4.1 \times $ & \hspace{-0.55cm} $10^{2}$ & $ 2.4 \times $ & \hspace{-0.55cm} $10^{3}$ & $- 4.9 \times $ & \hspace{-0.55cm} $10^{3}$ & $- 8.0 \times $ & \hspace{-0.55cm} $10^{4}$ & $- 2.2 \times $ & \hspace{-0.55cm} $10^{6}$ & $- 6.6 \times $ & \hspace{-0.55cm} $10^{7}$ & $- 2.0 \times $ & \hspace{-0.55cm} $10^{9}$ \\ 
\rule[-1.5mm]{0mm}{3.0mm} 5 & 4 & $ 5.2 \times $ & \hspace{-0.55cm} $10^{0}$ & $- 9.2 \times $ & \hspace{-0.55cm} $10^{1}$ & $ 2.4 \times $ & \hspace{-0.55cm} $10^{2}$ & $- 1.1 \times $ & \hspace{-0.55cm} $10^{3}$ & $- 2.0 \times $ & \hspace{-0.55cm} $10^{4}$ & $- 4.7 \times $ & \hspace{-0.55cm} $10^{5}$ & $- 1.2 \times $ & \hspace{-0.55cm} $10^{7}$ & $- 3.4 \times $ & \hspace{-0.55cm} $10^{8}$ \\ 
\hline 
\rule[-1.5mm]{0mm}{3.0mm} 6 & 0 & $ 1.9 \times $ & \hspace{-0.55cm} $10^{1}$ & $- 1.2 \times $ & \hspace{-0.55cm} $10^{3}$ & $ 2.3 \times $ & \hspace{-0.55cm} $10^{4}$ & $- 1.6 \times $ & \hspace{-0.55cm} $10^{5}$ & $ 2.2 \times $ & \hspace{-0.55cm} $10^{5}$ & $- 3.4 \times $ & \hspace{-0.55cm} $10^{6}$ & $- 1.3 \times $ & \hspace{-0.55cm} $10^{8}$ & $- 5.3 \times $ & \hspace{-0.55cm} $10^{9}$ \\ 
\rule[-1.5mm]{0mm}{3.0mm} 6 & 1 & $ 5.8 \times $ & \hspace{-0.55cm} $10^{1}$ & $- 3.2 \times $ & \hspace{-0.55cm} $10^{3}$ & $ 5.6 \times $ & \hspace{-0.55cm} $10^{4}$ & $- 3.3 \times $ & \hspace{-0.55cm} $10^{5}$ & $ 1.1 \times $ & \hspace{-0.55cm} $10^{5}$ & $- 1.0 \times $ & \hspace{-0.55cm} $10^{7}$ & $- 3.8 \times $ & \hspace{-0.55cm} $10^{8}$ & $- 1.5 \times $ & \hspace{-0.55cm} $10^{10}$ \\ 
\rule[-1.5mm]{0mm}{3.0mm} 6 & 2 & $ 7.9 \times $ & \hspace{-0.55cm} $10^{1}$ & $- 4.0 \times $ & \hspace{-0.55cm} $10^{3}$ & $ 5.8 \times $ & \hspace{-0.55cm} $10^{4}$ & $- 2.7 \times $ & \hspace{-0.55cm} $10^{5}$ & $- 2.3 \times $ & \hspace{-0.55cm} $10^{5}$ & $- 1.4 \times $ & \hspace{-0.55cm} $10^{7}$ & $- 4.9 \times $ & \hspace{-0.55cm} $10^{8}$ & $- 1.8 \times $ & \hspace{-0.55cm} $10^{10}$\\ 
\rule[-1.5mm]{0mm}{3.0mm} 6 & 3 & $ 6.6 \times $ & \hspace{-0.55cm} $10^{1}$ & $- 2.8 \times $ & \hspace{-0.55cm} $10^{3}$ & $ 3.2 \times $ & \hspace{-0.55cm} $10^{4}$ & $- 1.0 \times $ & \hspace{-0.55cm} $10^{5}$ & $- 2.9 \times $ & \hspace{-0.55cm} $10^{5}$ & $- 1.0 \times $ & \hspace{-0.55cm} $10^{7}$ & $- 3.5 \times $ & \hspace{-0.55cm} $10^{8}$ & $- 1.2 \times $ & \hspace{-0.55cm} $10^{10}$\\ 
\rule[-1.5mm]{0mm}{3.0mm} 6 & 4 & $ 3.1 \times $ & \hspace{-0.55cm} $10^{1}$ & $- 1.2 \times $ & \hspace{-0.55cm} $10^{3}$ & $ 8.9 \times $ & \hspace{-0.55cm} $10^{3}$ & $- 2.1 \times $ & \hspace{-0.55cm} $10^{4}$ & $- 1.6 \times $ & \hspace{-0.55cm} $10^{5}$ & $- 4.7 \times $ & \hspace{-0.55cm} $10^{6}$ & $- 1.4 \times $ & \hspace{-0.55cm} $10^{8}$ & $- 4.5 \times $ & \hspace{-0.55cm} $10^{9}$ \\ 
\rule[-1.5mm]{0mm}{3.0mm} 6 & 5 & $ 1.1 \times $ & \hspace{-0.55cm} $10^{1}$ & $- 2.2 \times $ & \hspace{-0.55cm} $10^{2}$ & $ 9.1 \times $ & \hspace{-0.55cm} $10^{2}$ & $- 2.7 \times $ & \hspace{-0.55cm} $10^{3}$ & $- 3.8 \times $ & \hspace{-0.55cm} $10^{4}$ & $- 9.2 \times $ & \hspace{-0.55cm} $10^{5}$ & $- 2.4 \times $ & \hspace{-0.55cm} $10^{7}$ & $- 7.0 \times $ & \hspace{-0.55cm} $10^{8}$  \\ 
\hline 
\rule[-1.5mm]{0mm}{3.0mm} 7 & 0 & $ 5.3 \times $ & \hspace{-0.55cm} $10^{1}$ & $- 3.7 \times $ & \hspace{-0.55cm} $10^{3}$ & $ 9.1 \times $ & \hspace{-0.55cm} $10^{4}$ & $- 9.3 \times $ & \hspace{-0.55cm} $10^{5}$ & $ 3.3 \times $ & \hspace{-0.55cm} $10^{6}$ & $- 1.0 \times $ & \hspace{-0.55cm} $10^{7}$ & $- 3.0 \times $ & \hspace{-0.55cm} $10^{8}$ & $- 1.2 \times $ & \hspace{-0.55cm} $10^{10}$\\ 
\rule[-1.5mm]{0mm}{3.0mm} 7 & 1 & $ 1.9 \times $ & \hspace{-0.55cm} $10^{2}$ & $- 1.2 \times $ & \hspace{-0.55cm} $10^{4}$ & $ 2.7 \times $ & \hspace{-0.55cm} $10^{5}$ & $- 2.4 \times $ & \hspace{-0.55cm} $10^{6}$ & $ 6.4 \times $ & \hspace{-0.55cm} $10^{6}$ & $- 2.9 \times $ & \hspace{-0.55cm} $10^{7}$ & $- 1.0 \times $ & \hspace{-0.55cm} $10^{9}$ & $- 4.1 \times $ & \hspace{-0.55cm} $10^{10}$\\ 
\rule[-1.5mm]{0mm}{3.0mm} 7 & 2 & $ 3.1 \times $ & \hspace{-0.55cm} $10^{2}$ & $- 1.9 \times $ & \hspace{-0.55cm} $10^{4}$ & $ 3.6 \times $ & \hspace{-0.55cm} $10^{5}$ & $- 2.7 \times $ & \hspace{-0.55cm} $10^{6}$ & $ 4.5 \times $ & \hspace{-0.55cm} $10^{6}$ & $- 4.3 \times $ & \hspace{-0.55cm} $10^{7}$ & $- 1.6 \times $ & \hspace{-0.55cm} $10^{9}$ & $- 6.2 \times $ & \hspace{-0.55cm} $10^{10}$ \\ 
\rule[-1.5mm]{0mm}{3.0mm} 7 & 3 & $ 3.0 \times $ & \hspace{-0.55cm} $10^{2}$ & $- 1.6 \times $ & \hspace{-0.55cm} $10^{4}$ & $ 2.7 \times $ & \hspace{-0.55cm} $10^{5}$ & $- 1.6 \times $ & \hspace{-0.55cm} $10^{6}$ & $ 1.0 \times $ & \hspace{-0.55cm} $10^{6}$ & $- 4.1 \times $ & \hspace{-0.55cm} $10^{7}$ & $- 1.5 \times $ & \hspace{-0.55cm} $10^{9}$ & $- 5.5 \times $ & \hspace{-0.55cm} $10^{10}$ \\ 
\rule[-1.5mm]{0mm}{3.0mm} 7 & 4 & $ 2.0 \times $ & \hspace{-0.55cm} $10^{2}$ & $- 9.3 \times $ & \hspace{-0.55cm} $10^{3}$ & $ 1.2 \times $ & \hspace{-0.55cm} $10^{5}$ & $- 5.3 \times $ & \hspace{-0.55cm} $10^{5}$ & $- 3.5 \times $ & \hspace{-0.55cm} $10^{5}$ & $- 2.6 \times $ & \hspace{-0.55cm} $10^{7}$ & $- 8.8 \times $ & \hspace{-0.55cm} $10^{8}$ & $- 3.1 \times $ & \hspace{-0.55cm} $10^{10}$\\ 
\rule[-1.5mm]{0mm}{3.0mm} 7 & 5 & $ 8.1 \times $ & \hspace{-0.55cm} $10^{1}$ & $- 3.2 \times $ & \hspace{-0.55cm} $10^{3}$ & $ 3.1 \times $ & \hspace{-0.55cm} $10^{4}$ & $- 9.4 \times $ & \hspace{-0.55cm} $10^{4}$ & $- 3.1 \times $ & \hspace{-0.55cm} $10^{5}$ & $- 1.0 \times $ & \hspace{-0.55cm} $10^{7}$ & $- 3.1 \times $ & \hspace{-0.55cm} $10^{8}$ & $- 1.0 \times $ & \hspace{-0.55cm} $10^{10}$\\ 
\rule[-1.5mm]{0mm}{3.0mm} 7 & 6 & $ 2.3 \times $ & \hspace{-0.55cm} $10^{1}$ & $- 5.5 \times $ & \hspace{-0.55cm} $10^{2}$ & $ 3.0 \times $ & \hspace{-0.55cm} $10^{3}$ & $- 8.5 \times $ & \hspace{-0.55cm} $10^{3}$ & $- 7.4 \times $ & \hspace{-0.55cm} $10^{4}$ & $- 1.9 \times $ & \hspace{-0.55cm} $10^{6}$ & $- 5.0 \times $ & \hspace{-0.55cm} $10^{7}$ & $- 1.5 \times $ & \hspace{-0.55cm} $10^{9}$ \\ 
\hline 
\rule[-1.5mm]{0mm}{3.0mm} 8 & 0 & $ 1.5 \times $ & \hspace{-0.55cm} $10^{2}$ & $- 1.2 \times $ & \hspace{-0.55cm} $10^{4}$ & $ 3.6 \times $ & \hspace{-0.55cm} $10^{5}$ & $- 4.7 \times $ & \hspace{-0.55cm} $10^{6}$ & $ 2.7 \times $ & \hspace{-0.55cm} $10^{7}$ & $- 7.3 \times $ & \hspace{-0.55cm} $10^{7}$ & $- 6.6 \times $ & \hspace{-0.55cm} $10^{8}$ & $- 2.9 \times $ & \hspace{-0.55cm} $10^{10}$ \\ 
\rule[-1.5mm]{0mm}{3.0mm} 8 & 1 & $ 6.2 \times $ & \hspace{-0.55cm} $10^{2}$ & $- 4.7 \times $ & \hspace{-0.55cm} $10^{4}$ & $ 1.3 \times $ & \hspace{-0.55cm} $10^{6}$ & $- 1.5 \times $ & \hspace{-0.55cm} $10^{7}$ & $ 7.1 \times $ & \hspace{-0.55cm} $10^{7}$ & $- 1.8 \times $ & \hspace{-0.55cm} $10^{8}$ & $- 2.7 \times $ & \hspace{-0.55cm} $10^{9}$ & $- 1.1 \times $ & \hspace{-0.55cm} $10^{11}$ \\ 
\rule[-1.5mm]{0mm}{3.0mm} 8 & 2 & $ 1.2 \times $ & \hspace{-0.55cm} $10^{3}$ & $- 8.2 \times $ & \hspace{-0.55cm} $10^{4}$ & $ 2.0 \times $ & \hspace{-0.55cm} $10^{6}$ & $- 2.1 \times $ & \hspace{-0.55cm} $10^{7}$ & $ 8.0 \times $ & \hspace{-0.55cm} $10^{7}$ & $- 2.1 \times $ & \hspace{-0.55cm} $10^{8}$ & $- 5.0 \times $ & \hspace{-0.55cm} $10^{9}$ & $- 2.0 \times $ & \hspace{-0.55cm} $10^{11}$ \\ 
\rule[-1.5mm]{0mm}{3.0mm} 8 & 3 & $ 1.4 \times $ & \hspace{-0.55cm} $10^{3}$ & $- 8.8 \times $ & \hspace{-0.55cm} $10^{4}$ & $ 1.9 \times $ & \hspace{-0.55cm} $10^{6}$ & $- 1.7 \times $ & \hspace{-0.55cm} $10^{7}$ & $ 4.7 \times $ & \hspace{-0.55cm} $10^{7}$ & $- 1.8 \times $ & \hspace{-0.55cm} $10^{8}$ & $- 5.7 \times $ & \hspace{-0.55cm} $10^{9}$ & $- 2.2 \times $ & \hspace{-0.55cm} $10^{11}$ \\ 
\rule[-1.5mm]{0mm}{3.0mm} 8 & 4 & $ 1.1 \times $ & \hspace{-0.55cm} $10^{3}$ & $- 6.2 \times $ & \hspace{-0.55cm} $10^{4}$ & $ 1.2 \times $ & \hspace{-0.55cm} $10^{6}$ & $- 8.2 \times $ & \hspace{-0.55cm} $10^{6}$ & $ 1.5 \times $ & \hspace{-0.55cm} $10^{7}$ & $- 1.2 \times $ & \hspace{-0.55cm} $10^{8}$ & $- 4.3 \times $ & \hspace{-0.55cm} $10^{9}$ & $- 1.6 \times $ & \hspace{-0.55cm} $10^{11}$ \\ 
\rule[-1.5mm]{0mm}{3.0mm} 8 & 5 & $ 5.9 \times $ & \hspace{-0.55cm} $10^{2}$ & $- 2.9 \times $ & \hspace{-0.55cm} $10^{4}$ & $ 4.5 \times $ & \hspace{-0.55cm} $10^{5}$ & $- 2.4 \times $ & \hspace{-0.55cm} $10^{6}$ & $ 1.7 \times $ & \hspace{-0.55cm} $10^{6}$ & $- 6.4 \times $ & \hspace{-0.55cm} $10^{7}$ & $- 2.2 \times $ & \hspace{-0.55cm} $10^{9}$ & $- 7.7 \times $ & \hspace{-0.55cm} $10^{10}$ \\ 
\rule[-1.5mm]{0mm}{3.0mm} 8 & 6 & $ 2.1 \times $ & \hspace{-0.55cm} $10^{2}$ & $- 8.9 \times $ & \hspace{-0.55cm} $10^{3}$ & $ 1.0 \times $ & \hspace{-0.55cm} $10^{5}$ & $- 3.9 \times $ & \hspace{-0.55cm} $10^{5}$ & $- 4.0 \times $ & \hspace{-0.55cm} $10^{5}$ & $- 2.2 \times $ & \hspace{-0.55cm} $10^{7}$ & $- 7.0 \times $ & \hspace{-0.55cm} $10^{8}$ & $- 2.3 \times $ & \hspace{-0.55cm} $10^{10}$\\ 
\rule[-1.5mm]{0mm}{3.0mm} 8 & 7 & $ 5.2 \times $ & \hspace{-0.55cm} $10^{1}$ & $- 1.4 \times $ & \hspace{-0.55cm} $10^{3}$ & $ 9.7 \times $ & \hspace{-0.55cm} $10^{3}$ & $- 3.0 \times $ & \hspace{-0.55cm} $10^{4}$ & $- 1.4 \times $ & \hspace{-0.55cm} $10^{5}$ & $- 3.9 \times $ & \hspace{-0.55cm} $10^{6}$ & $- 1.1 \times $ & \hspace{-0.55cm} $10^{8}$ & $- 3.2 \times $ & \hspace{-0.55cm} $10^{9}$\\ 
\hline 
\hline
\end{tabular}
\end{tiny}
\caption{\label{tableeee} The $e$-coefficients (instanton
coefficients) determine the resurgent expansion (\ref{ecomexp}) for the 
energy eigenvalues of the quantum mechanical double-well oscillator.
Here, all coefficients up to eighth order in the instanton interaction
(up to the ``eight-instanton order'') and up to seventh order in 
the coupling constant $g$ are considered. Calculations are carried
out for the ground state. We recall that $e_{0,nkl}$ is the coefficient
multiplying the term $\xi(g) \,\lambda(g) \, g^l$. 
The coefficient $e_{0,n(n-1)0}$
multiplies leading term in the $n$-instanton order.
There is a rapid growth of the absolute magnitude 
of the coefficients in higher orders in $g$ as well as in 
higher orders in $n$.}
\end{center}
\end{minipage}
\end{center}
\end{table}

\newpage

\clearpage\fancyhead[R]{}

\end{document}